\documentclass[onecolumn,aps,superscriptaddress,prl,longbibliography]{revtex4-2} 

\usepackage{graphicx}
\usepackage{bm}

\usepackage{xcolor}
\usepackage[colorlinks=true,citecolor=blue,linkcolor=magenta]{hyperref}
\usepackage{orcidlink} 

\usepackage[latin9]{inputenc}
\usepackage{amsmath}
\usepackage{amssymb}  
\usepackage{graphicx}
\usepackage{verbatim}  
\usepackage{bm} 
\usepackage{color}  

\usepackage{graphicx,epsfig,amsfonts,amssymb}
\usepackage{bm}
\usepackage{times}
\usepackage{lipsum}
\usepackage{verbatim}

\newcommand{\la}{\langle}
\newcommand{\ra}{\rangle}
\newcommand{\rar}{\rightarrow}

\newcommand{\bqe}{\begin{eqnarray}}
\newcommand{\eqe}{\end{eqnarray}}

\newcommand{\be}{\begin{equation}}
\newcommand{\ee}{\end{equation}}

\begin{document}
\title{Multivariable Painlev\'e-II equation:  connection formulas for arbitrary system size}
\author{Nikolai A. Sinitsyn\orcidlink{0000-0002-0746-0400}}\email{nsinitsyn@lanl.gov}
\affiliation{Theoretical Division, Los Alamos National Laboratory, Los Alamos, New Mexico 87545, USA} 
\begin{abstract}
Connection formulas for the asymptotic solutions of a system of $n> 1$ coupled Painlev\'e-II equations with symmetry-breaking parameters are written explicitly.  An asymptotically exact WKB approach to these formulas relies on the quantum-mechanical independent crossing approximation for an explicitly time-dependent Schr\"odinger equation.
\end{abstract}

\date{October 2025}
\maketitle
\section{Introduction}
Special functions of theoretical physics are usually solutions of certain differential equations. The most nontrivial among them are the solutions of the six Painlev\'e equations, called Painlev\'e transcendents \cite{Fokas2006}. They generally do not have representations as integrals of elementary functions \cite{note}. Nevertheless, methods for studying asymptotic solutions of Painlev\'e equations have relatively recently emerged \cite{Novokshenov1984DE,ItsKapaev1988ConnectionFormulasPII,BaikBuckinghamDiFrancoIts2009,Kokocki2020,BothnerBuckingham2017}. The major advance was the discovery of the {\it connection formulas}, which relate parameters of the Painlev\'e transcendents at large positive and large negative values of their argument \cite{Fokas2006}.

These discoveries sparked interest among physicists \cite{Itin2009b,Sadhasivam2024,Lee2011ViscousShocks,BanerjeeSinitsyn2023MesoscopicCriticalFluctuations,SinitsynPokrovsky2026QuasiAdiabaticEffects,ItsMezzadriMo2008,ZelenyiNeishtadtArtemyevVainchteinMalova2013, NeishtadtArtemyevTuraev2019,ClercDavilaKowalczykSmyrnelisVidalHenriquez2017,BothnerLiechty2013}. However, this interest has been overshadowed by the power of modern computers, which provide fast numerical solutions of low-order differential equations. Against this background, analytical results for Painlev\'e transcendents simplify certain topics  but do not break the limits of what can be understood numerically.
Mathematical physics should find a path to more complex phenomena. New special functions, described by compact analytical formulas, are needed to characterize processes beyond the reach of modern computers.

This article explores a system of arbitrary $n$ coupled nonlinear differential equations that may serve as an example:
\begin{eqnarray}
\nonumber u_1''(x)&=&xu_1(x)-2u_1(x)\sum_{k=1}^n u_k^2(x)- \varepsilon_1u_1(x),\\
\nonumber u_2''(x)&=&xu_2(x)-2u_2(x)\sum_{k=1}^n u_k^2(x)-\varepsilon_2 u_2(x),\\
\label{P2-n}
&\cdots& \\
\nonumber u_n''(x)&=&xu_n(x)-2u_n(x)\sum_{k=1}^n u_k^2(x)-\varepsilon_n u_n(x).
\end{eqnarray}
Here, $n$  real parameters are assumed to be ordered as
$$
\varepsilon_1<\varepsilon_2<\ldots <\varepsilon_n.
$$
A shift, $x\rightarrow x+\varepsilon_1$,  sets $\varepsilon_1=0$, which will be assumed here.  

The model in Eq.~(\ref{P2-n}) has been recognized as integrable from different perspectives \cite{ClaeysDoeraene2018,Tyagi2025}. However, for multivariable systems, integrability does not guarantee the existence of simple explicite connection formulas. In a Letter \cite{Sinitsyn2026-letter}, the author of the present article announced the existence of such formulas and provided an  example for $n=2$. The present article is a follow-up discussion of these formulas for arbitrary $n\ge 1$.

For $n=1$, the system~(\ref{P2-n}) degenerates into the famous uniform Painlev\'e-II (P-II) equation
\be
u''(x)=xu(x)-2u(x)^3.
\label{P2-def}
\ee
Solving Eq.~(\ref{P2-def}) numerically is not hard but already takes a noticeable amount of time on a laptop. The solution converges to its asymptotic form slowly, with a rapidly changing oscillation frequency.

The complexity of such simulations escalates swiftly as the number of coupled equations in the system~(\ref{P2-n}) increases. The asymptotic solution then becomes stiff, i.e., it contains important contributions that vary on different scales, including contributions that grow slowly with $x$, oscillate with rapidly growing frequencies, or oscillate with a fixed frequency but an additional logarithmically slow phase drift.

The solution of the system~(\ref{P2-n}) has points of singular dependence on the initial conditions and pockets of high sensitivity to the parameter values. There are also pseudo-saturation regimes \cite{Tyagi2025} that, during numerical simulations, may be misinterpreted as stable large-$x$ asymptotic dynamics but eventually switch to qualitatively different dynamics at larger $x$. The analytical solution handles this complexity and  provides insight even in the ``thermodynamic limit," $n\rightarrow \infty$, as will be discussed in Section~\ref{sec-physics}, revealing unusual effects such as the thermalization of nonadiabatic excitations in an integrable Hamiltonian dynamics.

 The derivation of the connection formulas is very involved, although the final result looks simple. Therefore, Section~\ref{sec-asympt} presents the asymptotic solutions right away. Section~\ref{sec-physics} discusses features of the solution that may be of interest to physicists. Section~\ref{sec-num} contains explicit connection formulas for $n=3$ and a discussion of the basic features of the solution. Only after this section is the derivation of the connection formulas addressed.
Section~\ref{sec-WKB} provides a bird's-eye view of the solution
strategy, explaining the WKB approach. The next two sections, \ref{sec-wkb-xm} and~\ref{sec-wkb-xp},
provide details for the asymptotic solution of the associated 
time-dependent Schr\"odinger equation as $x\rightarrow - \infty$ and $x\rightarrow +\infty$, respectively. Section~\ref{sec-compare} shows how by comparing results of Sections~\ref{sec-wkb-xm} and~\ref{sec-wkb-xp} the connection formulas finally follow.  
For technical details, the presentation in the main text often refers to numerous appendices containing tedious but straightforward calculations.

\section{Main Result: asymptotic solutions}
\label{sec-asympt}

\subsection{Parametrization as $x\rightarrow \pm\infty$}
Standard perturbative analysis \cite{Landau1980} fixes the asymptotic behavior of the solution of the system~(\ref{P2-n}) as $x\rightarrow -\infty$ up to the initial amplitudes $\alpha_k>0$ and phases $\varphi_k$, where $k=1,\ldots,n$. Appendix~\ref{app-minusx} shows that
\begin{eqnarray}
    \label{u1m} 
\nonumber     {\rm for} && x\rightarrow -\infty:\\
   u_k&\!=\!&\frac{\alpha_k}{(-x+\varepsilon_k)^{1/4}} \sin \left[ \frac{2}{3}(-x+\varepsilon_k)^{3/2}+\frac{(3\alpha_k^2+2\sum_{j\ne k}^{n}\alpha_j^2)}{4}\ln (-x) +\varphi_k \right], \quad  k=1,\ldots,n,
\end{eqnarray}
where $\alpha_k$ and $\varphi_k$ are parameters that will be referred to as, respectively, initial amplitudes and phases.  


As $x\rightarrow +\infty$, trivial perturbation theory \cite{Landau1980} easily fixes only the leading $x$-dependent terms in the oscillation phases, as explained in Appendix~\ref{app-plusx}, while the subleading terms, starting with those proportional to $\ln x$, are derived using the WKB approach described in Section~\ref{sec-WKB} and Appendix~\ref{app-log}. The solution at $x\rightarrow +\infty$ is parametrized by $n$ positive amplitudes, $\rho$ and $A_2, \ldots, A_n$; $n$ phases, $\phi_{1},\ldots, \phi_n$; one sign parameter, $\sigma=\pm 1$; and $n-1$ equation parameters, $\varepsilon_2,\ldots,\varepsilon_n$:

\begin{eqnarray}
\label{u1-largex}   
{\rm For}   &&  x\rightarrow +\infty:\\
u_1(x) &=& \sigma \sqrt{\frac{x}{2}-\sum_{j=2}^n u_j(x)^2 }+ 
\sigma \frac{\rho}{(2x)^{1/4}} \cos\left[ \frac{2\sqrt{2}}{3} x^{3/2} -\frac{3}{2}\rho^2\ln x +\phi_1\right],\\
 \label{u2-largex}
 u_j(x)&=&  A_j\cos\left[\sqrt{\varepsilon_j}x-\frac{A_j^2\sqrt{\varepsilon_j}}{2}\ln x+\phi_j\right], \quad j=2,\ldots,n.
\end{eqnarray}


\subsection{Connection formulas}

Introduce the following combinations of the final and initial
parameters:
\be
I_1\equiv \frac{\rho^2}{2}, \quad I_m\equiv \frac{A_m^2\sqrt{\varepsilon_m}}{2}, \quad m=2,\ldots,n,
\label{actions-def}
\ee
\be
p_j=e^{-\pi \alpha_j^2}, \quad q_j\equiv 1-p_j, \quad j=1,\ldots,n,
\label{inprob-def}
\ee
and a set of phases:
\begin{widetext}
\begin{equation}
\label{app-phase-final}
\Phi_j^{(n)}
\equiv
\frac{\pi}{4}
+{\rm arg}\Gamma\left(i\frac{\alpha_j^2}{2}\right)
+\varphi_j
-\frac{3\alpha_j^2}{2}\ln2
+
\frac12
\sum_{\substack{k=1\\k\ne j}}^n
\alpha_k^2
\ln\left(
\frac{|\varepsilon_j-\varepsilon_k|}{4}
\right),\quad j=1,\ldots,n,
\end{equation}
\end{widetext}
 Here and in what follows, the notation $X^{(n)}$ means that $X$ depends on the number of equations, $n$.

In terms of $p_j$, $q_j$, and $\Phi^{(n)}_j$, introduce the following
complex parameters $C_j^{(n)}$:
\begin{equation}
\label{uk0-2}
C_0^{(n)}\equiv 1, \quad C_j^{(n)}\equiv\prod_{k=1}^jp_k+\sum_{k=1}^jq_k\left(\prod_{s=k+1}^j p_s \right)e^{2i\Phi_k^{(n)}}, \quad j=1,\ldots,n,
\end{equation}
where the convention is assumed that $\prod_{s=j+1}^j p_s \equiv 1$. Note a useful identity: $C_j^{(n)}=p_jC_{j-1}^{(n)}+q_je^{2i\Phi_j^{(n)}} $. Not accidentally, positive coefficients at phase factors in $C_j^{(n)}$ coincide with transition probabilities from a solvable quantum mechanical Demkov-Osherov model \cite{SinitsynPokrovsky2026QuasiAdiabaticEffects}, which will appear  in discussion later.  

The {\it main result of this article} is the following set of connection formulas relating the final parameters
$
\{\sigma, I_1, \ldots, I_n, \phi_1, \ldots, \phi_n \}
$
to the initial parameters
$
\{p_1,\ldots, p_n, \Phi_1^{(n)}, \ldots \Phi_n^{(n)} \}
$ and the parameters of the equation, $\varepsilon_1,\ldots, \varepsilon_n$, where $\varepsilon_1=0$:

\begin{eqnarray}
\label{sigma-fin2}
 \sigma&=& {\rm sign}\left[ \sin \left(\Phi_1^{(n)}\right)\right],\\
 \label{rho-fin2}
  I_1&=&-\frac{1}{4\pi} \ln \left( 1-|C_{n}^{(n)}|^2\right),\\
 \label{phi1-fin2}
  \phi_1&=&-\frac{3\pi}{4}-7I_1\ln 2 +{\rm arg}\Gamma\left(2iI_1\right)-{\rm arg}C_n^{(n)}, \\
  \label{I2-fin2}
I_j
&=&
\frac{1}{2\pi}
\ln
\left[
\frac{
1-\left|C_j^{(n)}\right|^2
}{
p_j
\left(
1-\left|C_{j-1}^{(n)}\right|^2
\right)
}
\right],
\qquad
j=2,\ldots,n.\\
  \label{phi2-fin2}
  \phi_j&=&\frac{3\pi}{4}
-\frac{2}{3}\varepsilon_j^{3/2}
-I_j\ln\left(4\sqrt{\varepsilon_j}\right)
+\arg\Gamma(iI_j)
-
\arg\left[
e^{i\Phi_j^{(n)}}
-
e^{-i\Phi_j^{(n)}}C_{j-1}^{(n)}
\right] +
\sum \limits_{ \substack{k=2\\k\ne j}}^n I_k \ln \frac{ \left |\sqrt{\varepsilon_k} -\sqrt{\varepsilon_j} \right |}{\sqrt{\varepsilon_k} +\sqrt{\varepsilon_j}}.
\end{eqnarray}
Corrections to the connection formulas are of order $1/\sqrt{x}$. They will not be studied here.

The system in Eq.~(\ref{P2-n}) describes classical  dynamics \cite{Tyagi2025} with the Hamiltonian
\be
{\cal H}=\frac12\sum_{j=1}^n v_j^2
-\frac12\sum_{j=1}^n(x-\varepsilon_j)u_j^2
+\frac12\left(\sum_{j=1}^n u_j^2\right)^2,
\qquad v_j=u_j'.
\label{classical-Hamiltonian}
\ee
Hamilton's equations are $u_j'=\partial{\cal H}/\partial v_j$
and $v_j'=-\partial{\cal H}/\partial u_j$, with $x$ playing the role
of time. Although ${\cal H}$ depends on $x$, this evolution preserves
the phase space volume. Combinations of
$$
i_k \equiv \frac{\alpha_k^2}{2}, \quad k=1,\ldots,n
$$
and initial oscillation phases are asymptotically, as $x\rightarrow -\infty$, the action and angle variables of a harmonic oscillator. The pairs of final $I_k$ and oscillation phases are  similar action-angle variables as $x\rightarrow +\infty$ of adiabatically changing harmonic oscillations near one of the potential minima.

The phase space volume during the Hamiltonian dynamics is conserved. Consequently,  interpreting the 
connection formulas as the canonical transformation between those variables, the absolute value of the Jacobyan must be one. More precisely, in the notation of Ref.~\cite{Landau1980},
\be
\det\frac{\partial(I_1,\ldots,I_n,\phi_1,\ldots,\phi_n)}
{\partial(i_1,\ldots,i_n,\varphi_1,\ldots,\varphi_n)}=(-1)^n,
\label{main-det}
\ee
where the minus sign is due to the convention for the initial and final 
oscillation phases. Namely, the oscillation phase in Eq.~(\ref{u1m}) decreases with $x$, while
those in Eqs.~(\ref{u1-largex}) and~(\ref{u2-largex}) increase.
Consequently, an increase of $\varphi_k$ shifts the initial
oscillation backwards along its orbit, whereas an increase of
$\phi_k$ shifts the final oscillation forwards. This reverses
the orientation of each action and phase pair and gives one minus
sign per mode in Eq.~(\ref{main-det}). 

According to \cite{Landau1980}, for $n>1$, there are also lower dimensional conserved volumes. Such conservation laws were not helpful in derivation of the connection formulas but they served as an additional test. Using symbolic software, I verified that the connection formulas satisfied such conservation laws, including Eq.~(\ref{main-det}), up to $n=3$ (not shown).

Finally,  an interesting property of Eqs.~(\ref{sigma-fin2})--(\ref{phi2-fin2}) is that  they can be inverted. That is, given the final data $I_1,\ldots I_n$, $\phi_1,\ldots,\phi_n$, and $\sigma$, one can write explicit expressions
for the initial parameters $\alpha_j$ and $\varphi_j$. In certain theoretical physics applications of the conventional Painlev\'e-II equation such conditions were needed \cite{Itin2009b}. An arbitrary $n$ case is now discussed in Appendix~\ref{inversion}.

\section{Physics of the model in the limit $n\rightarrow \infty$}
\label{sec-physics}
The system in Eq.~(\ref{P2-n}) has already proven useful in practice. It describes the scattering of solitons in the nonlinear Schr\"odinger equation and the statistics of gaps between the energy levels of random matrices \cite{ClaeysDoeraene2018}. However, explicit quantitative predictions have been studied only for a system of two variables \cite{Tyagi2025,Suzuki2025,Sinitsyn2026-letter}. It is interesting to consider the regime of very large $n$, which is inaccessible to direct numerical simulations. In particular, if all equations were decoupled from one another, then each of them would produce a characteristic change, let's call it $\Delta I_j$, where $j=1,\ldots,n$. For similar initial conditions in different equations, the net change $\sum_{j=1}^n \Delta I_j$ would then likely diverge as $n\rightarrow \infty$. Does this property persist in the presence of the all-to-all interactions described by Eq.~(\ref{P2-n})? How do such changes depend on the initial oscillation amplitudes and the model parameters $\varepsilon_j$? The exact connection formulas provide elementary and somewhat unexpected answers to these questions.


\subsection{Production of quasiparticles in the limit $n\rightarrow \infty$}

The model in Eq.~(\ref{P2-n}) and the connection formulas describe competition between coherent chemical reactions and the production of excitations during passage through a second-order phase transition \cite{Tyagi2025,Suzuki2025,Sinitsyn2026-letter}. In this case, the quantities $I_j$ correspond to the adiabatic invariants of asymptotically harmonic oscillatory motion, and $I_j/\hbar$ can then be identified with semiclassical expressions for the number of produced quasiparticles \cite{Tyagi2025}. Thus, the changes of the adiabatic invariants characterize the number of elementary excitations, usually quasiparticles, produced after passage through the phase transition. Moreover, Refs.~\cite{Tyagi2025,Suzuki2025} identified $I_1$ with the number of Higgs bosons and $I_k$, $k\ge 2$, with the number of massive Goldstone bosons. In this section, this particular interpretation of the dynamics will be considered.

Following Ref.~\cite{Tyagi2025}, introduce the {\it excess of the adiabatic invariants} as
\be
\Delta I\equiv I-I_{\rm in},
\label{Delta-I-def}
\ee
where 
\be
I\equiv 2I_1+\sum_{j=2}^{n}I_j,
\label{total-final-I}
\ee
 and
\be
I_{\rm in}\equiv\frac{1}{2}\sum_{j=1}^{n}\alpha_j^2.
\label{total-initial-I}
\ee
The factor $2$ multiplying $I_1$ in Eq.~(\ref{total-final-I}) follows from the fact that, upon entering one of two identical disjoint phase-space regions, the initial adiabatic invariant is reduced by a factor of $2$. This is not counted as a purely nonadiabatic effect \cite{SinitsynPokrovsky2026QuasiAdiabaticEffects}.

Semiclassically, the initial quantum-vacuum conditions correspond to \cite{Tyagi2025,Sinitsyn2026-letter}
$$
\alpha_j^2/2 = {\cal I} \ll 1,
$$
and averaging over all initial phases $\varphi_j$ is needed to mimic averaging over the ground-state wave function \cite{SinitsynPokrovsky2026QuasiAdiabaticEffects}. Averaging over $\varphi_j$ is equivalent to averaging over $\Phi_j^{(n)}$ \cite{Sinitsyn2026-letter}. Hence, introduce the notation
\be
\left\langle F\right\rangle_{\Phi}
\equiv
\int D\Phi\,F,
\qquad
\int D\Phi
\equiv
\frac{1}{(2\pi)^n}
\int_0^{2\pi}d\Phi_1^{(n)}
\cdots
\int_0^{2\pi}d\Phi_n^{(n)}.
\label{phase-average-def}
\ee
According to Eq.~(\ref{net-inv}) derived later, 
\be
e^{-\pi I}
=
2\left(
q_1\prod_{j=1}^{n}p_j
\right)^{1/2}
\left|\sin\Phi_1^{(n)}\right|.
\label{total-I-survival}
\ee
On the other hand, since $p_j=e^{-\pi\alpha_j^2}$,
\be
-\frac{1}{2\pi}
\ln\left(\prod_{j=1}^{n}p_j\right)
=
\frac{1}{2}\sum_{j=1}^{n}\alpha_j^2
=
I_{\rm in}.
\label{initial-I-product}
\ee
Taking the logarithm of Eq.~(\ref{total-I-survival}) and subtracting
Eq.~(\ref{initial-I-product}) leads to
\be
\Delta I
=
-\frac{1}{\pi}
\ln\left(
2\left|\sin\Phi_1^{(n)}\right|
\right)
-\frac{1}{2\pi}\ln q_1.
\label{Delta-I-phase}
\ee
The phase-dependent term vanishes after averaging because
\be
\frac{1}{2\pi}
\int_0^{2\pi}
\ln\left(2|\sin\Phi|\right)d\Phi
=0.
\label{log-sine-average}
\ee
Consequently,
\be
\left\langle\Delta I\right\rangle_{\Phi}
=
-\frac{1}{2\pi}\ln q_1
=
-\frac{1}{2\pi}
\ln\left(1-e^{-\pi\alpha_1^2}\right).
\label{average-Delta-I}
\ee

Surprisingly, according to Eq.~(\ref{average-Delta-I}), even for a possibly infinite number of coupled differential equations, the change of the net adiabatic invariant, and hence the average number of produced quasiparticles, does not depend on $n$. Even more counterintuitively, $\left\langle\Delta I\right\rangle_{\Phi}$ is independent of the initial amplitudes $\alpha_j$ and equation parameters $\varepsilon_j$ for every $j\ge 2$ and every $n$, including in the limit $n\rightarrow\infty$. 

Consider now equal and small initial adiabatic invariants. This condition follows from the semiclassical assumption that, in the initial ground state, the adiabatic invariants are all identical and of the order of the Planck constant $\hbar$:
\be
\frac{\alpha_j^2}{2}={\cal I}\ll1,
\qquad
j=1,\ldots,n.
\label{equal-initial-invariants}
\ee
In this case,
\be
\left\langle\Delta I\right\rangle_{\Phi}
=
-\frac{1}{2\pi}
\ln\left(1-e^{-2\pi{\cal I}}\right)
\approx
\frac{1}{2\pi}
\ln\left(\frac{1}{2\pi{\cal I}}\right).
\label{average-Delta-I-small}
\ee

\subsection{Statistics of excitations}

The statistics of excitations produced during passage through a phase transition has also attracted some interest \cite{Sun2016,delcampo2018universal,gomezruiz2020full,bando2022statistics}. The connection formulas provide closed-form expressions for the moments of $I$, averaged over a uniform distribution of the initial phases. These statistics can then be related, for example, to the number of spin qubits pointing in the wrong direction after a quantum-annealing computation \cite{Itin2009a,Itin2009b}.
Equation~(\ref{total-I-survival}) taken to power $\lambda$ is
\be
e^{-\lambda\pi I}
=
2^\lambda
\left(
q_1\prod_{j=1}^{n}p_j
\right)^{\lambda/2}
\left|\sin\Phi_1^{(n)}\right|^\lambda.
\label{I-generating-start}
\ee
Therefore, for ${\rm Re}\,\lambda>-1$,
\be
\begin{aligned}
\left\langle e^{-\lambda\pi I}\right\rangle_{\Phi}
={}&
2^\lambda
\left(
q_1\prod_{j=1}^{n}p_j
\right)^{\lambda/2}
\frac{1}{2\pi}
\int_0^{2\pi}|\sin\Phi|^\lambda d\Phi.
\end{aligned}
\label{I-generating-integral}
\ee
The remaining integral is
\be
\frac{1}{2\pi}
\int_0^{2\pi}|\sin\Phi|^\lambda d\Phi
=
\frac{
\Gamma\left(\frac{\lambda+1}{2}\right)
}{
\sqrt{\pi}\,
\Gamma\left(1+\frac{\lambda}{2}\right)
}.
\label{sine-power-integral}
\ee
Hence, 
\be
\left\langle e^{-\lambda\pi I}\right\rangle_{\Phi}=\left(
q_1\prod_{j=1}^{n}p_j
\right)^{\lambda/2}
\frac{
\Gamma(1+\lambda)
}{
\Gamma^2\left(1+\frac{\lambda}{2}\right)
}.
\label{I-generating-final}
\ee

Consider again equal initial adiabatic invariants,
\be
\frac{\alpha_j^2}{2}={\cal I},
\qquad
j=1,\ldots,n.
\label{equal-initial-invariants-statistics}
\ee
In this case,
\be
p_j=e^{-2\pi{\cal I}},
\qquad
q_j=1-e^{-2\pi{\cal I}},
\qquad
\prod_{j=1}^{n}p_j=e^{-2\pi n{\cal I}},
\label{equal-pq-statistics}
\ee
and 
\be
\begin{aligned}
\left\langle e^{-\lambda\pi I}\right\rangle_{\Phi}
={}&
e^{-\lambda\pi n{\cal I}}
\left(1-e^{-2\pi{\cal I}}\right)^{\lambda/2}
\\
&\times
\frac{
\Gamma(1+\lambda)
}{
\Gamma^2\left(1+\frac{\lambda}{2}\right)
}.
\end{aligned}
\label{I-generating-equal}
\ee

  \begin{figure}[t!]
    \centering
    \includegraphics[width=0.45\textwidth]{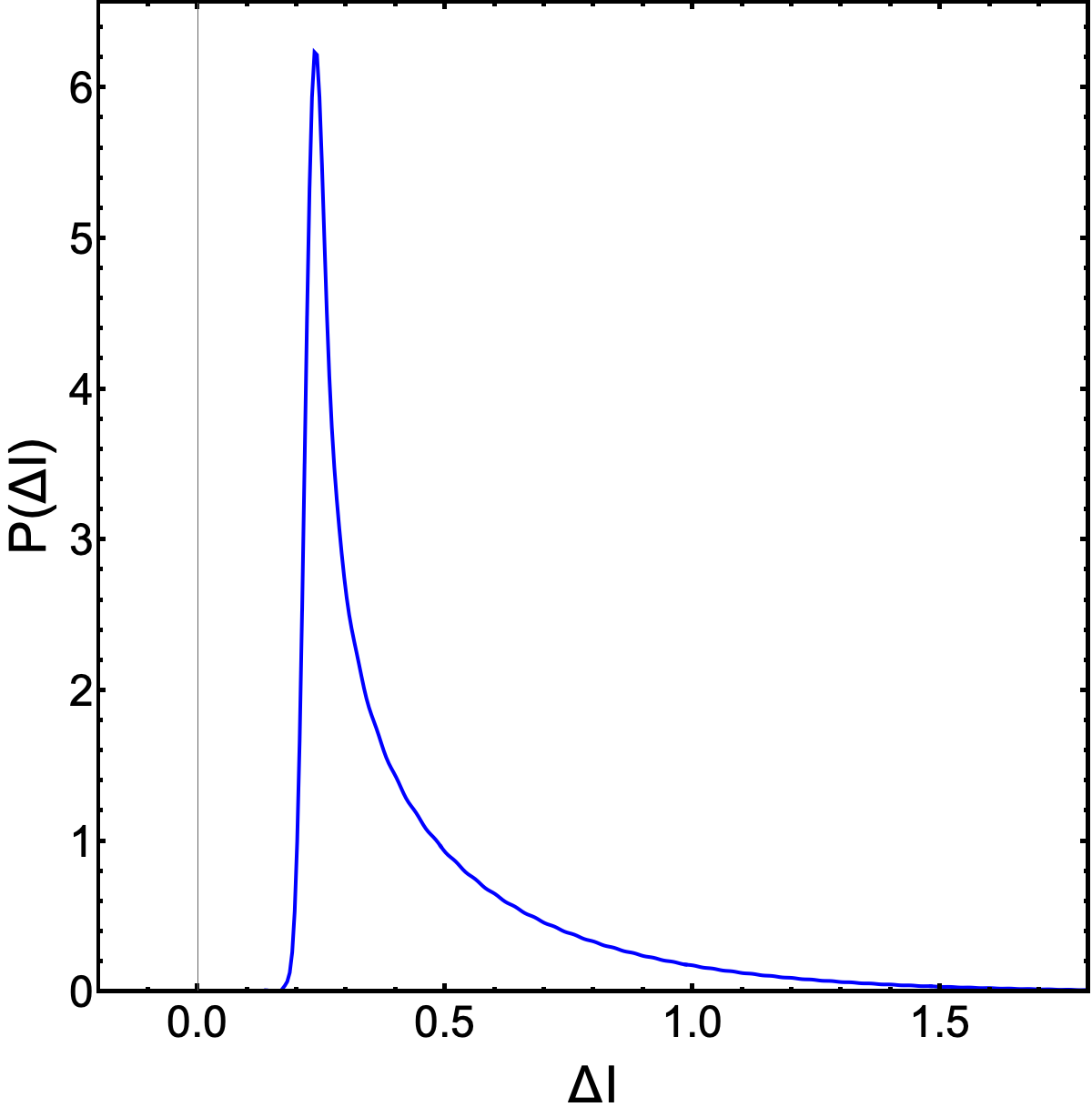}
    \caption{Probability distribution of $\Delta I$  obtained by numerical inverse Fourier transformation of $Z(i\lambda)$. The uniform distribution of initial phases $\varphi_1,\ldots, \varphi_n$, leads to the distribution of the changes of the adiabatic invariant,  $P(\Delta I)$, which is strongly skewed. It has a sharp cut-off at the left tail, and an exponentially decaying right tail. Here, ${\cal I}=0.01$.}
    \label{distribution-fig}
\end{figure} 
The nontrivial statistics of the produced excitations is described by
the excess invariant
\be
\Delta I=I-n{\cal I}.
\label{Delta-I-equal}
\ee
It follows from Eq.~(\ref{I-generating-equal}) that
\be
Z(\lambda)\equiv \left\langle e^{-\lambda\pi\Delta I}\right\rangle_{\Phi}
=
\left(1-e^{-2\pi{\cal I}}\right)^{\lambda/2}
\frac{
\Gamma(1+\lambda)
}{
\Gamma^2\left(1+\frac{\lambda}{2}\right)
}.
\label{Delta-I-generating}
\ee

All cumulants of $\Delta I$ can be obtained from
\be
\ln
\left\langle e^{-\lambda\pi\Delta I}\right\rangle_{\Phi}
=
\sum_{m=1}^{\infty}
\frac{(-\lambda\pi)^m}{m!}\,
\kappa_m,
\label{Delta-I-cumulants-def}
\ee
where $\kappa_m$ is the $m$th cumulant of $\Delta I$. The first
cumulant reproduces Eq.~(\ref{average-Delta-I}),
while all higher cumulants are independent of ${\cal I}$. For example, the variance of the produced adiabatic invariant is
\be
\left\langle
\left(
\Delta I-\left\langle\Delta I\right\rangle_{\Phi}
\right)^2
\right\rangle_{\Phi}
=
\frac{1}{12}.
\label{Delta-I-variance}
\ee

Listing the other cumulants, however, is not as informative as plotting the
probability distribution $P(\Delta I)$, shown in Fig.~\ref{distribution-fig}. This distribution is surprisingly universal. Only its mean depends on the initial value of $\alpha_1$ -- otherwise, its shape remains the same. 
The fact that $P(\Delta I)$ does not depend on $n$ can be understood for well-separated $\varepsilon_j$ using arguments from Ref.~\cite{Tyagi2025}. The first variable, $u_1$, passes through the critical point first, while the other variables are still rapidly oscillating. Effectively, this resonance is described by the standard P-II equation, which is responsible for the nonadiabatic excitation $\Delta I$. Later, other resonances are encountered, but $\Delta I$ does not change at any of them. Instead, the excitations are redistributed among all modes while $\Delta I$ remains invariant.
This behavior would be unexpected, however, when all $\varepsilon_j$ are small, so that all resonances are encountered almost simultaneously.

\subsection{Continuum limit}

The redistribution of excitations among different variables $u_j$, however, can substantially change amplitudes of oscillations of individual modes. 
For example, for $n=1$, all excitations belong to the mode $u_1$. Consider now a situation with small ${\cal I}$ but large $n$:
\[
\frac{\alpha_j^2}{2}={\cal I}\ll1,\qquad n\gg1,\qquad n{\cal I}=O(1),
\]

This regime allows the continuum limit to be obtained. Appendix~\ref{app-I1-continuum-limit} produces the following estimate of the final first adiabatic invariant 
\be
\left\langle I_1\right\rangle_{\Phi}
\approx
-\frac{1}{4\pi}
\ln\left(1-e^{-4\pi n{\cal I}}\right)
+
\frac{{\cal I}}
{4\left(1-e^{-4\pi n{\cal I}}\right)}.
\label{I1-continuum-main}
\ee
The first term is the strict continuum-limit result. The second term is the leading correction due to the phase-dependent fluctuations, which was derived assuming that the first term dominates. 
Equation~(\ref{I1-continuum-main}) shows that the characteristic power of $u_1$ oscillations decays with large $n$.  This means that excitations are spread among many oscillatory modes $u_j$ with $j=2,\ldots, n$.  Nevertheless, the mode $u_1$ retains an excitation of the order of the original small value ${\cal I}$. 

The phase-averaged redistribution in the same limit is best described by
\[
\widetilde{\Delta I}(j)
=\frac{\left\langle\Delta I_j\right\rangle_{\Phi}}{2{\cal I}}.
\]
To leading order in ${\cal I}$, Appendix~\ref{app-I1-continuum-limit} gives the universal scaling form
\be
\widetilde{\Delta I}(j)
=
\frac{1}{\exp\left(4\pi {\cal I}j\right)-1}.
\label{Delta-I-continuum-scaling}
\ee

This profile is similar to the Bose--Einstein distribution for particles, at temperature $T$ and zero chemical potential, occupying an equidistant spectrum with linear dispersion \cite{Sun2016}
\[
E_j=Dj.
\]
The Bose--Einstein occupation is
\[
N_j=
\frac{1}{\exp\left[E_j/(k_{\mathrm B}T)\right]-1}
=
\frac{1}{\exp\left[Dj/(k_{\mathrm B}T)\right]-1}.
\]

Thus, the phase-averaged profile of $\Delta I_j/(2{\cal I})$ has the same functional form as the Bose--Einstein distribution for bosons populating a nearly continuum equidistant spectrum of energy levels at finite temperature.
Surprisingly, despite the absence of stochastic terms in the system~(\ref{P2-n}), averaging over the initial angles $\varphi_j$ produces an excitation distribution that can be interpreted as thermalized.



\section{Connection formulas for $n=3$}
\label{sec-num}

For $n=1$ and $n=2$, the connection formulas coincide with previously derived formulas \cite{Fokas2006,Sinitsyn2026-letter}, which have already been discussed and tested numerically. The case $n=3$ has not been studied previously. Hence,  I provide a numerical test for it here.

Let 
\begin{eqnarray}
\nonumber u_1''(x)&=&xu_1(x)-2u_1(x)\sum_{k=1}^3 u_k^2(x),\\
\nonumber u_2''(x)&=&xu_2(x)-2u_2(x)\sum_{k=1}^3 u_k^2(x)-\varepsilon_2 u_2(x),\\
\label{P2-3}
 u_3''(x)&=&xu_3(x)-2u_3(x)\sum_{k=1}^3 u_k^2(x)-\varepsilon_3 u_3(x),
\end{eqnarray}
\begin{equation}
0=\varepsilon_1<\varepsilon_2<\varepsilon_3.
\label{n3-epsilon-order}
\end{equation}
The initial conditions are
\begin{eqnarray}
\nonumber u_1(x)
&=&
\frac{\alpha_1}{(-x)^{1/4}}
\sin\left[
\frac{2}{3}(-x)^{3/2}
+\frac{3\alpha_1^2+2\alpha_2^2+2\alpha_3^2}{4}
\ln(-x)
+\varphi_1
\right], 
\qquad x\rightarrow-\infty,\\
\nonumber u_2(x)
&=&
\frac{\alpha_2}{(-x+\varepsilon_2)^{1/4}}
\sin\left[
\frac{2}{3}(-x+\varepsilon_2)^{3/2}
+\frac{2\alpha_1^2+3\alpha_2^2+2\alpha_3^2}{4}
\ln(-x)
+\varphi_2
\right],\\
u_3(x)
&=&
\frac{\alpha_3}{(-x+\varepsilon_3)^{1/4}}
\sin\left[
\frac{2}{3}(-x+\varepsilon_3)^{3/2}
+\frac{2\alpha_1^2+2\alpha_2^2+3\alpha_3^2}{4}
\ln(-x)
+\varphi_3
\right].
\end{eqnarray}

At large positive $x$, the same solution has the asymptotic form
\begin{equation}
\begin{aligned}
u_1(x)
=
\sigma\sqrt{\frac{x}{2}-u_2^2(x)-u_3^2(x)}
+\sigma\frac{\rho}{(2x)^{1/4}}
\cos\left[
\frac{2\sqrt{2}}{3}x^{3/2}
-\frac{3\rho^2}{2}\ln x
+\phi_1
\right],
\qquad x\rightarrow+\infty,
\end{aligned}
\label{n3-final-u1}
\end{equation}
\begin{equation}
u_2(x)
=
 A_2
\cos\left[
\sqrt{\varepsilon_2}\,x
-\frac{A_2^2\sqrt{\varepsilon_2}}{2}\ln x
+\phi_2
\right],
\qquad x\rightarrow+\infty,
\label{n3-final-u2}
\end{equation}
\begin{equation}
u_3(x)
=
 A_3
\cos\left[
\sqrt{\varepsilon_3}\,x
-\frac{A_3^2\sqrt{\varepsilon_3}}{2}\ln x
+\phi_3
\right],
\qquad x\rightarrow+\infty.
\label{n3-final-u3}
\end{equation}
Here, $\sigma=\pm1$, while $\rho$, $A_2$, and $A_3$ are related to
the new parameters by
\begin{equation}
I_1=\frac{\rho^2}{2},
\qquad
I_2=\frac{A_2^2\sqrt{\varepsilon_2}}{2},
\qquad
I_3=\frac{A_3^2\sqrt{\varepsilon_3}}{2}.
\label{n3-final-actions}
\end{equation}

In addition, define
$$
p_j=e^{-\pi \alpha_j^2}, \quad q_j\equiv 1-p_j, \quad j=1,\ldots,3
.$$
Equation~(\ref{app-phase-final}) gives
\begin{widetext}
\begin{eqnarray}
\Phi_1^{(3)}
&=&
\frac{\pi}{4}
+{\rm arg}\Gamma\left(i\frac{\alpha_1^2}{2}\right)
+\varphi_1-\frac{3\alpha_1^2}{2}\ln2
+\frac{\alpha_2^2}{2}\ln\left(\frac{\varepsilon_2}{4}\right)
+\frac{\alpha_3^2}{2}\ln\left(\frac{\varepsilon_3}{4}\right),
\label{n3-Phi1}\\
\Phi_2^{(3)}
&=&
\frac{\pi}{4}
+{\rm arg}\Gamma\left(i\frac{\alpha_2^2}{2}\right)
+\varphi_2-\frac{3\alpha_2^2}{2}\ln2
+\frac{\alpha_1^2}{2}\ln\left(\frac{\varepsilon_2}{4}\right)
+\frac{\alpha_3^2}{2}
\ln\left(\frac{\varepsilon_3-\varepsilon_2}{4}\right),
\label{n3-Phi2}\\
\Phi_3^{(3)}
&=&
\frac{\pi}{4}
+{\rm arg}\Gamma\left(i\frac{\alpha_3^2}{2}\right)
+\varphi_3-\frac{3\alpha_3^2}{2}\ln2
+\frac{\alpha_1^2}{2}\ln\left(\frac{\varepsilon_3}{4}\right)
+\frac{\alpha_2^2}{2}
\ln\left(\frac{\varepsilon_3-\varepsilon_2}{4}\right).
\label{n3-Phi3}
\end{eqnarray}
\end{widetext}

  \begin{figure}[t!] 
    \centering
    \includegraphics[width=0.65\textwidth]{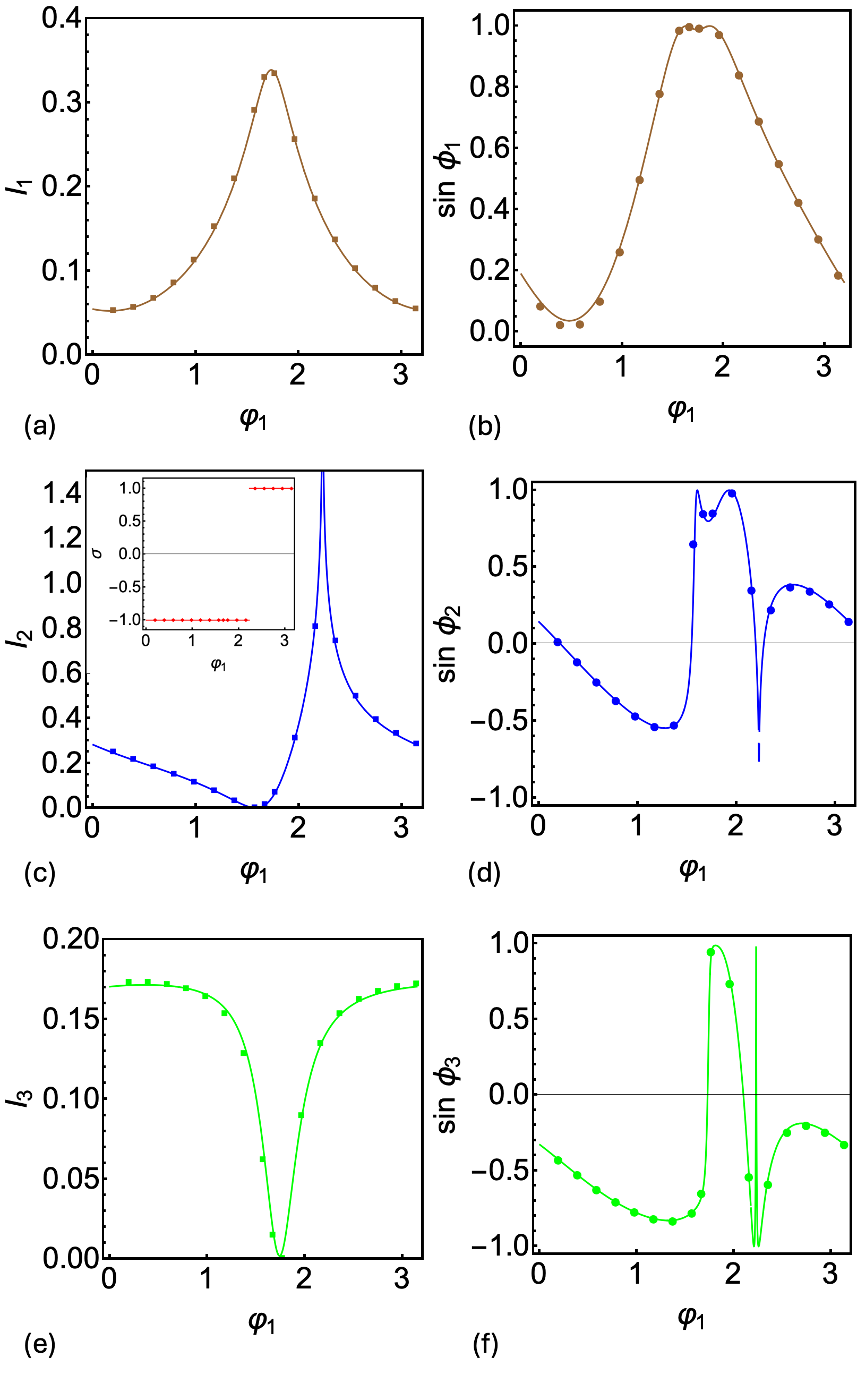}
    \caption{Numerical test of connection formulas for $n=3$. Solid curves are the theoretical predictions in Eqs.~(\ref{n3-sigma})--(\ref{n3-phi3}) for different initial angle $\varphi_1$. The dots of the same color are the results of the numerical solution at the same parameters  in the interval $x\in(-6000,6000)$. The numerical simulations used the second-order Verlet algorithm, described in \cite{Sinitsyn2025,Tyagi2025}, with a step $dx=0.00001$. The fixed parameters for the curves and numerical simulations are as follows: $\varepsilon_2=0.75$, $\varepsilon_3=1.1$, $\alpha_1=0.9$, $\alpha_2=0.3$, $\alpha_3=0.7$, $\varphi_2=\varphi_3=\pi/2$. The red line and points in the inset in (c) is the test for the sign factor from Eq.~(\ref{n3-sigma}). The data in (a,b; brown), (c,d;blue) and (e,f; green) show final adiabatic invariants and sines of angles for, respectively, $I_1$ and $\sin \phi_1$, $I_2$ and $\sin \phi_2$, $I_3$ and $\sin \phi_3$. 
    }
    \label{num-fig}
\end{figure} 
In terms of these phases and the probabilities (\ref{inprob-def}), the
connection formulas, after some regrouping of terms, are given by

\begin{eqnarray}
\label{n3-sigma}
\sigma
&=&
{\rm sign}\left[\sin\Phi_1^{(3)}\right],
\\
\label{n3-I1}
I_1
&=&
-\frac{1}{4\pi}\ln\left[
1-\left|
p_1p_2p_3
+q_1p_2p_3e^{2i\Phi_1^{(3)}}
+q_2p_3e^{2i\Phi_2^{(3)}}
+q_3e^{2i\Phi_3^{(3)}}
\right|^2
\right],
\\
I_2
&=&
\frac{1}{2\pi}\ln\left[
1+\frac{
q_2\left|
p_1\sin\Phi_2^{(3)}
+q_1e^{i\Phi_1^{(3)}}
\sin\left(\Phi_2^{(3)}-\Phi_1^{(3)}\right)
\right|^2
}{
p_1q_1\sin^2\Phi_1^{(3)}
}
\right],
\label{n3-I2}\\
\label{n3-I3}
I_3
&=&
\frac{1}{2\pi}\ln\left[
1+\frac{
q_3\left|
p_1p_2\sin\Phi_3^{(3)}
+q_1p_2e^{i\Phi_1^{(3)}}
\sin\left(\Phi_3^{(3)}-\Phi_1^{(3)}\right)
+q_2e^{i\Phi_2^{(3)}}
\sin\left(\Phi_3^{(3)}-\Phi_2^{(3)}\right)
\right|^2
}{
p_2\left[
p_1q_1\sin^2\Phi_1^{(3)}
+q_2\left|
p_1\sin\Phi_2^{(3)}
+q_1e^{i\Phi_1^{(3)}}
\sin\left(\Phi_2^{(3)}-\Phi_1^{(3)}\right)
\right|^2
\right]
}
\right],
\end{eqnarray}
\begin{eqnarray}
\label{n3-phi1}
\phi_1
&=&
-\frac{3\pi}{4}
-7I_1\ln2
+{\rm arg}\Gamma(2iI_1)
-{\rm arg}\left[
p_1p_2p_3
+q_1p_2p_3e^{2i\Phi_1^{(3)}}
+q_2p_3e^{2i\Phi_2^{(3)}}
+q_3e^{2i\Phi_3^{(3)}}
\right],\\
\phi_2
&=&
\frac{3\pi}{4}
-\frac{2}{3}\varepsilon_2^{3/2}
-I_2\ln\left(4\sqrt{\varepsilon_2}\right)
+{\rm arg}\Gamma(iI_2)
-{\rm arg}\left[
e^{i\Phi_2^{(3)}}
-e^{-i\Phi_2^{(3)}}
\left(p_1+q_1e^{2i\Phi_1^{(3)}}\right)
\right]
+I_3\ln
\frac{\sqrt{\varepsilon_3}-\sqrt{\varepsilon_2}}
{\sqrt{\varepsilon_3}+\sqrt{\varepsilon_2}},
\\
\label{n3-phi3}
\phi_3
&=&
\frac{3\pi}{4}
-\frac{2}{3}\varepsilon_3^{3/2}
-I_3\ln\left(4\sqrt{\varepsilon_3}\right)
+{\rm arg}\Gamma(iI_3)
\nonumber\\
&&
-{\rm arg}\left[
e^{i\Phi_3^{(3)}}
-e^{-i\Phi_3^{(3)}}
\left(
p_1p_2
+q_1p_2e^{2i\Phi_1^{(3)}}
+q_2e^{2i\Phi_2^{(3)}}
\right)
\right]
+I_2\ln
\frac{\sqrt{\varepsilon_3}-\sqrt{\varepsilon_2}}
{\sqrt{\varepsilon_3}+\sqrt{\varepsilon_2}}.
\end{eqnarray}

Figure~\ref{num-fig} shows predictions and numerical tests of the connection formulas for the dependence of the final adiabatic invariants and angles on the initial angle $\varphi_1$, with all other parameters being fixed. The agreement with the numerical solution of the system of equations is excellent, leaving no doubts that all relevant terms were identified correctly. This degree of numerical precision requires simulations during the interval $x\in (6000,6000)$ with a time step at least $dx=1\times 10^{-5}$, which is already time-consuming for a single processor. 

As in the case of two coupled equations, which was  discussed in  \cite{Sinitsyn2026-letter}, varying $\varphi_1$ produces a singularity of $I_2$ at $\Phi_1^{(3)}=0\pmod{\pi}$, provided $\Phi_2^{(3)}\ne0\pmod{\pi}$. Indeed, the denominator in Eq.~(\ref{n3-I2}) is
$
p_1q_1\sin^2\Phi_1^{(3)}
$.
 Hence, $I_2$ diverges logarithmically as $\varphi_1$ varies near this singularity, while $I_1$ and $I_3$ remain finite. In contrast, varying only $\varphi_2$ at a fixed $\Phi_1^{(3)}\ne0\pmod{\pi}$ does not produce a singularity of any of the three actions at a random choice of the other parameters. A singularity of $I_3$ requires simultaneously $\Phi_1^{(3)}=\Phi_2^{(3)}=0\pmod{\pi}$, with $\Phi_3^{(3)}\ne0\pmod{\pi}$.

This is a general feature of the connection formulas for arbitrary $n\ge2$. For nonzero finite initial amplitudes, the logarithmic singularity for $I_{j+1}$ appears when coefficients $C_j^{(n)}$ satisfy
$$
1-\left|C_j^{(n)}\right|^2=0, 
$$
which requires that 
\be
\Phi_1^{(n)}=\cdots=\Phi_j^{(n)}=0\pmod{\pi}.
\label{cond-sing}
\ee
Indeed,  $C_j^{(n)}$ in Eq.~(\ref{uk0-2})
is a sum of complex phase factors with positive coefficients.

For nonzero finite initial amplitudes, the sum of all these coefficients is one:
\[
\prod_{k=1}^j p_k
+\sum_{k=1}^j q_k\prod_{s=k+1}^j p_s=1.
\]
Hence, $|C_j^{(n)}|\le 1$, with equality only when all
terms contributing to $C_j^{(n)}$ have the same phase factors. The first term  in Eq.~(\ref{uk0-2}) is positive
and real, so equality requires
$e^{2i\Phi_k^{(n)}}=1$ for $k=1,\ldots,j$, which leads to Eq.~(\ref{cond-sing}).

Thus, varying $\varphi_1$ at  randomly chosen fixed values of the other initial angles produces only the singularity of $I_2$ at certain $\varphi_1$; by varying both $\varphi_1$ and $\varphi_2$, it is possible to encounter also a singularity in $I_3$, and so on. The parameter values at which such singularities appear in the connection formulas correspond to emergence of unstable solutions that have exponentially decaying one or several variables $u_1,\ldots, u_j$ as $x\rightarrow +\infty$. Such solutions will be discussed elsewhere.


\section{WKB path to connection formulas}
\label{sec-WKB}
 This section closely follows the analogous section
in Ref.~\cite{Sinitsyn2026-letter}, but now focuses on
arbitrary $n$ coupled Painlev\'e-II equations.

\subsection{Lax pair}
The integrability of the system in Eq.~(\ref{P2-n}) follows from the existence of a Lax pair of two $(n+1)$-dimensional Hermitian matrices, $H(t,x)$ and $H_1(t,x)$, that depend on $u_k(x)$, $k=1,\ldots,n$, and satisfy the consistency condition
\begin{equation}
\label{consHH}
\left(\frac{\partial H}{\partial x} - \frac{\partial H_1}{\partial t}\right) - i [H,H_1]=0,
\end{equation}
given that $u_k(x)$ satisfy Eq.~(\ref{P2-n}).
Explicitly, 
\begin{widetext}
\begin{eqnarray}
\label{HH-gen}
H &=& A - 4t B + 2C, \quad
H_1 \!=\! 
\left[\begin{array}{ccccc}
t & -u_1 & \cdots&-u_{n-1} & -u_n\\
-u_1 & -t & 0 &\cdots& 0\\
\vdots& \vdots& \ddots& \vdots& \vdots  \\
-u_{n-1} & 0 & \cdots &-t& 0\\
-u_n & 0 & \cdots & 0& -t
\end{array}\right],
\end{eqnarray}
where
\begin{eqnarray}
\nonumber &&A = \\
\nonumber &&\left[\begin{array}{ccccc}
4t^2+x-2\sum \limits_{k=1}^nu_k^2(x) & 0 & \cdots & \cdots& 0\\
0 & -(4t^2+x-2\varepsilon_1)+2u_1^2 &  2u_1u_2 & \cdots  &2u_1u_n \\
\vdots &2u_2u_1 & -(4t^2+x-2\varepsilon_2)+2u_2^2& \ddots & 2u_2 u_n\\
\vdots & \vdots & \ddots & \ddots & \vdots\\
0 & 2u_nu_1(x) & 2u_nu_2 &\cdots &  -(4t^2+x-2\varepsilon_n)+2u_n^2
\end{array}\right],\\
\nonumber \\
\nonumber \\
&&B =
\left[\begin{array}{cccc}
0 & u_1 & \cdots & u_n\\
u_1 & 0 & \cdots & 0\\
\vdots & \vdots & \ddots & \vdots\\
u_n & 0 & \cdots & 0
\end{array}\right],
\quad
C =
\left[\begin{array}{cccc}
0 & -i u_1' & \cdots & -i u_n'\\
i u_1' & 0 & \cdots & 0\\
\vdots & \vdots & \ddots & \vdots\\
i u_n' & 0 & \cdots & 0
\end{array}\right], \quad {\rm where } \quad  u_k\equiv u_k(x),\quad u_k' \equiv \frac{du_k(x)}{dx}.
\end{eqnarray}
\end{widetext}

Given this pair, the following calculations are tedious but straightforward. However, the pair of {\it Hermitian} operators $H$ and
$H_1$ for the system in Eq.~(\ref{P2-n}) was found by the author using a trial-and-error approach. Automation of this step for other systems of Painlev\'e equations thus remains an open problem. 

\subsection{Strategy}
The consistency conditions~(\ref{consHH}) allow one to define a state vector $|\Psi(t,x) \ra$ as a solution of the two-time Schr\"odinger equations \cite{Faddeev1987}
\begin{eqnarray}
\label{SE1}
i\frac{d|\Psi\ra}{dt} &=& H(t,x) |\Psi \ra,\\
\label{SE2}
i\frac{d|\Psi\ra}{dx} &=& H_1(t,x) |\Psi \ra,
\end{eqnarray}
where $H$ and $H_1$ are the Hermitian matrices given by Eq.~(\ref{HH-gen}). In what follows, they are referred to as Hamiltonians to stress that Eqs.~(\ref{SE1}) and~(\ref{SE2}), at real $t$ and $x$, describe unitary quantum-mechanical evolution. This allows application of the WKB approach based on the independent crossing approximation \cite{SinitsynPokrovsky2026QuasiAdiabaticEffects}, which never treats either $t$ or $x$ as a complex variable.

The evolution operator $U^{(n)}$  along a path $P$ in the two-time space $(t,x)$ is an $(n+1)\times(n+1)$ unitary matrix that can be written as a path-ordered exponent:
\be
U^{(n)}={\cal T}_{P} e^{-i\int_P \{H\,dt +H_1 \,dx \} },
\label{u-tx}
\ee
where ${\cal T}_{P}$ is the path-ordering operator, such that factors corresponding to earlier points along $P$ appear further to the right in the product of evolution exponents $e^{-i\{H\,dt +H_1 \,dx \} }$.

Let $|0\ra,|1\ra,\ldots,|n\ra$ be the diabatic states, i.e., the basis in which the Hamiltonians $H$ and $H_1$ are written, such that $H_{00}$ is the upper-left element of $H$, etc.
The elements of the evolution matrix are the transition amplitudes between the diabatic states. The full evolution matrix is not needed here. It is
sufficient to determine its first column,
\[
U_{00}^{(n)},\quad U_{10}^{(n)},\quad\ldots,\quad U_{n0}^{(n)},
\]
which is obtained by starting the evolution in state $|0\ra$. The amplitude of state $|k\rangle$ at the end is then $U_{k0}^{(n)}$.

\begin{figure}[t!]
    \centering
    \includegraphics[width=0.45\textwidth]{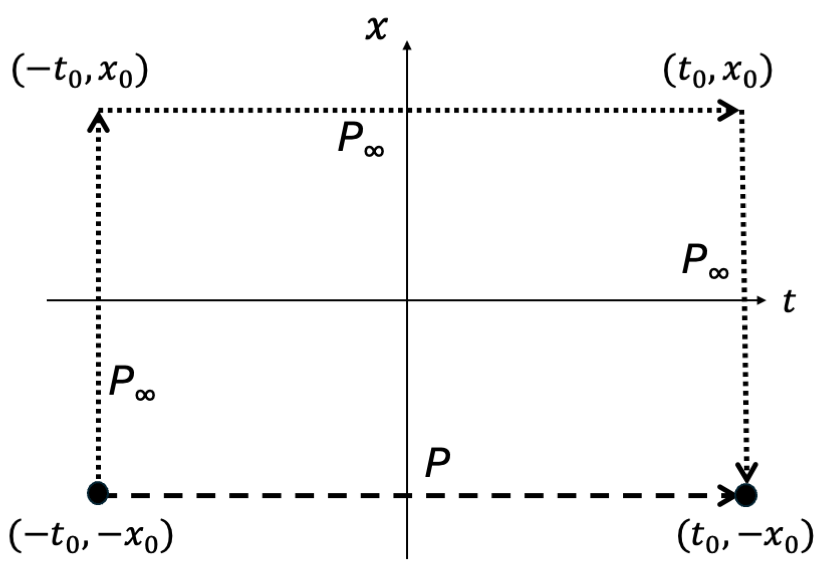}
    \caption{An integration path ${P}$ (dashed arrow) with $x_0\rightarrow -\infty$ and $t\in (-t_0,t_0)$, where $t_0\gg x_0$, is deformed into the path ${P}_{\infty}$, such that the horizontal segment of $ P_{\infty}$ lies at $x_0\rightarrow +\infty $ and
    $t\in (-t_0,t_0)$ (dotted arrows).  This deformation does not change the evolution operator in Eq.~(\ref{u-tx}), since the initial and final points of the path remain unchanged. The vertical legs of $P_{\infty}$ have $t=\pm t_0 \rightarrow \pm\infty$, which makes the evolution along them adiabatic. }
    \label{paths-fig}%
\end{figure}

 Equation~(\ref{u-tx}) can be written as 
\begin{equation}
U^{(n)}={\cal T}_{P} e^{\int_P {\bm A}\cdot d{\bm \tau}}, 
\label{gauge-exp}
\end{equation}
where ${\bm \tau} \equiv (t,x)$ is a point in the two-time space and ${\bm A} \equiv (-iH,-iH_1)$ is a non-Abelian field. This field is flat (has zero curvature) \cite{Sinitsyn2018}, so the evolution in Eq.~(\ref{gauge-exp}) depends only on the endpoints, which are $(-t_0,-x_0)$ and $(t_0,-x_0)$ in Fig.~\ref{paths-fig}, and not on the choice of path connecting them.
 
\vspace{0.2cm}
Thus, the evolution along the path $P$ shown in Fig.~\ref{paths-fig} can be calculated in two ways. First, one can consider $t$ varying over the interval $t\in (-t_0,t_0)$ at fixed negative $x=-x_0$. The evolution operator then takes the form
\be
U^{(n)}={\cal T}_{t} e^{-i\int_{-t_0}^{t_0} H(t,-x_0)\,dt }.
\label{u-tx2}
\ee

Alternatively, $U^{(n)}$ can be obtained as an evolution operator along the path $P_{\infty}$ in Fig.~\ref{paths-fig}. The vertical legs of $P_{\infty}$ correspond to evolution at fixed $t$. For example, evolution along the left vertical leg is described by the  operator 
$$
U_{vl}={\cal T}_{x} e^{-i\int_{-x_0}^{x_0} H_1(-t_0,x)\,dx },
$$
where the corner points of the paths are taken to $t_0,x_0 \rightarrow \infty$, with the limit $t_0 \rightarrow \infty$ taken first.

For $ t_0\rightarrow \infty$, the diabatic states coincide with the eigenstates of $H_1(-t_0,x)$. The evolution along
the left and right legs of $P_{\infty}$ is adiabatic, resulting only in phase factors $e^{\pm 2it_0x_0}$. Namely, the evolution matrix along the left leg starts at $x=-x_0\rightarrow -\infty$ and is given by
\be
U_{vL}=\left[
\begin{array}{cccc}
e^{2it_0x_0}&0&0&0\\
0&e^{-2it_0x_0}&0&0\\
\vdots&\vdots & \ddots& \vdots\\
0&0& \ldots&e^{-2it_0x_0}
\end{array}
\right].
\ee

Analogously, the evolution operator along the right leg of $P_{\infty}$ corresponds to $t=t_0$ and $x$ changing from large positive to large negative values, so
\be
U_{vR}=\left[
\begin{array}{cccc}
e^{2it_0x_0}&0&0&0\\
0&e^{-2it_0x_0}&0&0\\
\vdots&\vdots & \ddots& \vdots\\
0&0&\ldots&e^{-2it_0x_0}
\end{array}
\right].
\ee
Only the horizontal segment of $P_{\infty}$ contributes to the inter-state transitions. 

Both the horizontal part of $P_{\infty}$ and the path $P$ describe unitary quantum evolution over a pseudo-time $t\in (-t_0,t_0)$ at large positive and large negative $x$, respectively. Since the operator $H(t,x)$ depends on $u_{j}(x)$, the relations between the asymptotic values of $u_{j}(x)$ as $x\rightarrow \pm \infty$ are obtained by equating $U^{(n)}$ to $U_{{\infty}}^{(n)}$, where $U_{\infty}^{(n)}$ is the evolution matrix along the path $P_{\infty}$ in Fig.~\ref{paths-fig}. Since the interstate transition amplitudes are determined by the horizontal paths at either large positive or large negative $x$, both evolution operators can be evaluated using the WKB approach, which becomes exact in the limits $x\rightarrow \pm \infty$.

To calculate the scattering amplitudes $U^{(n)}_{j0}$ using this asymptotically exact WKB approach, the functions $u_{j}(x)$ in the Hamiltonians $H$ and $H_1$, as $x\rightarrow \pm\infty$, were assumed to take the form given by Eqs.~(\ref{u1-largex}) and~(\ref{u2-largex}), with unknown parameters including logarithmic phase contributions at positive $x$. In these limits, the evolution was found to be mostly adiabatic, except at a finite set of pairwise avoided crossings between energy levels of the Hamiltonian $H$.
Near each avoided crossing, the couplings were expressed in terms of the parameters of the asymptotic solutions, and the scattering amplitudes were estimated using the known scattering matrix for the Landau--Zener model.

After the amplitudes $U_{k0}^{(n)}$, where $k=0,1,\ldots,n$, are obtained separately for evolution along $P$ and $P_{\infty}$, the WKB results are compared order by order in powers of $x$ and $t_0$, down to the $O(1)$ contributions to the phases and the absolute values of the transition amplitudes. Terms that depend on $t_0$ and on high powers of $x$ then cancel between the two expressions, leaving the connection formulas in Eqs.~(\ref{sigma-fin2})--(\ref{phi2-fin2}) and fixing the logarithmic $x$ dependence in Eqs.~(\ref{u1-largex}) and~(\ref{u2-largex}).

\section{WKB analysis in the limit $x\rightarrow -\infty$}
\label{sec-wkb-xm}
\subsection{Adiabatic Hamiltonian}
 For large negative $x$, there are two resonant regions for $t\in (-t_0,t_0)$ near 
 $
 t=\pm \sqrt{|x|}/2,
 $
where all nonadiabatic transitions occur. For large $|x|$, these resonances are well
 separated. By rescaling time 
\be
t=\tau \sqrt{|x|},
\label{t-resc1}
\ee
the Schr\"odinger equation (\ref{SE1}) is transformed to 
$$
i\frac{d}{d\tau} |\Psi \ra =H(\tau) |\Psi \ra,
$$
where $|x|$ becomes a large parameter whose presence justifies the WKB analysis, which becomes asymptotically exact as $x\rightarrow -\infty$.
The resonances with nonadiabatic transitions shift in $H(\tau)$ to time points $\tau=\pm 1/2$.

\begin{figure}[t!]
    \centering
    \includegraphics[width=0.55\textwidth]{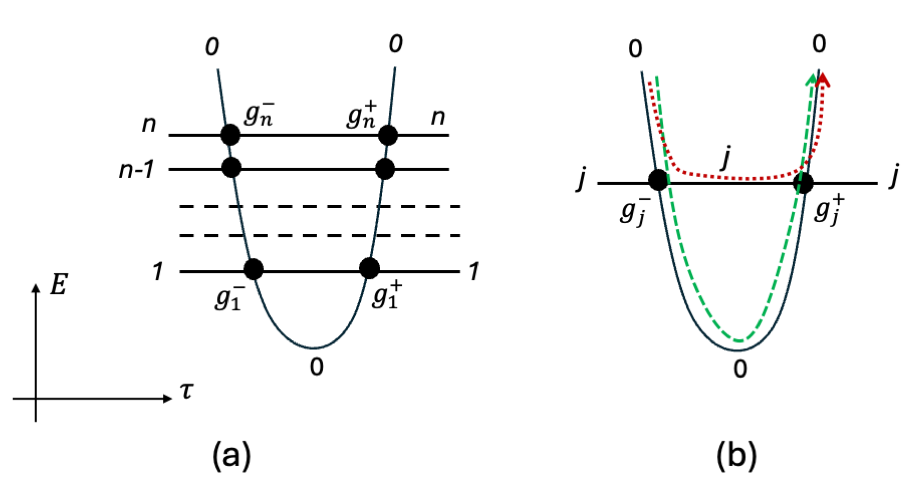}
    \caption{(a) Adiabatic energy levels of the Hamiltonian $H_a+|x|^{3/2}(4\tau^2-1)\hat{1}$, in which the parallel levels with indices $1,\ldots,n$ are shown as horizontal lines. The $0$-th level crosses all of them near the times $\tau=\pm 1/2$. At the crossing points (black circles), the adiabatic basis is not defined. Instead, the diabatic basis is used in small intervals around the crossing points, where levels with indices $j\ge 1$ are coupled to level $0$ with couplings $g_j^{\pm}$. (b) Only levels $j$ and $0$ are shown to illustrate the semiclassical trajectories. Both trajectories start at level $0$. The green arrow shows the trajectory that always stays on the $0$-th adiabatic level without turning to any other level, while the purple arrow shows a trajectory that turns to level $j$ at the first crossing point and then returns to level $0$ without turning to any other level. The phase difference between the amplitudes of these trajectories is $2\Phi_j^{(n)}$.}
    \label{scheme-fig}
\end{figure}

Adding to the Hamiltonian a term proportional to the unit matrix does not affect transition probabilities. Therefore, it is convenient 
to present the spectrum of $H(\tau)+|x|^{3/2}(4\tau^2-1)\hat{1}$, where $\hat{1}$ is the unit matrix, instead of that of $H(\tau)$ alone. This transformation makes the energy levels for $k=1,\ldots,n$ horizontal except in the vicinity of the time points $\tau=\pm 1/2$, as shown schematically in Fig.~\ref{scheme-fig}(a).

As $x\rightarrow-\infty$, there are two  regions near
$\tau=-1/2$ and $\tau=1/2$ with avoided level crossings (black circles in Fig.~\ref{scheme-fig}(a)).  Away from $\tau=\pm 1/2$ points, the adiabatic approximation applies, and the effective  Hamiltonian is diagonal. Its elements are the energies calculated to second order in perturbation theory, which is the relevant order here. Appendix~\ref{app-heff1} shows that the effective Hamiltonian
for the adiabatic regions is given by 
\be
\label{ha}
\small{H_{{\rm a}}(\tau) = \left[\begin{array}{cccc}
|x|^{3/2}(4\tau^2-1)+\frac{2\sum_{k=1}^n \alpha_k^2}{4\tau^2-1}&0&\cdots &0\\
0& -|x|^{3/2}(4\tau^2-1)-\frac{2\alpha_1^2}{4\tau^2-1} +2\sqrt{|x|}\varepsilon_1& 0&\ldots \\
\vdots &0&\ddots& \vdots\\
0&0&\cdots& -|x|^{3/2}(4\tau^2-1)-\frac{2\alpha_n^2}{4\tau^2-1}+2\sqrt{|x|}\varepsilon_n
\end{array}
\right].}
\ee

The {\it diabatic states} are the states in the basis in which the
original Hamiltonian $H$  is written in Eq.~(\ref{HH-gen}), while the {\it adiabatic states} are
the solutions of the time-dependent Schr\"odinger equation with the
effective Hamiltonian~(\ref{ha}). In each basis, states have indices
$0,1,\ldots,n$, such that
$$
\varepsilon_1<\ldots<\varepsilon_n,
$$
so,  away from the intervals around $\tau=\pm 1/2$ and for $j>k$, the $j$-th adiabatic energy is higher than the $k$-th adiabatic energy.

Note that, due to the large value of $|x|$, the diabatic and adiabatic basis states coincide, up to a time-dependent phase, almost everywhere except in
the vicinity of the points $\tau^{\pm}_j \approx \pm 1/2$, where $j=1,\ldots,n$. The independent crossing approximation \cite{SinitsynPokrovsky2026QuasiAdiabaticEffects} splits the evolution into pieces governed by the
adiabatic Hamiltonian (\ref{ha}), interrupted by intervals $\tau \in (-\eta +\tau^-_j,\tau_j^- +\eta)$, which will be called ${\rm Int}_j^-$ 
and $\tau\in (-\eta +\tau_j^+,\tau_j^++\eta)$, called ${\rm Int}_j^+$, that contain pairwise crossings of level $0$ with level $j$. 
Here, $\eta$ can be considered vanishingly small in the limit $|x|\rightarrow \infty$ but
sufficiently large in order to saturate pairwise transition amplitudes between levels $j$ and $0$ for evolution over each of these two intervals. In physics, this approach is usually used as an approximation but in the present case it becomes asymptotically exact as $x\rightarrow \pm \infty$. 

Let $|k\ra$, where $k=0,1,\ldots,n$, be the time-independent diabatic states.
The adiabatic states are defined as
\be
|k_a(\tau)\ra \equiv e^{-i{\rm  P}\int_{-T}^\tau E_{k}^{(n)}(\tau') \, d\tau'} |k\ra,
\label{adiab-def}
\ee
where $E_{k}^{(n)}(\tau)=[H_{{\rm a}}(\tau)]_{kk}$ is the adiabatic energy, e.g., 
$$E_{0}^{(n)}(\tau)=|x|^{3/2}(4\tau^2-1)+\frac{2\sum_{k=1}^n \alpha_k^2}{4\tau^2-1};
$$
$$
T\equiv  t_0 /\sqrt{|x|}, 
$$
and 
``${\rm P}$" in Eq.~(\ref{adiab-def}) means the principal part of the integral that excludes integration
over the intervals 
\be
{\rm Int}_k^{-}=(\tau_k^{-}-\eta,\tau_k^{-}+\eta ),\qquad
{\rm Int}_k^{+}=(\tau_k^{+}-\eta,\tau_k^{+}+\eta ),
\label{interv-mk}
\ee
where $\tau^{\pm}_k$ are the degeneracy times at which, to the leading order in $|x|$, 
$$
E_{0}^{(n)}(\tau^{\pm}_k)=E_{k}^{(n)}(\tau^{\pm}_k),
$$
and the sign $\pm$ corresponds to points near $\tau=\pm 1/2$.

The parameter $\eta$ in Eq.~(\ref{interv-mk}) is chosen to be much smaller than the time interval between successive crossing points, i.e., $0<\eta\ll|\tau^{\pm}_k-\tau^{\pm}_{k+1}|$, but sufficiently large to saturate the Landau--Zener transitions. In the limit $|x|\rightarrow \infty$, such a value of $\eta$ can always be found.
The evolution over the intervals in Eq.~(\ref{interv-mk}) is then described by the textbook Landau-Zener scattering matrix \cite{SinitsynPokrovsky2026QuasiAdiabaticEffects}. 

\subsection{Hamiltonians near $\tau=\pm 1/2$}
Near the points $\tau=\pm 1/2$, one has to retain the off-diagonal elements and thus obtains Hamiltonians of the Demkov--Osherov model \cite{SinitsynPokrovsky2026QuasiAdiabaticEffects}. Namely, after redefining time as
$$
s=\tau+1/2, 
$$
the effective Hamiltonian near $s=0$, to leading order in $|x|$, is given in the diabatic basis by
\be
\label{hm1d}
\small{H_{-1/2}(s) = \left[\begin{array}{cccc}
|x|^{3/2}(-4s) &2|x| u_1-2i\sqrt{|x|}u_1'&\cdots &2|x| u_n-2i\sqrt{|x|}u_n'\\
2|x| u_1+2i\sqrt{|x|}u_1'& |x|^{3/2}4s +2\sqrt{|x|}\varepsilon_1& 0&\ldots \\
\vdots &0&\ddots& \vdots\\
2|x| u_n+2i\sqrt{|x|}u_n' &0&\cdots& |x|^{3/2}4s+2\sqrt{|x|}\varepsilon_n
\end{array}
\right]},
\ee
and for 
$$
s= \tau-1/2, 
$$
the effective Hamiltonian near $s=0$, to leading order in $|x|$, is given by

\be
\label{hp12}
H_{+1/2}(s) = \left[\begin{array}{cccc}
|x|^{3/2} 4s &-2|x| u_1-2i\sqrt{|x|}u_1'&\cdots &-2|x| u_n-2i\sqrt{|x|}u_n'\\
-2|x| u_1+2i\sqrt{|x|}u_1'& -|x|^{3/2}4s +2\sqrt{|x|}\varepsilon_1& 0&\ldots \\
\vdots &0&\ddots& \vdots\\
-2|x| u_n+2i\sqrt{|x|}u_n' &0&\cdots& -|x|^{3/2}4s+2\sqrt{|x|}\varepsilon_n
\end{array}
\right].
\ee

Both $H_{-1/2}(s)$ and $H_{+1/2}(s)$ describe a single energy level crossing a band of $n$ parallel levels linearly, with time-independent pairwise couplings. The scattering matrix for the DOM is known analytically, but only one of its properties is used here. Namely, the exact DOM scattering matrix can be reconstructed using the independent crossing approximation, in which the individual pairwise avoided crossings are treated independently, disregarding the presence of other energy levels, while the evolution between successive crossings is considered adiabatic.

In the independent crossing approximation, each encountered pairwise diabatic level crossing is treated as an independent two-state Landau--Zener transition between the crossing energy levels, followed by adiabatic evolution between successive level crossings \cite{SinitsynPokrovsky2026QuasiAdiabaticEffects}. Therefore, the calculation of the elements of the evolution matrix $U^{(n)}$ reduces to identifying semiclassical trajectories connecting the initial and final states and then finding the corresponding quantum amplitudes. Two semiclassical trajectories connecting the initial state $|0\ra$ to the final state $|0\ra$ are shown by the green and purple arrows in Fig.~\ref{scheme-fig}(b).

By comparing the Hamiltonians $H_{\pm 1/2}$ with Fig.~\ref{scheme-fig}(a), the pairwise couplings can be expressed through the asymptotic values of the functions $u_k(x)$:

\be
\label{gmp-def}
g_j^-
\equiv
(H_{-1/2})_{0j}
=
2\left(
|x|u_j-i\sqrt{|x|}\,u_j'
\right), \quad 
g_j^+
\equiv
(H_{+1/2})_{0j}
=
-2\left(
|x|u_j+i\sqrt{|x|}\,u_j'
\right).
\ee
Substituting the asymptotic values of $u_j$ from Eq.~(\ref{u1m}) gives the absolute values and phases of the couplings:
\be
\label{g-abs}
|g_j^-|
=
|g_j^+|
=
2\alpha_j|x|^{3/4}
\ee
\be
{\rm arg}\,g_j^-
=
\frac{\pi}{2}-\theta_j,
\qquad
{\rm arg}\,g_j^+
=
\frac{\pi}{2}+\theta_j,
\label{app-lz-gamma-phases}
\ee
where
\begin{equation}
\theta_j(x)=
\frac{2}{3}|x|^{3/2}
+
\varepsilon_j\sqrt{|x|}
+
\frac{
3\alpha_j^2+
2\sum_{\substack{k=1\\k\ne j}}^n\alpha_k^2
}{4}\ln|x|
+
\varphi_j.
\label{app-lz-theta-j}
\end{equation}

The transition amplitudes between diabatic states are obtained using the scattering matrix for the Landau--Zener model.
The scattering and evolution matrices for the Landau--Zener model are
listed in Appendix~\ref{sec-LZ-diab}. Their derivation can be found in textbooks on quantum mechanics, e.g., in \cite{SinitsynPokrovsky2026QuasiAdiabaticEffects}. Between such pairwise scatterings, the system undergoes adiabatic evolution, as already discussed.

For the level slopes $\beta =\pm 4|x|^{3/2}$, the Landau--Zener probability to remain on the same diabatic level after the crossing is \cite{SinitsynPokrovsky2026QuasiAdiabaticEffects}
\be
p_j=e^{-\pi |g^{\pm}_j|^2/|\beta|}=e^{-\pi \alpha_j^2},
\label{pj-xm}
\ee
while the probability to transfer to the other level is $1-p_j$. This explains the appearance of the parameters $p_k$ and $q_k$ in Eq.~(\ref{inprob-def}) and in the connection formulas.

Landau--Zener scattering is usually considered in the diabatic basis, but the adiabatic basis is preferable for the present problem. The adiabatic states~(\ref{adiab-def}) coincide with the diabatic states up to the adiabatic phase, which modifies the off-diagonal elements of the scattering matrix, as described in Appendix~\ref{sec-LZ-diab}. One complication is that the adiabatic phase is formally singular precisely at the avoided-crossing point. Therefore, when Landau--Zener evolution is considered over an interval of size $2\eta$ around an avoided crossing, the off-diagonal elements of the evolution matrix over this pseudo-time interval depend on $\ln \eta$. The parameter $\eta$, however, cancels when the subsequent contribution from purely adiabatic evolution is included. Hence, the final scattering phases for evolution over $t\in (-t_0,t_0)$ are independent of $\eta$.

\subsection{Amplitude to stay on the $0$-th adiabatic level}
\label{inductive-minusx}
There are interfering semiclassical trajectories for any transition starting at the $0$-th diabatic state and ending in any state $j=0,1,\ldots,n$.
Consider first the amplitude of the green trajectory in Fig.~\ref{scheme-fig}(b), which always stays on the $0$-th adiabatic level.
In absolute value, the Landau--Zener transition amplitude at the crossing of levels $0$ and $j$ is $\sqrt{p_j}$. Since there are two such crossings along this trajectory, level $j$ contributes a factor $(\sqrt{p_j})^2=p_j$, and the absolute value of the entire green-trajectory amplitude is $\prod_{j=1}^{n} p_j$. Analogously, the absolute value of the amplitude of the trajectory that starts and ends on level $0$ but propagates through level $j$ is $q_j\prod_{k=j+1}^{n}p_k$.

However, different trajectories with the same initial and final conditions have different phases and interfere.  This interference is contained in the set of phases 
$\Phi_j^{(n)}$. By definition, $\Phi_j^{(n)}$ is half the phase difference between the green and purple trajectories in Fig.~\ref{scheme-fig}(b), i.e., between the amplitude of the trajectory that turns to level $j$ and then returns to level $0$ and the amplitude of the trajectory that always stays on level $0$. Their values are derived in Appendix~\ref{app-phases} and listed in Eq.~(\ref{app-phase-final}).

Let $\phi_{00}(t_0)$ be the adiabatic phase of the trajectory that starts at $\tau=-T$ and ends at $\tau=T$, where $T\equiv t_0/\sqrt{|x|}$, while always staying on the $0$-th adiabatic energy level. It is evaluated as an integral over the adiabatic energy $E_0^{(n)}$ in Appendix~\ref{app-phi00}, while the nonadiabatic intervals ${\rm Int}_j^{\pm}$ make a negligible contribution to this phase in the limit $|x| \rightarrow \infty$:
\begin{equation}
\phi_{00}(t_0;x)
=
-\frac{8}{3}t_0^3
+
2|x|t_0.
\label{phi00-neg}
\end{equation}
It does not depend on $n$ but depends only on large $t_0$ and $x$ values that will eventually cancel later when compared with similar amplitudes as $x\rightarrow +\infty$.

This part of the phase is the same for all semiclassical trajectories that start and end on the $0$-th level. Thus, using the definition of the phases $\Phi_j^{(n)}$, the amplitude to remain on the $0$-th level is given by the sum of $n+1$ trajectory amplitudes:
\be
U_{00}^{(n)} =e^{i\phi_{00}} \left[\prod_{j=1}^n p_j
+
\sum_{k=1}^n
q_k
\left(
\prod_{j=k+1}^n p_j
\right) e^{2i\Phi_j^{(n)}}   \right].
\label{u00-xm}
\ee

\subsection{Transition amplitudes to levels for $k>0$}

  \begin{figure}[t!]
    \centering
    \includegraphics[width=0.55\textwidth]{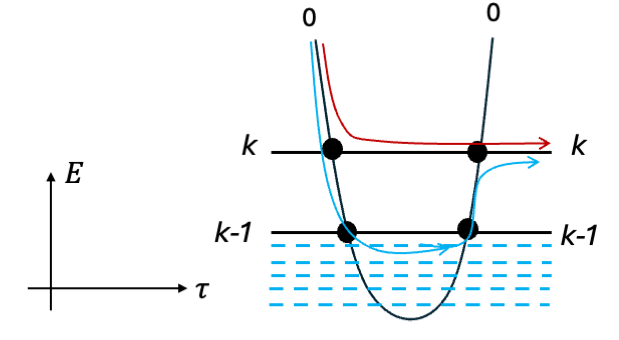}
    \caption{The amplitude of the transition from $0$ to $k$ can be viewed as the sum of the amplitude of the elementary (purple) trajectory that starts on $0$, turns to $k$ and then stays there. 
    The blue color represents a block of trajectories that start as a single path at $0$ until it crosses level $k$ without turning. This block represents all trajectories that reemerge on level 0 before crossing level $k$ second time and then staying on $k$. }
    \label{recursion-fig}
\end{figure} 

The term with index $k$
corresponds to the trajectory that propagates through level $k$.
Introduce amplitudes that resemble $U_{00}^{(k-1)}$ in Fig.~\ref{recursion-fig} but with phases $\Phi_j^{(n)}$ taken for $n$ levels:
\be
C_0^{(n)}=1,
\label{ind-C0}
\ee
and, for $k\geq2$,
\be
C_{k-1}^{(n)}
=
\prod_{j=1}^{k-1}p_j
+
\sum_{r=1}^{k-1}
q_r
\left(
\prod_{m=r+1}^{k-1}p_m
\right)
e^{2i\Phi_r^{(n)}}.
\label{ind-Ck}
\ee
This is the survival amplitude constructed from the first $k-1$
levels, with all phases evaluated at their final $n$-level values. 

The amplitude of a transition to level $k$ is the sum of two amplitudes, illustrated in Fig.~\ref{recursion-fig}.
First, there is a single trajectory (purple) that passes through all levels with indices $j>k$, then turns on 
the $k$-th level and stays there. The amplitude of this trajectory can be written as 

$$
e^{i\phi_{k0}^{(n)}}\sqrt{p_kq_k\prod_{j=k+1}^n p_j}e^{2i\Phi_k^{(n)}}.
$$

Second, the system can initially pass through the levels with $j\ge k$, return to level $0$ before the second crossing of levels $k$ and $0$, and turn to level $k$ at this crossing (blue block in Fig.~\ref{recursion-fig}).
This turning amplitude acquires an extra minus sign in comparison to the turning amplitude of the previous trajectory as discussed in Appendix~\ref{app-phik0-minus-sign}. 

Therefore, the transition amplitude to level $k$ can be written as
\begin{widetext}
\be
U_{k0}^{(n)}
=
e^{i\phi_{k0}^{(n)}}\sqrt{
p_kq_k
\prod_{j=k+1}^n p_j
}
\left(
e^{2i\Phi_k^{(n)}}-C_{k-1}^{(n)}
\right), 
\label{ind-Uk0-general}
\ee
\end{widetext}
where, according to Appendix~\ref{app-phik0},
\begin{equation}
\phi_{k0}^{(n)}=-2\varepsilon_kt_0-\Phi_k^{(n)} -\frac{\pi}{2},
    \label{phi-k0}
\end{equation}
is the common phase of all trajectories starting at level $0$ and ending at level $k$. This phase is derived  by calculating the adiabatic phases as integrals of the adiabatic energies over time and also accounting for the off-diagonal phase of the Landau--Zener evolution matrix through the intervals ${\rm Int}_j^{\pm}$.

This completes derivation of the required amplitudes $U_{j0}^{(n)}$, $j=0,1,\ldots,n$, for the time evolution along the path $P$ in Fig.~\ref{paths-fig}.
Later, when  comparing amplitudes as $x\rightarrow \pm \infty$, the terms that depend on $t_0$ in phases will cancel but the contributions $O(1)$ will eventually contribute to the final connection formulas. 

\section{WKB analysis as $x\rightarrow +\infty$}
\label{sec-wkb-xp}
\subsection{Radial basis for $x\rightarrow+\infty$}
\label{plusx-radial-sec}

\begin{figure}[t!]
    \centering
    \includegraphics[width=0.55\textwidth]{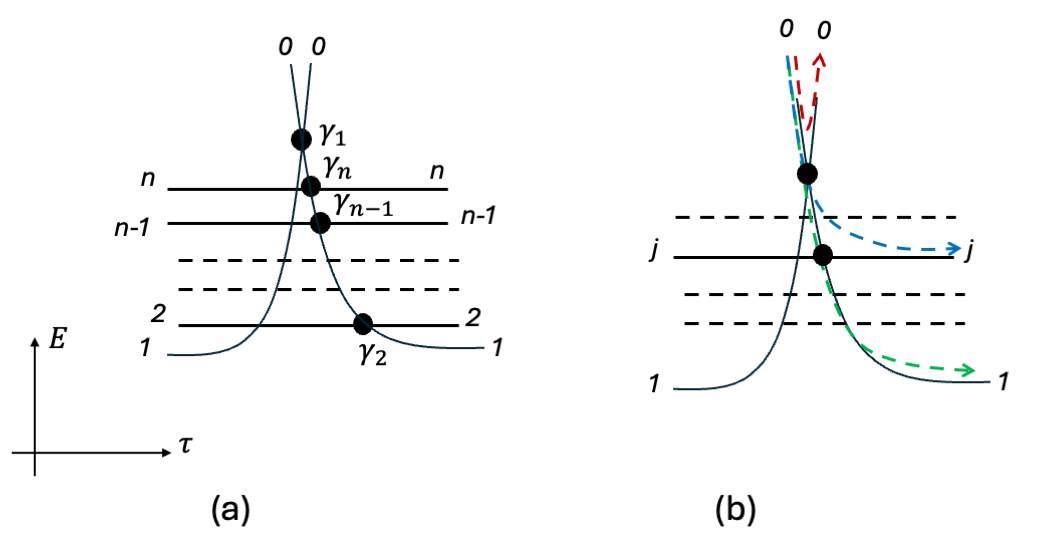}
    \caption{(a) Adiabatic energy levels of the Hamiltonian $H(x)$ as $x\rightarrow +\infty$, in which parallel levels with indices $2,\ldots,n$ are shown as horizontal lines.
    This time, level $1$ crosses all other levels, including level 0, near $t=0$. The crossing pattern shows two Demkov-Osherov regions, in which level $1$ crosses linearly $(n-1)$ parallel levels. 
    (b) Only levels $j$, $1$, and $0$ are shown to illustrate the absence of interference between semiclassical trajectories that start at level $0$ and end on a given level. Purple shows the only semiclassical trajectory that starts at $0$ and, after crossing level $1$, ends at $0$.
    The blue dashed arrow shows the trajectory that starts at level $0$, then turns to $1$, crosses levels $j+1,\ldots, n $ and then turns to level $j$. The green arrow corresponds to level $1$ that becomes populated after crossing with level $0$ and then stays at $1$ after crossing all levels $2,\ldots n$. The amplitude of this particular trajectory is used to infer the sign parameter $\sigma$ in the connection formulas.}
    \label{scheme-xp-fig}
\end{figure}

The complication in the positive limit, $x\rightarrow+\infty$, is that the off-diagonal elements of $H_{nm}(x,t)$ for $n\ne m$ are no longer irrelevant. In the original diabatic basis, they couple all states to one another.
Therefore, it is convenient to choose a basis in which, as $x\rightarrow+\infty$, the states with indices $k=2,\ldots,n$ decouple from one another. Only after switching to this basis does the pattern of the new diabatic energies, i.e., the diagonal elements of the Hamiltonian, resemble the pattern formed by the exact eigenvalues, as shown in Fig.~\ref{scheme-xp-fig}(a).

Introduce the new state vector $|R\ra$,
\be
|R\ra
\equiv
\sum_{j=1}^{n}\frac{u_j(x)}{R(x)}|j\ra,
\ee
where $|j\ra$ are the original diabatic basis states, and
\be
R(x)\equiv
\sqrt{\sum_{j=1}^{n}u_j^2(x)},
\label{main-ra-R}
\ee
Let $|a\ra$, where $a=2,\ldots,n$, be real orthonormal vectors constructed from the original diabatic states $|j\ra$, where $j=1,\ldots,n$, and orthogonal to $|R\ra$. Their signs are fixed by requiring that each $|a\ra$ in the new basis tends to the corresponding original diabatic state $|j\ra$ at $j=a$ as $x\rightarrow+\infty$, for either value of $\sigma$.
Thus, the new basis is
\be
\{|0\ra,|2\ra,\ldots,|n\ra,|R\ra\},
\label{main-ra-basis}
\ee
where $|0\ra$ is the original level-$0$ state, which is left unchanged.

This change of basis can be represented by the matrix
\be
V(x)=
\left[
\begin{array}{cc}
1&0\\
0&{\cal V}(x)
\end{array}
\right],
\qquad
H^{(+)}(t,x)=V(x)H(t,x)V^{\dagger}(x).
\label{main-ra-H-transform}
\ee
A convenient explicit choice of ${\cal V}$, with rows ordered as
$|2\ra,\ldots,|n\ra,|R\ra$, is
\be
{\cal V}_{a-1,b}
=
\delta_{ab}
-
\frac{u_a(u_b+\sigma R\delta_{b1})}{R(R+\sigma u_1)},
\qquad
a=2,\ldots,n,\quad b=1,\ldots,n,
\qquad
{\cal V}_{nb}=\frac{u_b}{R},
\label{main-ra-Vblock}
\ee
where $\delta_{b1}$ is the Kronecker symbol.
Here $\sigma$ is the asymptotic sign of $u_1$ that will be defined in
Eq.~(\ref{app-ra-sigma}), so that $R+\sigma u_1\sim2R$.
This matrix is orthogonal and rotates $(u_1,\ldots,u_n)$ exactly
to $(0,\ldots,0,R)$. In particular, the rows for states $|a\rangle$  tend to the
original states without a sign factor, whereas the radial row
tends to $\sigma |1\rangle^T$.

Let's Introduce $n-1$ components of the ``angular momentum":
\be
L_a=u_1u_a'-u_au_1', \quad a=2,\ldots,n.
\label{ldef}
\ee
and let
$
Q(t,x)\equiv4t^2+x. 
$
From
the asymptotic solution in Appendix~\ref{app-plusx}, 
$R\sim\sqrt{x/2}$. In the basis~(\ref{main-ra-basis}), the Hamiltonian $H$ 
for $x\rightarrow+\infty$ becomes
\be
H^{(+)}
=
\left[
\begin{array}{c|cccc|c}
Q-2R^2
&
-2i\sigma L_2/R
&
-2i\sigma L_3/R
&
\cdots
&
-2i\sigma L_n/R
&
-4tR-2iR'
\\ \hline
2i\sigma L_2/R
&
-Q+2\varepsilon_2
&
0
&
\cdots
&
0
&
0
\\
2i\sigma L_3/R
&
0
&
-Q+2\varepsilon_3
&
\cdots
&
0
&
0
\\
\vdots
&
\vdots
&
\vdots
&
\ddots
&
\vdots
&
\vdots
\\
2i\sigma L_n/R
&
0
&
0
&
\cdots
&
-Q+2\varepsilon_n
&
0
\\ \hline
-4tR+2iR'
&
0
&
0
&
\cdots
&
0
&
-Q+2R^2
\end{array}
\right].
\label{app-ra-H-plus-expanded}
\ee
Thus, to leading order, the states $|a\ra$, where $a=2,\ldots,n$, form parallel diabatic energy levels that are uncoupled from one another and have energies $-Q+2\varepsilon_a$. Each state
$|a\ra$
couples to the zeroth state through its own coupling
$\propto L_a/R$.  The radial state couples strongly to the zeroth
state through $R$ and $R'$ in the upper-right and bottom-left matrix elements of $H^{(+)}$. 

In the Hamiltonian~(\ref{app-ra-H-plus-expanded}), the state $|R\ra$ appears to be decoupled from the states $|a\ra$, where $a=2,\ldots,n$. However, near the times at which their diabatic energies cross, the relative level slopes are of order $1/x$. Therefore, even when the coupling between them is of order $\gamma_k\sim 1/\sqrt{x}$, the probability of a nonadiabatic transition is substantial. At this order, both the direct matrix element $2\varepsilon_a u_a/R$ omitted from Eq.~(\ref{app-ra-H-plus-expanded}) and the virtual transition through state $|0\ra$ must be retained, as derived in Appendix~\ref{app-Hm-eff}. Therefore, the energy-level diagram
can be schematically depicted as in Fig.~\ref{scheme-xp-fig}, where the state $|R\ra$ is associated with energy level $1$. 

It is important to note that, as $x\rightarrow+\infty$, the radial
state becomes the original diabatic state $|1\ra$ up to a sign:
\be
|R\ra
=
\sum_{j=1}^{n}\frac{u_j}{R}|j\ra
\longrightarrow
\sigma |1\ra,
\qquad
\sigma\equiv
\lim_{x\rightarrow+\infty}
\frac{u_1(x)}{\sqrt{x/2}}
=\pm1.
\label{app-ra-sigma}
\ee
The factor $\sigma$ must therefore be retained when the final
radial state $|R\rangle$ is expressed through the original state $|1\ra$. In addition, the same factor appears when the radial oscillation
is expressed through $u_1$. Hence, both the slowly growing and the oscillating
contributions to $u_1(x)$ carry the same factor $\sigma$ in Eq.~(\ref{u1-largex}).

For a trajectory ending on state $j>1$, the radial state $|R\rangle$ is
only an intermediate state. Its sign convention cannot
produce an additional overall sign: changing
$|R\ra\rightarrow-|R\ra$ changes the signs of both the
entering and leaving matrix elements, leaving their product
unchanged. In the convention~(\ref{main-ra-Vblock}), neither the
angular coupling phase nor the final angular-state conversion
introduces an extra $\sigma$. This agrees with the definition of
$u_j(x)$ at $j>1$ in Eq.~(\ref{u2-largex}).
As a check, changing $u_1\rightarrow-u_1$ conjugates the original
Hamiltonian by ${\rm diag}(1,-1,1,\ldots,1)$. Consequently,
$U_{10}^{(n)}$ changes sign, whereas $U_{j0}^{(n)}$ for $j>1$
does not, as required by the formulas below.

\subsection{Pairwise avoided crossings}
The limit $x\rightarrow +\infty$ is also simpler in one respect because, as illustrated in Fig.~\ref{scheme-xp-fig}, there is no interference between different semiclassical trajectories that start at level $0$ and end at any given level.
There are regions of nonadiabatic transitions between pairs of levels. Such avoided crossings are well separated from one another both in energy and in time $t$. The first region is at $t=0$, where levels $0$ and $1$ undergo a single avoided crossing. In the ordered basis $(|{-}\ra,|{+}\ra)$, where $|\mp\ra=(|0\ra\mp|R\ra)/\sqrt{2}$, the effective Hamiltonian is the Landau--Zener Hamiltonian:
\be
H_0^+ =\left[
\begin{array}{cc}
\beta_1 t& \gamma_1\\
\gamma_1^* &-\beta_1 t
\end{array}
\right].
\label{lz-tp}
\ee
Appendix~\ref{LZ01} shows that 
\be
\beta_1=4\sqrt{x/2}, \quad \gamma_1=2(2x)^{1/4} \rho e^{-i(\pi+\theta_1)},
\label{betag0}
\ee
where 
\be
\theta_1(x) =\frac{2\sqrt{2}}{3} x^{3/2} - \frac{3\rho^2}{2} \ln x +\phi_1.
\label{theta1p}
\ee
This defines the Landau--Zener survival probability at the first avoided crossing:
\be
P_1=e^{-\pi |\gamma_1|^2/\beta_1} = e^{-2\pi \rho^2} = e^{-4\pi I_1}.
\label{P10}
\ee

The other crossing points occur at much later times,
\[
t_k^{+} \approx x/(4\sqrt{\varepsilon_k}),\quad k=2,\ldots,n.
\]
Here, after turning from level $0$, level $1$ crosses the $(n-1)$ parallel levels with indices $k=2,\ldots,n$.
Let $t=s+t_{k}^+$.
The region near $s=0$ is described by the two-state Landau--Zener model in the ordered basis $(|R\ra,|k\ra)$:
\be
H_k^+ =
\left[
\begin{array}{cc}
-\beta_k s& \gamma_k^*\\
\gamma_k &\beta_k s
\end{array}
\right].
\label{do-tp}
\ee
Here $\gamma_k=(H_k^+)_{kR}$. Appendix~\ref{app-Hm-eff} estimates that
\be
\beta_k = \frac{8\varepsilon_k^{3/2}}{x},\quad
\gamma_k =\frac{2\sqrt{2}\varepsilon_kA_k}{\sqrt{x}} e^{-i\theta_k},\quad k=2,\ldots,n,
\label{betak}
\ee
where
\be
\theta_k(x) = \sqrt{\varepsilon_k}x -I_k \ln x +\phi_k,\qquad I_k=\frac{A_k^2\sqrt{\varepsilon_k}}{2}, 
\label{thetak}
\ee
and the corresponding Landau--Zener survival probability at the crossing of levels $1$ and $k$ is

\be 
P_k=e^{-\pi |\gamma_k|^2/\beta_k} = e^{-2\pi I_k}.
\label{Pkn}
\ee

Interestingly, the crossings of the levels with indices $k=2,\ldots,n$ by level $1$ are not described by the Demkov--Osherov model. Rather, the effective Hamiltonian restricted to the state $|R\ra$ and the parallel states $|a\ra$, $a=2,\ldots,n$, has the structure of the multistate Coulomb model solved by Ostrovsky in \cite{Ostrovsky2003}. In this model, one diabatic level with Coulombic time dependence crosses several parallel levels and couples to them by constant matrix elements. For fixed distinct $\varepsilon_j$, however, the exact multistate solution is not needed because the independent crossing approximation applies. Indeed, the crossing times satisfy $|t_k^+-t_{k+1}^+|\propto x$, whereas the saturation time of an individual Landau--Zener transition is of order $\gamma_k/\beta_k\propto\sqrt{x}$ \cite{SinitsynPokrovsky2026QuasiAdiabaticEffects}. Thus, individual Landau--Zener transitions have sufficient time to saturate before another crossing is encountered. Ostrovsky's solution may become useful when several $\varepsilon_j$ are close enough that the corresponding crossings form a cluster of levels with energy spacings $\sim 1/\sqrt{x}$; such situations are not considered here.

\subsection{Transition amplitudes as $x\rightarrow +\infty$}
The amplitude of a transition from level $0$ to level $0$ along the contour $P_{\infty}$ can now be written as
\be
U_{00}^{(n)}=e^{i\phi_{00}^+}\sqrt{1-P_{1}},
\label{U00-xp}
\ee
where $P_{1}$ is defined in Eq.~(\ref{P10}), and $\phi_{00}^+$ is the sum of a Landau--Zener off-diagonal phase and the adiabatic phase along the purple trajectory in Fig.~\ref{scheme-xp-fig}(b). Its calculation in Appendix~\ref{app-phase00p} gives
\be
\phi_{00}^+=-\frac{8t_0^3}{3}
+2xt_0
-\frac{7\rho^2}{2}\ln2
-\frac{3\pi}{4}
+{\rm arg}\Gamma(i\rho^2)
-\phi_1.
\label{ph00-xp}
\ee

To end on level $j>0$, the system must pass to level $1$ after its crossing with level $0$, remain on level $1$ while crossing levels $k=j+1,\ldots,n$, and then turn to level $j$. The amplitude of this transition can be written as
\be
U_{j0}^{(n)}=e^{i\phi_{j0}^+}\sqrt{P_{1}(1-P_j)\prod_{k=j+1}^n P_k}, \quad j=2,\ldots,n
\label{Uk0-xp}
\ee
where $P_k$ are defined in Eq.~(\ref{Pkn}). The final angular state
tends to the original state $|j\ra$ without a sign factor.

The phase $\phi_{j0}^+$ combines the Landau--Zener phase
at the transition to state $j$ with the adiabatic phase
along the blue trajectory in Fig.~\ref{scheme-xp-fig}(b).
Its calculation in Appendix~\ref{app-phik0-xp} includes
a contribution that first appears at $n=3$, and therefore is not found in previously published connection formulas for $n=2$ in \cite{Sinitsyn2026-letter}. 
Namely, couplings of the radial state $|R\ra$ to the other
states $|k\ra$, where $k>1$ and $k\ne j$, produce energy
shifts of order $1/x$ that follow from the second-order
perturbative corrections to adiabatic energy. Although formally small, as derived in Appendix~\ref{app-phik0-xp}, they accumulate over time intervals
of order $x$ before encountering avoided crossings and therefore give finite contributions to
the adiabatic phase. The final result is
\begin{eqnarray}
\phi_{j0}^+
=
-2\varepsilon_jt_0
+\frac{\pi}{4}
+\arg\Gamma(iI_j)
-\phi_j
-\frac{2}{3}\varepsilon_j^{3/2}
-I_j\ln\left(4\sqrt{\varepsilon_j}\right)
+\sum_{\substack{k=2\\k\ne j}}^n
I_k\ln
\frac{|\sqrt{\varepsilon_k}-\sqrt{\varepsilon_j}|}
{\sqrt{\varepsilon_k}+\sqrt{\varepsilon_j}}.
\label{phik0-xp}
\end{eqnarray}
The simplest to find is the scattering amplitude from level $0$ to level $1$, which,  up to $\sigma$-sign, is the amplitude to stay at the radial state $|R\rangle$. All phases of this amplitude cancel, leaving only the sign $\sigma$. Thus,

\be
U_{10}^{(n)}=\sigma \sqrt{\prod_{k=1}^{n}P_k}.
\label{u10-xp}
\ee

\section{Connection formulas from comparison of transition amplitudes as $x\rightarrow \pm \infty$}
\label{sec-compare}

Due to the zero-curvature condition, the transition amplitudes
$U_{j0}^{(n)}$, $j=0,\ldots,n$, calculated along the two paths
$P$ and $P_{\infty}$ in Fig.~\ref{paths-fig}, must coincide.
The connection formulas are therefore obtained by comparing the
large-negative-$x$ amplitudes in Eqs.~(\ref{u00-xm}) and
(\ref{ind-Uk0-general}) with the large-positive-$x$ amplitudes in
Eqs.~(\ref{U00-xp}), (\ref{Uk0-xp}), and (\ref{u10-xp}).

First consider the amplitude to return to level $0$. Equation
(\ref{U00-xp}) gives
\[
\left|U_{00}^{(n)}\right|^2=1-P_1.
\]
Using Eq.~(\ref{P10}) and $I_1=\rho^2/2$, this becomes
\[
\left|U_{00}^{(n)}\right|^2=1-e^{-4\pi I_1}.
\]
On the other hand, Eq.~(\ref{u00-xm}) gives the same probability as
\[
\left|U_{00}^{(n)}\right|^2
=
\left|
\prod_{j=1}^n p_j
+
\sum_{k=1}^n
q_k
\left(
\prod_{j=k+1}^n p_j
\right)e^{2i\Phi_k^{(n)}}
\right|^2.
\]
The expression inside the absolute value is precisely
$C_n^{(n)}$ from Eq.~(\ref{uk0-2}). Hence
\[
1-e^{-4\pi I_1}=\left|C_n^{(n)}\right|^2,
\]
which is equivalent to Eq.~(\ref{rho-fin2}).

The remaining actions $I_j$, $j=2,\ldots,n$, are most directly
obtained from  probabilities
\be
W_j
\equiv
P_1\prod_{k=j+1}^n P_k
=
e^{-4\pi I_1-2\pi\sum_{k=j+1}^n I_k},
\qquad j=1,\ldots,n.
\label{w-def}
\ee
$W_j$ is the probability not to end
on level $0$ and not to end on any level $k>j$. By unitarity, this is
the same as the probability to end on one of the levels
$1,\ldots,j$:
\be
W_j=\sum_{k=1}^j\left|U_{k0}^{(n)}\right|^2 .
\label{w-sum}
\ee
The ratio of two consecutive probabilities isolates the $j$-th
adiabatic invariant:
\be
\frac{W_j}{W_{j-1}}
=
P_j^{-1}
=
e^{2\pi I_j},
\qquad j=2,\ldots,n.
\label{prob-ratioj}
\ee

It remains to express the same $W_j$ through the parameters
$p_k$, $q_k$, and $\Phi_k^{(n)}$ that appear at $x\rightarrow-\infty$.
The useful identity, proved in Appendix~\ref{app-proof-actions-W-minus2}, is
\be
W_j
=
\left(
\prod_{k=j+1}^{n}p_k
\right)
\left(
1-\left|C_j^{(n)}\right|^2
\right).
\label{actions-W-minus2}
\ee
Substituting Eq.~(\ref{actions-W-minus2}) into Eq.~(\ref{prob-ratioj})
gives
\[
e^{2\pi I_j}
=
\frac{
1-\left|C_j^{(n)}\right|^2
}{
p_j\left(
1-\left|C_{j-1}^{(n)}\right|^2
\right)
},
\]
which is Eq.~(\ref{I2-fin2}).

The connection formulas~(\ref{phi1-fin2}) and~(\ref{phi2-fin2}) for the phases follow straightforwardly by equating the phases~(\ref{ph00-xp}) and~(\ref{phik0-xp}) to ${\rm arg}U_{j0}^{(n)}$ in Eqs.~(\ref{u00-xm}) and~(\ref{ind-Uk0-general}).

Finally, the expression for the sign parameter $\sigma$ follows from equating the amplitudes $U_{10}^{(n)}$. It turns out that various dynamic phases for this amplitude have the same expression for $x\rightarrow \pm \infty$, so only the real components of this transition amplitude should be compared. This leads to equation
\be
\sigma \sqrt{\prod_{j=1}^{n}P_j}=2\sqrt{
p_1q_1
\prod_{j=2}^{n}p_j
}
\sin\Phi_1^{(n)}.
\label{U10-mp}
\ee
Comparing the signs on both sides of Eq.~(\ref{U10-mp}) leads to the connection formula
in Eq.~(\ref{sigma-fin2}).

Comparing the absolute values in Eq.~(\ref{U10-mp}) gives a relation that is not independent of the other connection formulas but provides a very simple expression for the net adiabatic invariant:
\be
e^{-\pi\left(\rho^2+\sum_{m=2}^{n}I_m\right)}
=
2\sqrt{
p_1q_1
\prod_{j=2}^{n}p_j
}
\left|\sin\Phi_1^{(n)}\right|.
\label{net-inv-2}
\ee
Consequently,
\be
2I_1+\sum_{m=2}^{n}I_m
=
-\frac{1}{\pi}
\ln\left[
2\sqrt{
p_1q_1
\prod_{j=2}^{n}p_j
}
\left|\sin\Phi_1^{(n)}\right|
\right].
\label{net-inv}
\ee

This completes the derivation of the main results.
\section{Conclusion}

The systems of Painlev\'e-like equations attract growing interest in mathematics \cite{ItsProkhorov2020,AdlerSokolov2021,MobasheraminiBertola2021,BershteinGrigorevShchechkin2023,GaiurRubtsov2025}, but to physicists this topic has remained obscure.  Physicists often deal with scattering problems. This is why the existence of simple connection formulas, in addition to practical relevance, is what makes a solution of a differential equation for them a valuable special function. The solution of the system~(\ref{P2-n}) now satisfies these criteria.

The connection formulas in Eqs.~(\ref{sigma-fin2})--(\ref{phi2-fin2}) look simple despite the considerable complexity of the dynamics described by Eq.~(\ref{P2-n}). This gives hope that similar extensions of other Painlev\'e equations to considerably more complex systems with simple connection formulas will be found in the future. Potentially, this progress may lead to  understanding of truly complex behavior, not reachable to purely numerical investigations, as it was done for combinatorially complex quantum mechanical systems \cite{Sinitsyn2016,Sun2016,Li2018-2,Yuzbashyan2018,Chernyak2019,Suzuki2025,Barik-PRB26,78xb-5lmw}.  

Although various applications of the system~(\ref{P2-n}) have already been discussed, it should be interesting to explore more of them. For example, the interaction terms in Eq.~(\ref{P2-n}) are typical of interactions induced by noise averaging in replica partition functions. In fact, standard Painlev\'e equations have previously emerged in the context of disordered systems and replica trick \cite{claeys2008multicritical,kanzieper2002replica,kanzieper2003exact,bachmann2012disordered}, suggesting that applications of the system~(\ref{P2-n}) in this context are possible. Painlev\'e equations are also known to describe correlation functions in spin-chain models \cite{nijhoff1995bilinear,gorsky2024kpz}, which may produce the system ~(\ref{P2-n}). Another interesting direction is the application of connection formulas in numerical algorithms. The system~(\ref{P2-n}) describes a genuinely nonperturbative passage through a critical point, away from which the dynamics is usually simple and can be simulated using the adiabatic approximation, considerably reducing the complexity. By using exact formulas that relate the incoming and outgoing states of a critical region, such numerical simulations may spend very little time on the most complex intervals of the evolution. On the mathematical side, by analogy with standard Painlev\'e equations, the system~(\ref{P2-n}) is expected to have a discrete counterpart, and progress should be possible in solving the associated Riemann--Hilbert problems \cite{Fokas2006}. 


\newpage

\appendix
\setlength{\abovedisplayskip}{5pt plus 2pt minus 2pt}
\setlength{\belowdisplayskip}{5pt plus 2pt minus 2pt}
\setlength{\abovedisplayshortskip}{3pt plus 2pt minus 2pt}
\setlength{\belowdisplayshortskip}{3pt plus 2pt minus 2pt}
\setlength{\jot}{2pt}

\section{Elementary perturbation theory for asymptotic solutions}
\subsection{Asymptotic solution as $x\rightarrow -\infty$}
\label{app-minusx}

 For $x\rightarrow-\infty$, all functions $u_k(x)$
oscillate rapidly, with the leading frequency $\sim\sqrt{-x}$. Seek the solution in the form
\be
u_k(x)=\frac{\alpha_k}{(-x)^{1/4}}\sin\theta_k(x),
\qquad k=1,\ldots,n.
\label{uk-leading}
\ee
The amplitudes change slowly on the scale of one period of the oscillations.

Substitute Eq.~(\ref{uk-leading}) into the nonlinear terms in
Eq.~(\ref{P2-n}). For the term that contains $u_k^3$, use
\be
\sin^3\theta_k=
\frac{3}{4}\sin\theta_k-\frac{1}{4}\sin(3\theta_k).
\ee
The second term oscillates three times faster than $u_k$ and becomes zero
after averaging over the period of oscillations. Hence, for the purpose of
calculating the slowly accumulated correction to the phase, it is disregarded, so that
\be
2u_k^3\rightarrow
\frac{3\alpha_k^2}{2\sqrt{-x}}u_k.
\label{self-average}
\ee
Similarly, for $j\ne k$,
\be
u_j^2=
\frac{\alpha_j^2}{\sqrt{-x}}\sin^2\theta_j.
\ee
Since
$
\left\langle\sin^2\theta_j\right\rangle=\frac{1}{2},
$
it follows
\be
2u_ku_j^2\rightarrow
\frac{\alpha_j^2}{\sqrt{-x}}u_k.
\label{cross-average}
\ee
After averaging the rapidly oscillating terms, the equation for the
$k$-th function becomes
\be
u_k''+
\left[
(-x+\varepsilon_k)
+\frac{1}{\sqrt{-x}}
\left(
\frac{3}{2}\alpha_k^2+
\sum_{j\ne k}^n\alpha_j^2
\right)
\right]u_k=0.
\label{uk-effective}
\ee

Equation~(\ref{uk-effective}) describes an oscillator with a slowly changing
frequency
\begin{equation}
\omega_k(x)
=
\sqrt{-x+\varepsilon_k}
+
\frac{3\alpha_k^2+
2\sum_{j\ne k}^n\alpha_j^2}
{4(-x)}
+o(|x|^{-1}).
\label{omega-k}
\end{equation}
The corresponding slowly changing amplitude is proportional to
$\omega_k^{-1/2}$, and therefore
\be
\omega_k^{-1/2}\sim(-x+\varepsilon_k)^{-1/4}.
\ee
Integration of Eq.~(\ref{omega-k}) gives the phase
\begin{eqnarray}
\nonumber
\int \omega_k\,d(-x)
=
\frac{2}{3}(-x+\varepsilon_k)^{3/2}
+
\frac{3\alpha_k^2+
2\sum_{j\ne k}^n\alpha_j^2}{4}\ln(-x)
+O(1).
\end{eqnarray}
The $O(1)$ integration constant is denoted by $\varphi_k$. This gives Eq.~(\ref{u1m}).

\subsection{Perturbative analysis of $x\rightarrow+\infty$}
\label{app-plusx}

Let $\varepsilon_1=0$, and suppose that $u_1(x)$ develops a regular
contribution of order $\sqrt{x}$, whereas all other components remain
bounded. Separate the bounded components as
\be
u_k(x)=v_k(x)+\delta u_k(x),
\qquad k=2,\ldots,n,
\label{app-plusx-decomposition}
\ee
where $v_k(x)$ contains only finite-frequency oscillations and
$\delta u_k(x)$ belongs to a common rapidly oscillating mode. Here
\be
\sigma\equiv
\lim_{x\rightarrow+\infty}
\frac{u_1(x)}{\sqrt{x/2}}
=\pm1.
\label{app-plusx-sigma}
\ee

\subsubsection{Slowly changing contributions}

Disregarding the rapidly oscillating terms and the second derivative
of the regular part of $u_1$, its equation gives
\be
\left[u_1^{\rm reg}(x)\right]^2
+\sum_{k=2}^n v_k^2(x)
=
\frac{x}{2}.
\label{app-plusx-algebraic}
\ee
The sign in Eq.~(\ref{app-plusx-sigma}) then fixes
\be
u_1^{\rm reg}(x)
=
\sigma\sqrt{
\frac{x}{2}-\sum_{k=2}^n v_k^2(x)
}.
\label{app-plusx-u1-regular}
\ee
Although the bounded terms under the square root change
$u_1^{\rm reg}$ only by $O(x^{-1/2})$, they must be retained.
Substitution of Eq.~(\ref{app-plusx-algebraic}) into the equations for
the remaining components gives
\be
v_k''(x)=-\varepsilon_kv_k(x),
\qquad k=2,\ldots,n.
\label{app-plusx-vk}
\ee
Consequently,
\be
v_k(x)=A_k\cos\theta_k(x),
\qquad
\theta_k(x)=\sqrt{\varepsilon_k}\,x+o(x),
\label{app-plusx-vk-solution}
\ee
where $A_k>0$. The notation $\theta_k(x)$ includes subleading phase
corrections that cannot be fixed by the present perturbative analysis.

\subsubsection{Rapidly oscillating contributions}

Write
\be
u_1(x)=u_1^{\rm reg}(x)+\sigma\delta u_1(x).
\label{app-plusx-u1-decomposition}
\ee
Linearization in the rapidly oscillating corrections gives
\begin{eqnarray}
(\delta u_1)''
&=&
\left[
-2x+4\sum_{j=2}^n v_j^2
\right]\delta u_1
-4
\sqrt{
\frac{x}{2}-\sum_{j=2}^n v_j^2
}
\sum_{j=2}^n v_j\delta u_j,
\label{app-plusx-delta1}
\\
(\delta u_k)''
&=&
-4
\sqrt{
\frac{x}{2}-\sum_{j=2}^n v_j^2
}\,
v_k\delta u_1
+O(\delta u_2,\ldots,\delta u_n).
\label{app-plusx-deltak}
\end{eqnarray}
The omitted terms in Eq.~(\ref{app-plusx-deltak}) have coefficients of
order one and are subleading relative to the rapidly oscillating
second derivatives, whose effective coefficient is of order $x$.
Hence,
\be
\delta u_k(x)
=
\sqrt{\frac{2}{x}}\,
v_k(x)\delta u_1(x)
+o(x^{-3/4}),
\qquad k=2,\ldots,n.
\label{app-plusx-deltak-result}
\ee
Substitution into Eq.~(\ref{app-plusx-delta1}) cancels the bounded-mode
contributions to the leading frequency and yields
\be
(\delta u_1)''
=
-2x\,\delta u_1+o(1)\delta u_1.
\label{app-plusx-delta1-final}
\ee
Its approximate solution is
\be
\delta u_1(x)
=
\frac{\rho}{(2x)^{1/4}}\cos\theta_1(x),
\qquad
\theta_1(x)
=
\frac{2\sqrt{2}}{3}x^{3/2}
+o(x^{3/2}),
\label{app-plusx-delta1-solution}
\ee
where $\rho>0$, and all subleading phase corrections are included in
$\theta_1(x)$.

Since $u_k^2=v_k^2$ to the order relevant under the square root,
the resulting asymptotic form can be written directly in terms of the
original components:
\begin{eqnarray}
\nonumber
u_1(x)
&=&
\sigma\sqrt{
\frac{x}{2}-\sum_{j=2}^n u_j^2(x)
}
+
\sigma\frac{\rho}{(2x)^{1/4}}
\cos\theta_1(x)
+\ldots,
\\
u_k(x)
&=&
A_k\cos\theta_k(x)
+\ldots,
\qquad k=2,\ldots,n.
\label{app-plusx-uk-final}
\end{eqnarray}
Here ``$\ldots$'' denotes higher-order terms in $1/x$ that are not
retained in the connection formulas. 

\section{Inverted connection formulas}
\label{inversion}
Explicitly, the inverted formulas look much more complex than the original ones. Therefore, they will be presented here as a recursive path to their construction for arbitrary $n$.
\subsection{Initiation of recursive procedure}

Introduce
\be
P_1=e^{-4\pi I_1},\qquad
P_j=e^{-2\pi I_j},\quad j=2,\ldots,n.
\label{inv-P}
\ee
The connection formulas for $I_1$ and $\phi_1$ in Eqs.~(\ref{rho-fin2}) and (\ref{phi1-fin2}) give the complex number
\be
C_n^{(n)}
=\sqrt{1-P_1}\,
\exp\left\{i\left[
-\frac{3\pi}{4}-7I_1\ln2
+\arg\Gamma(2iI_1)-\phi_1
\right]\right\}.
\label{inv-Cn}
\ee

Define
\be
\chi_j\equiv\arg\left[
e^{i\Phi_j^{(n)}}-C_{j-1}^{(n)}e^{-i\Phi_j^{(n)}}
\right].
\label{inv-chi-meaning}
\ee
For each $j=2,\ldots,n$, Eqs.~(\ref{I2-fin2}) and (\ref{phi2-fin2}) determine 
\begin{eqnarray}
\chi_j
=\frac{3\pi}{4}
-\frac{2}{3}\varepsilon_j^{3/2}
-I_j\ln(4\sqrt{\varepsilon_j})
+\arg\Gamma(iI_j)-\phi_j
+\sum_{\substack{k=2\\k\ne j}}^n
I_k\ln\frac{|\sqrt{\varepsilon_k}-\sqrt{\varepsilon_j}|}
{\sqrt{\varepsilon_k}+\sqrt{\varepsilon_j}}.
\label{inv-chi}
\end{eqnarray}

Thus, although $\Phi_j^{(n)}$ and $C_{j-1}^{(n)}$ are initially
unknown, $\chi_j$ is known for all $j$.
\subsection{Backward recursion}

Let, at some step, the quantity $C_j^{(n)}$ be known, in addition to the already discussed known values of $P_j$, and $\chi_j$. The recursive construction is based on the following formulas for $\Phi_j^{(n)}$, $p_j$, and
$C_{j-1}^{(n)}$ that will be derived in next subsection:
\begin{eqnarray}
e^{i\Phi_j^{(n)}}
&=&\frac{e^{i\chi_j}+C_j^{(n)}e^{-i\chi_j}}{\left|e^{i\chi_j}+C_j^{(n)}e^{-i\chi_j}\right|},
\label{inv-Phi-step}\\
p_j
&=&\frac{1-|C_j^{(n)}|^2}
{1-|C_j^{(n)}|^2+(1-P_j)\left|e^{i\chi_j}+C_j^{(n)}e^{-i\chi_j}\right|^2},
\qquad q_j=1-p_j,
\label{inv-p-step}\\
C_{j-1}^{(n)}
&=&\frac{C_j^{(n)}-q_j e^{2i\Phi_j^{(n)}}}{p_j}.
\label{inv-C-step}
\end{eqnarray}
Starting from Eq.~(\ref{inv-Cn}), these equations are applied
successively for $j=n,n-1,\ldots,2$. They determine both
$p_j$ and $\Phi_j^{(n)}$ at each step, as well as the complex
number needed for the next step.

The final step of the recursion should be performed with extra care. After $C_1^{(n)}$ is known, and using that
$C_0^{(n)}=1$, it satisfies
\be
C_1^{(n)}=p_1+(1-p_1)e^{2i\Phi_1^{(n)}}.
\label{inv-first-C}
\ee
Taking the modulus of $C_1^{(n)}-p_1$ gives
$|C_1^{(n)}-p_1|^2=(1-p_1)^2$, and hence
\begin{eqnarray}
p_1&=&\frac{1-|C_1^{(n)}|^2}
{2[1-{\rm Re}\,C_1^{(n)}]},
\label{inv-first-p}\\
e^{2i\Phi_1^{(n)}}
&=&\frac{C_1^{(n)}-p_1}{1-p_1}.
\label{inv-first-double-phase}
\end{eqnarray}
However, Eq.~(\ref{inv-first-double-phase}) determines
$\Phi_1^{(n)}$ only modulo $\pi$.

The remaining ambiguity is resolved by using the formula for $\sigma$:
\be
{\rm sign}\left[\sin\Phi_1^{(n)}\right]=\sigma.
\label{inv-sign}
\ee
From knowledge of $p_j$ and  $\Phi_j^{(n)}$, Eqs.~(\ref{inprob-def}) and (\ref{app-phase-final}) can be used to reconstruct the initial adiabatic invariants and angles as
\be
\frac{\alpha_j^2}{2}=-\frac{\ln p_j}{2\pi},
\qquad j=1,\ldots,n,
\label{inv-initial-actions}
\ee
\begin{eqnarray}
\varphi_j
=\Phi_j^{(n)}-\frac{\pi}{4}
-\arg\Gamma\left(i\frac{\alpha_j^2}{2}\right)
+\frac{3\alpha_j^2}{2}\ln2
-\frac12\sum_{\substack{k=1\\k\ne j}}^n
\alpha_k^2\ln\left(\frac{|\varepsilon_j-\varepsilon_k|}{4}\right),
\qquad j=1,\ldots,n.
\label{inv-initial-phases}
\end{eqnarray}

For example, for $n=1$, 
\be
\chi=-\frac{3\pi}{4}-7I_1\ln2+\arg\Gamma(2iI_1)-\phi_1,
\label{inv-n1-r-chi}
\ee
so that $C_1^{(1)}=\sqrt{1-e^{-4\pi I_1}} e^{i\chi}$, from which
\begin{eqnarray}
p_1&=&\frac{e^{-4\pi I_1}}{2(1-\sqrt{1-e^{-4\pi I_1}}\cos\chi)},
\label{inv-n1-p}\\
\frac{\alpha_1^2}{2}
&=&2I_1+\frac{1}{2\pi}\ln\left[2(1-\sqrt{1-e^{-4\pi I_1}}\cos\chi)\right],
\label{inv-n1-action}\\
e^{i\Phi_1^{(1)}}
&=&\sigma\frac{\sqrt{1-e^{-4\pi I_1}}\sin\chi+i(1-\sqrt{1-e^{-4\pi I_1}}\cos\chi)}
{\sqrt{1-2\sqrt{1-e^{-4\pi I_1}}\cos\chi+(1-e^{-4\pi I_1}))}}.
\label{inv-n1-Phi}
\end{eqnarray}
The initial phase is then
\be
\varphi_1=\Phi_1^{(1)}-\frac{\pi}{4}
-\arg\Gamma\left(i\frac{\alpha_1^2}{2}\right)
+\frac{3\alpha_1^2}{2}\ln2.
\label{inv-n1-varphi}
\ee

The limit $I_1\rightarrow 0$ in Eq.~(\ref{inv-n1-action}) gives the famous result \cite{Itin2009b,Sun2016,SinitsynPokrovsky2026QuasiAdiabaticEffects}
\be
\frac{\alpha_1^2}{2} \rightarrow \frac{\ln 2}{2\pi}.
\label{itin-res}
\ee
It is easy to trace the recursion formulas for the analog of this result at arbitrary $n$. The condition mimicking the quantum mechanical ground state with $I_j={\cal I}\ll 1$ for all $j$ leads for $j>1$ to that all adiabatic invariants $\alpha_j^2/2$ remain on average the same as ${\cal I}$, i.e. negligible. However, the adiabatic invariant of the first mode in the limit ${\cal I}\rightarrow 0$ is then given by Eq.~(\ref{itin-res}).  For the process of inverse transition through a critical point, discussed in \cite{Itin2009b,SinitsynPokrovsky2026QuasiAdiabaticEffects}, this means that presence of the additional modes does not influence the number and type of the produced nonadiabatic excitaions. 

\subsection{Derivation of Eqs.~(\ref{inv-Phi-step})-(\ref{inv-C-step})}

Define the combination:
\be
D_j\equiv\left|e^{i\chi_j}+C_j^{(n)}e^{-i\chi_j}\right|.
\label{inv-D}
\ee
\subsubsection{Derivation of Eq.~(\ref{inv-Phi-step}) for $\Phi_j^{(n)}$}
The difficulty is that the known phase $\chi_j$ in
Eq.~(\ref{inv-chi-meaning}) involves the unknown
$C_{j-1}^{(n)}$. The first step is to eliminate this unknown
using recursion relation for $C_{j}^{(n)}$. Introduce 
\be
B_j\equiv e^{i\Phi_j^{(n)}}-C_{j-1}^{(n)}e^{-i\Phi_j^{(n)}}.
\label{inv-B}
\ee
By Eq.~(\ref{inv-chi-meaning}), $B_j=|B_j|e^{i\chi_j}$:
its phase is known, although its modulus is not yet known.
The  recursion for $C_{j}^{(n)}$ is
\be
C_j^{(n)}=p_jC_{j-1}^{(n)}+q_je^{2i\Phi_j^{(n)}}.
\qquad p_j+q_j=1,
\label{inv-forward-recursion}
\ee
Multiplying this equation by $e^{-i\Phi_j^{(n)}}$ gives
\[
C_j^{(n)}e^{-i\Phi_j^{(n)}}
=p_jC_{j-1}^{(n)}e^{-i\Phi_j^{(n)}}
+q_je^{i\Phi_j^{(n)}}.
\]
Subtract this expression from $e^{i\Phi_j^{(n)}}$. Since
$1-q_j=p_j$, the result is
\begin{eqnarray*}
e^{i\Phi_j^{(n)}}-C_j^{(n)}e^{-i\Phi_j^{(n)}}
&=&(1-q_j)e^{i\Phi_j^{(n)}}
-p_jC_{j-1}^{(n)}e^{-i\Phi_j^{(n)}}\\
&=&p_j\left[e^{i\Phi_j^{(n)}}
-C_{j-1}^{(n)}e^{-i\Phi_j^{(n)}}\right].
\end{eqnarray*}
Thus, $C_{j-1}^{(n)}$ has been eliminated in favor of a positive
unknown multiplier $p_j|B_j|$:
\be
e^{i\Phi_j^{(n)}}-C_j^{(n)}e^{-i\Phi_j^{(n)}}
=p_jB_j=p_j|B_j|e^{i\chi_j}.
\label{inv-phase-equation}
\ee
To solve for $e^{i\Phi_j^{(n)}}$, write the complex conjugate
of Eq.~(\ref{inv-phase-equation}) explicitly:
\[
e^{-i\Phi_j^{(n)}}
-\left(C_j^{(n)}\right)^*e^{i\Phi_j^{(n)}}
=p_j|B_j|e^{-i\chi_j}.
\]
Multiplying it by $C_j^{(n)}$ gives
\[
C_j^{(n)}e^{-i\Phi_j^{(n)}}
-|C_j^{(n)}|^2e^{i\Phi_j^{(n)}}
=p_j|B_j|C_j^{(n)}e^{-i\chi_j}.
\]
Now add this equation to Eq.~(\ref{inv-phase-equation}).
The two terms proportional to $e^{-i\Phi_j^{(n)}}$ cancel,
leaving
\be
\left(1-|C_j^{(n)}|^2\right)e^{i\Phi_j^{(n)}}
=p_j|B_j|\left(e^{i\chi_j}+C_j^{(n)}e^{-i\chi_j}\right).
\label{inv-phase-solution}
\ee
For the generic solutions considered here, $|C_j^{(n)}|<1$, which is proved at the end of  section~\ref{sec-num}.
Consequently, the coefficient on the left is positive. Also,
$p_j|B_j|>0$, so the complex number
$e^{i\chi_j}+C_j^{(n)}e^{-i\chi_j}$ has precisely the phase
$\Phi_j^{(n)}$, with no additional $\pi$ ambiguity. This explains
why its modulus was introduced as $D_j$ in Eq.~(\ref{inv-D}).
In particular,
\[
D_j\geq 1-|C_j^{(n)}|>0.
\]
Taking the absolute value of Eq.~(\ref{inv-phase-solution})
and using $|e^{i\Phi_j^{(n)}}|=1$ gives
\[
1-|C_j^{(n)}|^2=p_j|B_j|D_j,
\]
or
\be
p_j|B_j|=\frac{1-|C_j^{(n)}|^2}{D_j}.
\label{inv-B-modulus}
\ee
Finally, divide Eq.~(\ref{inv-phase-solution}) by its modulus:
\[
e^{i\Phi_j^{(n)}}
=\frac{e^{i\chi_j}+C_j^{(n)}e^{-i\chi_j}}
{|e^{i\chi_j}+C_j^{(n)}e^{-i\chi_j}|}.
\]
This is Eq.~(\ref{inv-Phi-step}). Thus the phase is reconstructed
before $p_j$ or $|B_j|$ is separately determined.

\subsubsection{Derivation of Eq.~(\ref{inv-p-step}) for probability}
The next step uses the known action $I_j$. Its standard connection
formula is
\[
I_j=\frac{1}{2\pi}\ln\left[
\frac{1-|C_j^{(n)}|^2}
{p_j(1-|C_{j-1}^{(n)}|^2)}\right].
\]
Exponentiating $-2\pi I_j$ gives
\be
P_j=\frac{p_j(1-|C_{j-1}^{(n)}|^2)}{1-|C_j^{(n)}|^2}.
\label{inv-action-relation}
\ee
Consider the squared modulus of Eq.~(\ref{inv-forward-recursion}) and of Eq.~(\ref{inv-B}): 
\begin{eqnarray*}
|C_j^{(n)}|^2
&=&p_j^2|C_{j-1}^{(n)}|^2+q_j^2
+2p_jq_j{\rm Re}\left[C_{j-1}^{(n)}e^{-2i\Phi_j^{(n)}}\right],\\
|B_j|^2
&=&1+|C_{j-1}^{(n)}|^2
-2{\rm Re}\left[C_{j-1}^{(n)}e^{-2i\Phi_j^{(n)}}\right].
\end{eqnarray*}
The real parts can be eliminated between these two expressions:
\be
1-|C_j^{(n)}|^2
=p_j(1-|C_{j-1}^{(n)}|^2)+p_jq_j|B_j|^2
\label{inv-defect-identity}
\ee
Equation~(\ref{inv-action-relation}) identifies the first term
on the right as $P_j(1-|C_j^{(n)}|^2)$. Moving this term to
the left gives
\be
(1-P_j)(1-|C_j^{(n)}|^2)=p_jq_j|B_j|^2.
\label{inv-action-identity}
\ee
The phase calculation has already determined $p_j|B_j|$.
It is therefore useful to write
\[
p_jq_j|B_j|^2
=\frac{q_j}{p_j}\left(p_j|B_j|\right)^2
=\frac{q_j}{p_j}\frac{(1-|C_j^{(n)}|^2)^2}{D_j^2}.
\]
Substitution into Eq.~(\ref{inv-action-identity}) and cancellation
of one positive factor $1-|C_j^{(n)}|^2$ give
\[
\frac{q_j}{p_j}
=\frac{(1-P_j)D_j^2}{1-|C_j^{(n)}|^2}.
\]
Finally, $q_j=1-p_j$ implies $q_j/p_j=1/p_j-1$, so
\begin{eqnarray*}
\frac{1}{p_j}
=1+\frac{(1-P_j)D_j^2}{1-|C_j^{(n)}|^2},
\end{eqnarray*}
which is equivalent to Eq.~(\ref{inv-p-step}).

\subsubsection{Derivation of Eq.~(\ref{inv-C-step}) for $C_{j-1}^{(n)}$}
Both $p_j$ and $e^{i\Phi_j^{(n)}}$ are now known. Subtract
$q_je^{2i\Phi_j^{(n)}}$ from Eq.~(\ref{inv-forward-recursion})
and divide by $p_j$:
\[
C_{j-1}^{(n)}
=\frac{C_j^{(n)}-q_je^{2i\Phi_j^{(n)}}}{p_j}.
\]
This is Eq.~(\ref{inv-C-step}). It provides the complex number
needed to repeat the same calculation with $j$ replaced by $j-1$.
The sequence is therefore: normalize a known complex number to
find the phase, use the action to find $p_j$, and solve the linear
recursion for $C_{j-1}^{(n)}$.

\section{Phase average of $I_1$ in the continuum limit}
\label{app-I1-continuum-limit}

From Eq.~(\ref{rho-fin2}),
\be
I_1
=
-\frac{1}{4\pi}
\ln\left(
1-\left|C_n^{(n)}\right|^2
\right).
\label{app-I1-Cn}
\ee
For equal initial adiabatic invariants,
\be
p_j=p=e^{-2\pi{\cal I}} \sim 1,
\qquad
q_j=q=1-p \ll 1,
\qquad
j=1,\ldots,n,
\label{app-I1-pq}
\ee
the recurrence relation for $C_j^{(n)}$ gives
\be
C_n^{(n)}
=
p^n
+
q\sum_{r=0}^{n-1}
p^r e^{2i\Phi_{n-r}^{(n)}} .
\label{app-I1-Cn-sum}
\ee
Using smallness of $q$, the last term in Eq.~(\ref{app-I1-Cn-sum}) can be treated as small. Thus, it can be moved away from the logarithm before averaging Eq.~(\ref{app-I1-Cn}) over $\Phi_j^{(n)}$.
In the continuum scaling limit
${\cal I}\ll1$, $n\gg1$, and $n{\cal I}=O(1)$, the oscillating part of
$C_n^{(n)}$ is of order $\sqrt{\cal I}$. Averaging over independent
phases gives
\be
\left\langle
q\sum_{r=0}^{n-1}p^r e^{2i\Phi_{n-r}^{(n)}}
\right\rangle_{\Phi}
=0,
\label{app-I1-deltaC-average}
\ee
and
\be
\left\langle
\left|
q\sum_{r=0}^{n-1}p^r e^{2i\Phi_{n-r}^{(n)}}
\right|^2
\right\rangle_{\Phi}
=
q^2\sum_{r=0}^{n-1}p^{2r}
=
\pi{\cal I}
\left(
1-e^{-4\pi n{\cal I}}
\right)
+O({\cal I}^2).
\label{app-I1-deltaC2}
\ee
Expanding the logarithm in Eq.~(\ref{app-I1-Cn}) to first order in
${\cal I}$, while keeping $n{\cal I}$ finite, gives
\be
\left\langle I_1\right\rangle_{\Phi}
=
-\frac{1}{4\pi}
\ln\left(
1-e^{-4\pi n{\cal I}}
\right)
+
\frac{{\cal I}}
{4\left(
1-e^{-4\pi n{\cal I}}
\right)}
+O({\cal I}^2).
\label{app-I1-average-final}
\ee

The same continuum limit gives the phase-averaged changes
$\Delta I_j=I_j-{\cal I}$ for $j=2,\ldots,n$. To leading order in
${\cal I}$, the phase-dependent contributions cancel from the difference
of the two averaged logarithms in Eq.~(\ref{I2-fin2}), so that
\be
\left\langle I_j\right\rangle_{\Phi}
=
{\cal I}
+
\frac{1}{2\pi}
\ln
\left[
\frac{
1-e^{-4\pi j{\cal I}}
}{
1-e^{-4\pi (j-1){\cal I}}
}
\right]
+O({\cal I}^2).
\label{app-Ij-continuum-start}
\ee
Keeping ${\cal I}j=O(1)$ gives
\be
\ln
\left[
\frac{
1-e^{-4\pi j{\cal I}}
}{
1-e^{-4\pi (j-1){\cal I}}
}
\right]
=
\frac{4\pi{\cal I}}
{\exp(4\pi{\cal I}j)-1}
+O({\cal I}^2).
\label{app-Ij-continuum-log}
\ee
Hence
\be
\left\langle\Delta I_j\right\rangle_{\Phi}
=
\frac{2{\cal I}}
{\exp(4\pi{\cal I}j)-1}
+O({\cal I}^2),
\label{app-Ij-continuum-result}
\ee
which gives Eq.~(\ref{Delta-I-continuum-scaling}).

\section{Derivation of Eq.~(\ref{ha})}
\label{app-heff1}

 In the original Hamiltonian $H$, Eq.~(\ref{HH-gen}),  after the rescaling 
$$
t=\tau\sqrt{|x|},
$$
the leading diagonal energies in the $A$-term of Eq.~(\ref{HH-gen})  are
\begin{eqnarray}
E_{0;(0)}&=&|x|^{3/2}(4\tau^2-1),\\
E_{k;(0)}&=&-|x|^{3/2}(4\tau^2-1)+2\sqrt{|x|}\varepsilon_k ,
\label{E0-leading}
\end{eqnarray}
where the lower index $(0)$ means that the expression is written in the zeroth order of perturbation theory. 

The coupling between the diabatic states $0$ and $k$ is
\be
V_{0k}=-4|x|\tau u_k-2i\sqrt{|x|}u_k'.
\label{V0k-app}
\ee
For $x\rightarrow-\infty$, Eq.~(\ref{u1m}) gives, to the leading order,
\be
u_k\simeq\frac{\alpha_k}{|x|^{1/4}}\sin\theta_k,\qquad
u_k'\simeq-\alpha_k|x|^{1/4}\cos\theta_k,
\label{uk-app}
\ee
where $\theta_k$ denotes the phase of the oscillations in
Eq.~(\ref{u1m}). Hence,
\be
|V_{0k}|^2=
4\alpha_k^2|x|^{3/2}
\left(4\tau^2\sin^2\theta_k+\cos^2\theta_k\right).
\label{V0k2-app}
\ee

Let us first consider the energy of the $0$-th state. Apart from the
leading term in Eq.~(\ref{E0-leading}), its diagonal matrix element
contains the correction
\be
-2\sqrt{|x|}u_k^2\simeq-2\alpha_k^2\sin^2\theta_k
\ee
from each $k$-th mode. At the same order, the coupling in
Eq.~(\ref{V0k-app}) produces the standard quantum mechanical second-order energy correction
\begin{eqnarray}
\delta E_{0k;(2)}
=\frac{|V_{0k}|^2}{E_{0;(0)}-E_{k;(0)}}=
\frac{2\alpha_k^2
\left(4\tau^2\sin^2\theta_k+\cos^2\theta_k\right)}
{4\tau^2-1}.
\label{dE0-app}
\end{eqnarray}
Here, the term $2\sqrt{|x|}\varepsilon_k$ in the denominator was
disregarded, and the upper index $(2)$ means that this is the second-order correction. 
Adding the two corrections, it follows
\begin{equation}
-2\alpha_k^2\sin^2\theta_k+\delta E_{0k;(2)}
=
\frac{2\alpha_k^2}{4\tau^2-1}.
\label{cancel0-app}
\end{equation}
Thus, the dependence on the rapidly oscillating phase $\theta_k$
cancels in the diagonal elements of the effective Hamiltonian. 

The calculation for the $k$-th state is analogous. Its diagonal
matrix element contains
\be
2\sqrt{|x|}u_k^2\simeq2\alpha_k^2\sin^2\theta_k,
\ee
whereas the second-order correction to energy due to the coupling to the
$0$-th state has the opposite sign:
\begin{eqnarray}
\delta E_{k;(2)}
=\frac{|V_{0k}|^2}{E_{k;(0)}-E_{0;(0)}}
=-
\frac{2\alpha_k^2
\left(4\tau^2\sin^2\theta_k+\cos^2\theta_k\right)}
{4\tau^2-1}.
\label{dEk-app}
\end{eqnarray} 
Therefore,
\be
2\alpha_k^2\sin^2\theta_k+\delta E_{k;(2)}
=
-\frac{2\alpha_k^2}{4\tau^2-1}.
\label{cancelk-app}
\ee
Together with the leading energy and the term
$2\sqrt{|x|}\varepsilon_k$, this reproduces the diagonal element for
the $k$-th state in Eq.~(\ref{ha}).

Finally, the matrix elements between two different states $j$ and
$k$, with $j,k\ne0$, are of order
\be
2\sqrt{|x|}u_ju_k=O(1).
\ee
For $\varepsilon_j\ne\varepsilon_k$, the difference between their
diagonal energies is of order $\sqrt{|x|}$. Hence, the correction to
their energies due to these matrix elements vanishes in the limit
$x\rightarrow-\infty$. They can therefore be disregarded when
deriving Eq.~(\ref{ha}).

\section{Evolution matrix over the interval with Landau-Zener transitions}

\subsection{Landau-Zener (LZ) model in the diabatic basis}
\label{sec-LZ-diab}
Let $\beta$ and $g$ be constant parameters, and let $a_{1,2}$ be time-dependent amplitudes of two quantum states. 
The LZ model  describes the interaction between two states according to the  Schr\"odinger equation with a linearly time-dependent  Hamiltonian:
\begin{equation}
i\frac{d}{dt} \left( \begin{array}{l}
a_1\\
a_2
\end{array} \right)=
\left( \begin{array}{cc}
\beta t & g\\
g^* & -\beta t
\end{array} \right)
\left( \begin{array}{l}
a_1\\
a_2
\end{array} \right),
\label{lz1}
\end{equation}
where the 2$\times$2  matrix in Eq.~(\ref{lz1}) is the LZ Hamiltonian, $H_{LZ}(t)$. Assume that $\beta>0$, as the case $\beta<0$ is reproduced by redefining diabatic state indices.

Asymptotically, as $t\rightarrow \pm \infty$, there are solutions of this model that behave as
\begin{eqnarray}
\nonumber |\psi^{a}_{-} (t) \ra &= & e^{i \Phi^a_{-} (t)} |a\ra  +O(1/t)\quad  {\rm as} \,\,\,\, t\rar - \infty, \quad a=1,2, \\
|\psi^{a}_{+} (t) \ra &=& e^{i \Phi^a_{+} (t)} |a\ra +O(1/t) \quad  {\rm as}  \,\,\,\, t\rar + \infty, \quad a=1,2.
\label{state-adiab-asympt}
\end{eqnarray}
where $|1 \ra$ and $|2\ra$ are  two diabatic states -- the basis in which the Hamiltonian in Eq.~(\ref{lz1}) is written -- and $\Phi_{\pm}^a(t)$ are time-dependent phases,
\begin{eqnarray}
{\Phi}_{\pm}^{a} = -\frac{\beta_{a}t^{2}}{2} -  \frac{|g|^2}{2\beta_a} \ln (\sqrt{2\beta} |t|)+O(1/|t|), \,\,\, a\!=\!1,2, 
\label{adiab-ph-LZ}
\end{eqnarray}
where  $\beta_{1} =\beta$ and $\beta_2=-\beta$.

Equation~(\ref{lz1}) has only two linearly independent solutions, whereas  Eq.~(\ref{state-adiab-asympt}) defines four state vectors (with indices $\pm$ and $a=1,2$). If continued to arbitrary time, any two states in  (\ref{state-adiab-asympt}) can be expressed as a linear combination of the other two.
Let $|\psi(t) \ra$  obey the  initial condition 
\be
| \psi(t) \ra_{t\rar -\infty } =|\psi_{-}^{b}(t)\ra, 
\label{def-init-lz}
\ee 
where $b \in \{1,2 \}$ is the index of the initial diabatic state. The {\it scattering matrix} relates the pairs of the solutions as $t\rar -\infty$ to the pair as $t\rar +\infty$:
\begin{eqnarray}
\label{define-S-MLZ} |\psi(t)\ra_{t\rar +\infty} =\sum_{a=1,2} {S}_{ab} |\psi_{+}^{a}(t)\ra.
\end{eqnarray}

Let $U(t_2,t_1)$ be the evolution matrix for  Eq.~(\ref{lz1}) from time $t_1$ to time $t_2$.  The  scattering matrix in the basis of diabatic states is related to the evolution matrix over the time interval $t\in (-T,T)$ as
\be
S_{ab}=\lim_{T\rar \infty} U_{ab}(T,-T)e^{-i({\Phi}_{+}^{a}(T)-{\Phi}_{-}^{b}(-T))},
\label{scatt-lim1}
\ee
where $\Phi_{\pm}$ are defined in (\ref{adiab-ph-LZ}).

For the choice of  states in Eq.~(\ref{state-adiab-asympt}), and by the property ${\Phi}_{+}^{a} (-T) = {\Phi}_{-}^{a} (T)$, Eq.~(\ref{scatt-lim1}) also relates the diagonal elements 
\be
S_{aa}=\lim_{T\rar \infty} U_{aa}(T,-T), \quad a=1,2.
\label{s-u-rel1}
\ee

Equation~(\ref{lz1}) is reducible to the parabolic cylinder equation, whose asymptotics are well characterized, so the scattering matrix is known exactly \cite{SinitsynPokrovsky2026QuasiAdiabaticEffects}: 
\begin{widetext}
\begin{eqnarray}
\label{lz-scatt-mt}
&&S=
\\
\nonumber &&\left(\begin{array}{cc}
e^{-\pi \gamma^2} & \sqrt{1-e^{-2\pi \gamma^2}}e^{i\pi/4+i\arg[\Gamma(i\gamma^2)]+i{\arg(g)}} \\
- \sqrt{1-e^{-2\pi \gamma^2}}e^{-i\pi/4-i\arg[\Gamma(i\gamma^2)]-i\arg (g)} & e^{-\pi \gamma^2}
\end{array} \right), \quad \gamma=\frac{|g|}{\sqrt{2\beta}}.
\end{eqnarray}
\end{widetext}

Directly related to ${S}$ is the matrix of transition probabilities ${P}$, with elements 
\be
P_{ab}=|S_{ab}|^2,
\label{pabs}
\ee 
where $P_{ab}$ is the probability  of finding the system in the diabatic state $a$ at $t=+\infty$ if  at $t=-\infty$ the state was $b$.

The probability of remaining in the initial diabatic state after the evolution  from $t\rar -\infty$ to $t\rar +\infty$ is given by the {\it Landau-Zener formula}:
\be
P_{11}\equiv |S_{11}|^2 = e^{-2\pi \gamma^2}=e^{-\pi |g|^2 / \beta}.
\label{trans-prob}
\ee

\subsection{Landau-Zener evolution matrix in the adiabatic basis}
\label{sec-LZ-adiab}
\subsubsection{Crossing points as $x\rightarrow -\infty$}
Let 
$
\tau_j^{\pm}
$
be the moments of crossing of levels $0$ and $j$, with $\pm$ corresponding to $\tau_{j}^{\pm} \approx \pm 1/2$. Consider an interval $\tau \in (\tau_j^{-}-\eta, \tau_j^{-}+\eta)$,  where $\eta>0$, which  encloses the crossing point of levels $0$ and $j$, $j\ge 1$,  but does not overlap  with  all other crossing points of level $0$. Assume also that $\eta$ is sufficiently large to saturate the elements of the Landau-Zener scattering matrix for scattering between $0$ and $j$ at this crossing point. In the limit $|x|\rightarrow \infty$, it can also be assumed that $\eta$ is small in the sense that any nonsingular in $\eta$ contribution to scattering phases can be considered negligible in comparison to other adiabatic phases in this calculation. 

Although the scattering matrix for Landau-Zener model is provided in many textbooks, here it is useful to write the evolution matrix differently because
the required evolution matrix is written in the adiabatic basis, whose states have different time-dependent phases. For example, if $\hat{U}_{-1/2}^{(j)}$ is the evolution matrix for the two crossing states, $0$ and $j$, in the interval $(\tau_j^{-}-\eta,\tau_j^{-}+\eta)$, then the off-diagonal matrix elements in the adiabatic basis are given by 
\begin{eqnarray}
(U_{-1/2}^{(j)})_{j0} &\equiv& \la j_a(\tau_j^{-}+\eta) | \hat{U}_{-1/2}^{(j)} |0_a(\tau_j^{-}-\eta)\ra, \\
(U_{-1/2}^{(j)})_{0j} &\equiv& \la 0_a(\tau_j^{-}+\eta) | \hat{U}_{-1/2}^{(j)} |j_a(\tau_j^{-}-\eta)\ra
\end{eqnarray}

Analogously, let $\hat{U}_{+1/2}^{(j)}$ be the evolution matrix for the two crossing states, $j$ and $0$, in the interval $(\tau_j^{+}-\eta,\tau_j^{+}+\eta)$. Then the matrix elements in the adiabatic basis are given by 
\begin{eqnarray}
(U_{+1/2}^{(j)})_{j0} &\equiv& \la j_a(\tau_j^{+}+\eta) | \hat{U}_{+1/2}^{(j)} |0_a(\tau_j^{+}-\eta)\ra, \\
(U_{+1/2}^{(j)})_{0j} &\equiv& \la 0_a(\tau_j^{+}+\eta) | \hat{U}_{+1/2}^{(j)} |j_a(\tau_j^{+}-\eta)\ra. 
\end{eqnarray}

The diagonal matrix elements in the adiabatic basis of this operator are the same as in the diabatic one, but the off-diagonal elements acquire an additional $j$-dependent phase, which has a logarithmic singularity, $\sim \ln \eta$. Such singular terms, however, cancel with analogous terms that emerge from the subsequent evolution.  

\subsubsection{Evolution across the interval ${\rm Int}_j^-$}
\label{app-lz-Iminus-section}

Let
\be
{\rm Int}_j^-
=
\left(
\tau_j^--\eta,\tau_j^-+\eta
\right)
\label{app-lz-Iminus-j}
\ee
be an interval that contains only the crossing between levels $0$
and $j$ near $\tau=-1/2$. The crossing point is
\be
\tau_j^-
=
-\frac12-\frac{\varepsilon_j}{4|x|}
+
o(|x|^{-1}).
\label{app-lz-tau-minus-j}
\ee

In what follows, disregard from the beginning all contributions
linear in $\eta$. In particular, disregard the common phase
$2\varepsilon_j\sqrt{|x|}\eta$ and the linear term in the expansion
of $\ln(1+\eta)$. Retain the term
$|x|^{3/2}\eta^2$, because it cancels an identical term obtained from
the adiabatic phases, and retain the logarithm $\ln\eta$.

Introducing
\be
s=\tau-\tau_j^-,
\ee
and subtracting the common diagonal energy
$\varepsilon_j\sqrt{|x|}$, one obtains the two-state Hamiltonian in the
diabatic basis $(|0\ra,|j\ra)$:
\be
H_-^{(j)}(s)
=
\left[
\begin{array}{cc}
-\beta s&g_j^-\\
(g_j^-)^*&\beta s
\end{array}
\right],
\qquad
\beta=4|x|^{3/2}.
\label{app-lz-Hminus-j}
\ee
The coupling $g_j^-$ and its phase are given by
Eqs.~(\ref{gmp-def}) and~(\ref{app-lz-gamma-phases}).

The evolution matrix in the diabatic basis is defined by
\be
\left(\hat U_{{\rm Int}_j^-}\right)_{rs}
\equiv
\la r|
\hat U(\tau_j^-+\eta,\tau_j^--\eta)
|s\ra,
\qquad
r,s\in\{0,j\}.
\label{app-lz-Uhat-Iminus-def}
\ee
Using Eq.~(\ref{g-abs}), it follows
\be
\frac{|g_j^-|^2}{2\beta}
=
\frac{\alpha_j^2}{2}.
\ee
Hence,
\be
p_j=e^{-\pi\alpha_j^2},
\qquad
q_j=1-p_j.
\label{app-lz-pq-minus-j}
\ee

The scattering matrix in Eq.~(\ref{lz-scatt-mt}) is defined after
removing the time-dependent asymptotic phases in
Eq.~(\ref{adiab-ph-LZ}). Restoring these phases for evolution from
$s=-\eta$ to $s=\eta$ produces
\be
\Omega_j^-(\eta)
=
\beta\eta^2
+
\frac{|g_j^-|^2}{\beta}
\ln\left(\sqrt{2\beta}\,\eta\right).
\label{app-lz-Omega-minus-general}
\ee
For the parameters of the present problem,
\be
\Omega_j^-(\eta)
=
4|x|^{3/2}\eta^2
+
\alpha_j^2
\ln\left(
2\sqrt{2}\,|x|^{3/4}\eta
\right).
\label{app-lz-Omega-minus-j}
\ee

In the fixed diabatic order $(0,j)$, the evolution matrix is
\be
\hat U_{{\rm Int}_j^-}
=
\left[
\begin{array}{cc}
\sqrt{p_j}&-\sqrt{q_j}\,e^{-i\delta_j^-}\\
\sqrt{q_j}\,e^{i\delta_j^-}&\sqrt{p_j}
\end{array}
\right]
+
o(1),
\label{app-lz-Uhat-Iminus}
\ee
where
\be
\delta_j^-
=
\frac{\pi}{4}
+
{\rm arg}\Gamma\left(i\frac{\alpha_j^2}{2}\right)
-
{\rm arg}\,g_j^-
-
\Omega_j^-(\eta).
\label{app-lz-delta-minus-first}
\ee
Using Eq.~(\ref{app-lz-gamma-phases}), this becomes
\begin{eqnarray}
\nonumber
\delta_j^-
=
-\frac{\pi}{4}
+
{\rm arg}\Gamma\left(i\frac{\alpha_j^2}{2}\right)
+
\theta_j
-
4|x|^{3/2}\eta^2
-
\alpha_j^2
\ln\left(
2\sqrt{2}\,|x|^{3/4}\eta
\right).
\label{app-lz-delta-minus-j}
\end{eqnarray}

Now transform the matrix from the diabatic basis
$(|0\ra,|j\ra)$ to the adiabatic basis
$(|0_a\ra,|j_a\ra)$. The matrix in the adiabatic basis is defined by
\begin{eqnarray}
\left(U_{{\rm Int}_j^-}\right)_{rs}
\equiv
\la r_a(\tau_j^-+\eta)|
\hat U(\tau_j^-+\eta,\tau_j^--\eta)
\times
|s_a(\tau_j^--\eta)\ra,
\qquad
r,s\in\{0,j\}.
\label{app-lz-UIminus-def}
\end{eqnarray}

The interval ${\rm Int}_j^-$ is excluded from the adiabatic phase
integral. Hence, at either boundary of this interval,
\be
|m_a(\tau_j^-\pm\eta)\ra
=
 \exp\left[
-i\,{\rm P}\int_{-T}^{\tau_j^-\pm\eta}E_m^{(n)}(\tau)\,d\tau
\right]|m\ra,
\qquad
m=0,j.
\label{app-lz-adiabatic-boundary-states}
\ee
It follows that
\be
\left(U_{{\rm Int}_j^-}\right)_{rs}
=
\exp\left[
i\,{\rm P}\int_{-T}^{\tau_j^-+\eta}E_r^{(n)}(\tau)\,d\tau
\right]
\left(\hat U_{{\rm Int}_j^-}\right)_{rs}
\exp\left[
-i\,{\rm P}\int_{-T}^{\tau_j^--\eta}E_s^{(n)}(\tau)\,d\tau
\right].
\label{app-lz-basis-change-minus}
\ee

Therefore, in the adiabatic basis,
\be
U_{{\rm Int}_j^-}
=
\left[
\begin{array}{cc}
\sqrt{p_j}&-\sqrt{q_j}\,e^{-i\Lambda_j^-}\\
\sqrt{q_j}\,e^{i\Lambda_j^-}&\sqrt{p_j}
\end{array}
\right]
+
o(1),
\label{app-lz-UIminus}
\ee
where
\be
\Lambda_j^-
=
\delta_j^-
+
{\rm P}\int_{-T}^{\tau_j^-+\eta}E_j^{(n)}(\tau)\,d\tau
-
{\rm P}\int_{-T}^{\tau_j^--\eta}E_0^{(n)}(\tau)\,d\tau .
\label{app-lz-Lambda-minus-j}
\ee

According to Eq.~(\ref{ha}), outside the interval ${\rm Int}_j^-$,
\be
E_j^{(n)}-E_0^{(n)}
=
-2|x|^{3/2}(4\tau^2-1)
-
\frac{
2\left(
\sum_{\ell=1}^n\alpha_\ell^2+\alpha_j^2
\right)
}{4\tau^2-1}
+
2\sqrt{|x|}\varepsilon_j
+
o(1).
\label{app-lz-energy-diff-minus-j}
\ee
After dropping all terms linear in $\eta$, integration gives
\begin{eqnarray}
\nonumber
{\rm P}\int_{-T}^{\tau_j^-+\eta}E_j^{(n)}(\tau)\,d\tau
&-&
{\rm P}\int_{-T}^{\tau_j^--\eta}E_0^{(n)}(\tau)\,d\tau
=
-\frac{8}{3}|x|^{3/2}T^3
+
2|x|^{3/2}T
-
\frac{2}{3}|x|^{3/2}+
\\
&&
4|x|^{3/2}\eta^2
+
2\varepsilon_j\sqrt{|x|}
\left(T-\frac12\right)
+
\frac{
\sum_{\ell=1}^n\alpha_\ell^2+\alpha_j^2
}{2}\ln\eta
+
\frac{
\sum_{\ell=1}^n\alpha_\ell^2+\alpha_j^2
}{2}
\ln\left(\frac{2T+1}{2T-1}\right).
\label{app-lz-A-diff-minus-j-full}
\end{eqnarray}

Since $T\gg1$,
\be
\ln\left(\frac{2T+1}{2T-1}\right)
=
\frac{1}{T}
+
O(T^{-3}).
\label{app-lz-large-T-log}
\ee
This term vanishes in the large-$T$ limit. Hence,
\begin{eqnarray}
\nonumber
{\rm P}\int_{-T}^{\tau_j^-+\eta}E_j^{(n)}(\tau)\,d\tau
&-&
{\rm P}\int_{-T}^{\tau_j^--\eta}E_0^{(n)}(\tau)\,d\tau
=
-\frac{8}{3}|x|^{3/2}T^3
+
2|x|^{3/2}T
-
\frac{2}{3}|x|^{3/2}+
\\
&&
4|x|^{3/2}\eta^2
+
2\varepsilon_j\sqrt{|x|}
\left(T-\frac12\right)
+
\frac{
\sum_{\ell=1}^n\alpha_\ell^2+\alpha_j^2
}{2}\ln\eta.
\label{app-lz-A-diff-minus-j}
\end{eqnarray}

Using Eq.~(\ref{app-lz-theta-j}) for $\theta_j$, it follows that the
terms proportional to $|x|^{3/2}\eta^2$, $|x|^{3/2}$, and
$\varepsilon_j\sqrt{|x|}$ cancel in Eq.~(\ref{app-lz-Lambda-minus-j}).
The remaining phase is
\begin{eqnarray}
\nonumber
\Lambda_j^-
&=&
-\frac{8}{3}|x|^{3/2}T^3
+
2|x|^{3/2}T
+
2\varepsilon_j\sqrt{|x|}\,T\\
\nonumber 
&-&
\frac{\pi}{4}
+
{\rm arg}\Gamma\left(i\frac{\alpha_j^2}{2}\right)
+
\varphi_j
-
\frac{3\alpha_j^2}{2}\ln2
+
\frac12
\sum_{\substack{k=1\\k\ne j}}^n
\alpha_k^2\ln|x|
+
\frac12
\sum_{\substack{k=1\\k\ne j}}^n
\alpha_k^2\ln\eta
+
o(1).
\label{app-lz-Lambda-minus-full}
\end{eqnarray}

Using $T=t_0/\sqrt{|x|}$, it is useful to write
\be
\Lambda_j^-
=
\Lambda_{j,0}^-
+
\frac12
\sum_{\substack{k=1\\k\ne j}}^n
\alpha_k^2\ln\eta
+
o(1),
\label{app-lz-Lambda-minus-split}
\ee
where the $\eta$-independent part is
\begin{eqnarray}
\nonumber
\Lambda_{j,0}^-
=
-\frac{8}{3}t_0^3
+
2|x|t_0
+
2\varepsilon_jt_0
-
\frac{\pi}{4}
+
{\rm arg}\Gamma\left(i\frac{\alpha_j^2}{2}\right)
+
\varphi_j
-
\frac{3\alpha_j^2}{2}\ln2
+
\frac12
\sum_{\substack{k=1\\k\ne j}}^n
\alpha_k^2\ln|x|.
\label{app-lz-Lambda-minus-zero}
\end{eqnarray}

The final result is therefore
\begin{widetext}
\be
U_{{\rm Int}_j^-}
=
\left[
\begin{array}{cc}
\sqrt{p_j}
&
-\sqrt{q_j}
\exp\left\{
-i\left[
\Lambda_{j,0}^-
+
\frac12
\sum\limits_{\substack{k=1\\k\ne j}}^n
\alpha_k^2\ln\eta
\right]
\right\}
\\
\sqrt{q_j}
\exp\left\{
i\left[
\Lambda_{j,0}^-
+
\frac12
\sum\limits_{\substack{k=1\\k\ne j}}^n
\alpha_k^2\ln\eta
\right]
\right\}
&
\sqrt{p_j}
\end{array}
\right]
+
o(1).
\label{app-lz-UIminus-final}
\ee
\end{widetext}

\subsubsection{Evolution across the interval ${\rm Int}_j^+$}
\label{app-lz-Iplus-section}

Consider the interval
\be
{\rm Int}_j^+
=
\left(
\tau_j^+-\eta,\tau_j^++\eta
\right),
\qquad
\tau_j^+
=
\frac12+\frac{\varepsilon_j}{4|x|}
+
o(|x|^{-1}),
\label{app-lz-Iplus-j}
\ee
which contains only the crossing between levels $0$ and $j$ near
$\tau=1/2$. The same fixed ordering of the diabatic states,
$(|0\ra,|j\ra)$, is used as for ${\rm Int}_j^-$.

According to Eqs.~(\ref{g-abs}) and
(\ref{app-lz-gamma-phases}), the two couplings have the same absolute
value but different phases:
\be
|g_j^+|=|g_j^-|,
\qquad
{\rm arg}\,g_j^+
=
\frac{\pi}{2}+\theta_j,
\qquad
{\rm arg}\,g_j^-
=
\frac{\pi}{2}-\theta_j.
\label{app-lz-gamma-plus-difference}
\ee
Moreover, the slopes of the two crossing levels are reversed. After
introducing $s=\tau-\tau_j^+$ and subtracting the common diagonal
energy, the two-state Hamiltonian is
\be
H_+^{(j)}(s)
=
\left[
\begin{array}{cc}
\beta s&g_j^+\\
(g_j^+)^*&-\beta s
\end{array}
\right],
\qquad
\beta=4|x|^{3/2}.
\label{app-lz-Hplus-j}
\ee

As in the derivation for ${\rm Int}_j^-$, disregard all terms linear in
$\eta$. The parameters $p_j$, $q_j$, and the finite-time phase
$\Omega_j(\eta)$ are the same as in
Eqs.~(\ref{app-lz-pq-minus-j}) and
(\ref{app-lz-Omega-minus-j}). Therefore, in the diabatic basis,
\be
\hat U_{{\rm Int}_j^+}
=
\left[
\begin{array}{cc}
\sqrt{p_j}&\sqrt{q_j}\,e^{i\delta_j^+}\\
-\sqrt{q_j}\,e^{-i\delta_j^+}&\sqrt{p_j}
\end{array}
\right]
+
o(1),
\label{app-lz-Uhat-Iplus}
\ee
where
\be
\delta_j^+
=
\frac{\pi}{4}
+
{\rm arg}\Gamma\left(i\frac{\alpha_j^2}{2}\right)
+
{\rm arg}\,g_j^+
-
\Omega_j(\eta).
\label{app-lz-delta-plus-first}
\ee
Using Eq.~(\ref{app-lz-gamma-plus-difference}), one obtains
\begin{eqnarray}
\delta_j^+
=
\frac{3\pi}{4}
+
{\rm arg}\Gamma\left(i\frac{\alpha_j^2}{2}\right)
+
\theta_j
-
4|x|^{3/2}\eta^2
-
\alpha_j^2
\ln\left(
2\sqrt{2}\,|x|^{3/4}\eta
\right).
\label{app-lz-delta-plus-j}
\end{eqnarray}

To transform this matrix to the adiabatic basis
$(|0_a\ra,|j_a\ra)$, exclude ${\rm Int}_j^+$ from the adiabatic phase
integral. Then
\be
\left(U_{{\rm Int}_j^+}\right)_{rs}
=
\exp\left[
+i\,{\rm P}\int_{-T}^{\tau_j^++\eta}E_r^{(n)}(\tau)\,d\tau
\right]
\left(\hat U_{{\rm Int}_j^+}\right)_{rs}
\exp\left[
-i\,{\rm P}\int_{-T}^{\tau_j^+-\eta}E_s^{(n)}(\tau)\,d\tau
\right].
\label{app-lz-basis-change-plus}
\ee
Hence,
\be
U_{{\rm Int}_j^+}
=
\left[
\begin{array}{cc}
\sqrt{p_j}&\sqrt{q_j}\,e^{i\Lambda_j^+}\\
-\sqrt{q_j}\,e^{-i\Lambda_j^+}&\sqrt{p_j}
\end{array}
\right]
+
o(1),
\label{app-lz-UIplus}
\ee
where
\be
\Lambda_j^+
=
\delta_j^+
+
{\rm P}\int_{-T}^{\tau_j^++\eta}E_0^{(n)}(\tau)\,d\tau
-
{\rm P}\int_{-T}^{\tau_j^+-\eta}E_j^{(n)}(\tau)\,d\tau .
\label{app-lz-Lambda-plus-j}
\ee

Using Eq.~(\ref{app-lz-energy-diff-minus-j}), dropping all terms
linear in $\eta$, and taking $T\gg1$, one obtains
\begin{eqnarray}
{\rm P}\int_{-T}^{\tau_j^++\eta}E_0^{(n)}(\tau)\,d\tau
&-&
{\rm P}\int_{-T}^{\tau_j^+-\eta}E_j^{(n)}(\tau)\,d\tau
=
\frac{8}{3}|x|^{3/2}T^3
-
2|x|^{3/2}T
-
\frac{2}{3}|x|^{3/2}+
\\
\nonumber
&&
4|x|^{3/2}\eta^2
-
2\varepsilon_j\sqrt{|x|}
\left(T+\frac12\right)
+
\frac{
\sum_{\ell=1}^n\alpha_\ell^2+\alpha_j^2
}{2}\ln\eta.
\label{app-lz-A-diff-plus-j}
\end{eqnarray}

Combining this result with Eqs.~(\ref{app-lz-delta-plus-j}) and
(\ref{app-lz-theta-j}), the terms proportional to
$|x|^{3/2}\eta^2$, $|x|^{3/2}$, and
$\varepsilon_j\sqrt{|x|}$ cancel. Using
$T=t_0/\sqrt{|x|}$, it follows
\be
\Lambda_j^+
=
\Lambda_{j,0}^+
+
\frac12
\sum_{\substack{k=1\\k\ne j}}^n
\alpha_k^2\ln\eta
+
o(1),
\label{app-lz-Lambda-plus-split}
\ee
where
\begin{eqnarray}
\Lambda_{j,0}^+
=
\frac{8}{3}t_0^3
-
2|x|t_0
-
2\varepsilon_jt_0
+
\frac{3\pi}{4}
+
{\rm arg}\Gamma\left(i\frac{\alpha_j^2}{2}\right)
+
\varphi_j
-
\frac{3\alpha_j^2}{2}\ln2
+
\frac12
\sum_{\substack{k=1\\k\ne j}}^n
\alpha_k^2\ln|x|.
\label{app-lz-Lambda-plus-zero}
\end{eqnarray}

The final result is
\begin{widetext}
\be
U_{{\rm Int}_j^+}
=
\left[
\begin{array}{cc}
\sqrt{p_j}
&
\sqrt{q_j}
\exp\left\{
i\left[
\Lambda_{j,0}^+
+
\frac12
\sum\limits_{\substack{k=1\\k\ne j}}^n
\alpha_k^2\ln\eta
\right]
\right\}
\\
-\sqrt{q_j}
\exp\left\{
-i\left[
\Lambda_{j,0}^+
+
\frac12
\sum\limits_{\substack{k=1\\k\ne j}}^n
\alpha_k^2\ln\eta
\right]
\right\}
&
\sqrt{p_j}
\end{array}
\right]
+
o(1).
\label{app-lz-UIplus-final}
\ee
\end{widetext}

\section{Derivation of adiabatic phases of trajectories}

\subsection{The phase $\phi_{00}$}
\label{app-phi00}

Consider the evolution at a fixed large negative value of $x$. The
original Schr\"odinger equation is written in terms of the pseudo-time $t$,
with endpoints
\begin{equation}
-t_0<t<t_0.
\end{equation}
Introduce the rescaled time
\begin{equation}
t=\sqrt{|x|}\,\tau,
\qquad
T\equiv\frac{t_0}{\sqrt{|x|}}.
\label{phi00-time-rescaling}
\end{equation}
Thus, $t_0$ is the cutoff in the original pseudo-time, whereas $T$ is the
dimensionless cutoff after the rescaling. The two DO regions are centered at
$\tau=-1/2$ and $\tau=1/2$.

Away from these two regions, the adiabatic energy connected to level $0$ is
\begin{equation}
E_0^{(n)}(\tau)
=
|x|^{3/2}(4\tau^2-1)
+
\frac{2A}{4\tau^2-1}
+
o(1),
\qquad
A\equiv\sum_{k=1}^{n}\alpha_k^2.
\label{phi00-energy}
\end{equation}
The phase accumulated along this level is therefore
\begin{equation}
\phi_{00}(T;x)
=
-{\rm P}\int_{-T}^{T}E_0^{(n)}(\tau)\,d\tau,
\label{phi00-principal-value}
\end{equation}
where ${\rm P}$ indicates the principal value at both singular points.

The nonsingular part is evaluated directly:
\begin{equation}
-\int_{-T}^{T}
|x|^{3/2}(4\tau^2-1)\,d\tau
=
-\frac{8}{3}|x|^{3/2}T^3
+
2|x|^{3/2}T.
\label{phi00-regular-part}
\end{equation}
Consider the singular integral
\begin{equation}
{\rm P}\int_{-T}^{T}\frac{d\tau}{4\tau^2-1}.
\label{phi00-singular-integral}
\end{equation}
Introduce identical small intervals of width $2\eta$ around the two
singular points:
\begin{widetext}
\begin{eqnarray}
\nonumber
{\rm P}\int_{-T}^{T}\frac{d\tau}{4\tau^2-1}
=
\lim_{\eta\rightarrow0^+}
\left[
\int_{-T}^{-1/2-\eta}\frac{d\tau}{4\tau^2-1}
+
\int_{-1/2+\eta}^{1/2-\eta}
\frac{d\tau}{4\tau^2-1}
\right.
\left.
+
\int_{1/2+\eta}^{T}\frac{d\tau}{4\tau^2-1}
\right].
\label{phi00-P-definition}
\end{eqnarray}
\end{widetext}
To evaluate these integrals, it is useful to write
\begin{equation}
\frac{1}{4\tau^2-1}
=
\frac{1}{2}
\left(
\frac{1}{2\tau-1}
-
\frac{1}{2\tau+1}
\right).
\label{phi00-partial-fractions}
\end{equation}
Using
\begin{equation}
\int\frac{d\tau}{2\tau-1}
=
\frac{1}{2}\ln|2\tau-1|,
\qquad
\int\frac{d\tau}{2\tau+1}
=
\frac{1}{2}\ln|2\tau+1|,
\label{phi00-log-integrals}
\end{equation}
the contribution from the first interval is
\begin{eqnarray}
\nonumber
\int_{-T}^{-1/2-\eta}\frac{d\tau}{4\tau^2-1}
=
\frac{1}{4}\ln\left(\frac{1+\eta}{\eta}\right)
-\frac{1}{4}\ln\left(\frac{2T+1}{2T-1}\right).
\label{phi00-left-interval}
\end{eqnarray}
The contribution from the interval between the two DO regions is
\begin{equation}
\int_{-1/2+\eta}^{1/2-\eta}
\frac{d\tau}{4\tau^2-1}
=
-\frac{1}{2}\ln\left(\frac{1-\eta}{\eta}\right),
\label{phi00-middle-interval}
\end{equation}
and the contribution from the last interval is
\begin{eqnarray}
\nonumber
\int_{1/2+\eta}^{T}\frac{d\tau}{4\tau^2-1}
=
\frac{1}{4}\ln\left(\frac{2T-1}{2T+1}\right)
+\frac{1}{4}\ln\left(\frac{1+\eta}{\eta}\right).
\label{phi00-right-interval}
\end{eqnarray}
Adding the three contributions gives
\begin{eqnarray}
{\rm P}\int_{-T}^{T}\frac{d\tau}{4\tau^2-1}
=
\lim_{\eta\rightarrow0^+}
\left[
\frac{1}{2}\ln\left(\frac{2T-1}{2T+1}\right)
\right.
\left.
+
\frac{1}{2}\ln\left(\frac{1+\eta}{1-\eta}\right)
\right].
\label{phi00-eta-cancellation}
\end{eqnarray}
All terms proportional to $\ln\eta$ have canceled. Moreover,
\begin{equation}
\lim_{\eta\rightarrow0^+}
\ln\left(\frac{1+\eta}{1-\eta}\right)
=
0.
\end{equation}
Hence,
\begin{equation}
{\rm P}\int_{-T}^{T}\frac{d\tau}{4\tau^2-1}
=
\frac{1}{2}\ln\left(\frac{2T-1}{2T+1}\right).
\label{phi00-P-result}
\end{equation}

Combining Eqs.~(\ref{phi00-regular-part}) and
(\ref{phi00-P-result}), one obtains
\begin{equation}
\phi_{00}(T;x)
=
-\frac{8}{3}|x|^{3/2}T^3
+
2|x|^{3/2}T
+
A\ln\left(\frac{2T+1}{2T-1}\right)
+
o(1).
\label{phi00-T-result}
\end{equation}
Returning to the original pseudo-time cutoff
$t_0=\sqrt{|x|}\,T$, it follows
\begin{eqnarray}
\phi_{00}(t_0;x)
=
-\frac{8}{3}t_0^3
+
2|x|t_0
+
A
\ln\left(
\frac{2t_0+\sqrt{|x|}}
     {2t_0-\sqrt{|x|}}
\right)
+
o(1).
\label{phi00-t0-result}
\end{eqnarray}
The limit $t_0\rightarrow\infty$ is taken before the limit
$|x|\rightarrow\infty$. For large $t_0$,
\begin{equation}
A
\ln\left(
\frac{2t_0+\sqrt{|x|}}
     {2t_0-\sqrt{|x|}}
\right)
=
A\frac{\sqrt{|x|}}{t_0}
+
O\left(
\frac{|x|^{3/2}}{t_0^3}
\right).
\label{phi00-large-t0}
\end{equation}
This term vanishes when the pseudo-time cutoff is removed. Therefore,
retaining all contributions through zero power in $t_0$ and $x$,
\begin{equation}
\phi_{00}(t_0;x)
=
-\frac{8}{3}t_0^3
+
2|x|t_0
+
o(1).
\label{phi00-final}
\end{equation}
Thus, the phase $\phi_{00}$ has no finite $O(1)$ contribution and
is independent of $n$ to the required accuracy, so it has no upper index $(n)$.

\section{Derivation of the phase variables $\Phi_j^{(n)}$}
\label{app-phases}

Fix $j\in\{1,\ldots,n\}$ and consider the two trajectories shown in
Fig.~\ref{scheme-fig}. The reference trajectory remains on adiabatic level
$0$. The second trajectory changes from level $0$ to level $j$ in
${\rm Int}_j^-$, propagates on level $j$, and returns to level $0$ in ${\rm Int}_j^+$.
The phase of the second trajectory relative to the first one is, by
definition, $2\Phi_j^{(n)}$.

From Eqs.~(\ref{app-lz-UIminus}) and
(\ref{app-lz-UIplus}), the two off-diagonal matrix elements used
by the second trajectory are
\begin{equation}
\left(U_{{\rm Int}_j^-}\right)_{j0}
=
\sqrt{q_j}\,e^{i\Lambda_j^-},
\qquad
\left(U_{{\rm Int}_j^+}\right)_{0j}
=
\sqrt{q_j}\,e^{i\Lambda_j^+}.
\label{app-phase-two-vertices}
\end{equation}
The diagonal matrix elements followed by the reference trajectory are
real and positive. Therefore,
\begin{equation}
2\Phi_j^{(n)}
=
\Lambda_j^-+\Lambda_j^+
=
\delta_j^-+\delta_j^++{\cal S}_j,
\label{app-phase-sum}
\end{equation}
where
\begin{equation}
{\cal S}_j
=
-{\rm P}\!\int_{\tau_j^-+\eta}^{\tau_j^+-\eta}
\left[E_j^{(n)}(\tau)-E_0^{(n)}(\tau)\right]d\tau .
\label{app-phase-propagation}
\end{equation}
The symbol ${\rm P}$ means that the intervals around other crossing
points are omitted symmetrically. Terms linear in $\eta$ are dropped,
whereas $|x|^{3/2}\eta^2$ and $\ln\eta$ terms are retained until they
cancel.

The transition phases in Eqs.~(\ref{app-lz-delta-minus-j}) and
(\ref{app-lz-delta-plus-j}) give
\begin{eqnarray}
\delta_j^-+\delta_j^+
=
\frac{\pi}{2}
+2\,\arg\Gamma\!\left(i\frac{\alpha_j^2}{2}\right)
+2\theta_j
-2\Omega_j(\eta)=
\frac{\pi}{2}
+2\,\arg\Gamma\!\left(i\frac{\alpha_j^2}{2}\right)
+2\theta_j
-8|x|^{3/2}\eta^2
-2\alpha_j^2
\ln\!\left(2\sqrt{2}\,|x|^{3/4}\eta\right).
\label{app-phase-transition-sum}
\end{eqnarray}
Here $\Omega_j(\eta)$ and the phases of $g_j^\pm$ are given in
Eqs.~(\ref{app-lz-Omega-minus-j}) and
(\ref{app-lz-gamma-phases}).

Evaluate Eq.~(\ref{app-phase-propagation}) directly. To the
required order, the result following from the Hamiltonian
(\ref{ha}) can be separated as
\begin{equation}
{\cal S}_j
=
{\cal S}_{j;(0)}
+4\alpha_j^2
\int_{\tau_j^-+\eta}^{\tau_j^+-\eta}
\frac{d\tau}{4\tau^2-1-\varepsilon_j/|x|}
+\sum_{\substack{k=1\\k\ne j}}^n {\cal S}_{jk}.
\label{app-phase-propagation-parts}
\end{equation}
The leading diabatic gap gives
\begin{equation}
{\cal S}_{j;(0)}
=
-\frac{4}{3}|x|^{3/2}
-2\varepsilon_j\sqrt{|x|}
+8|x|^{3/2}\eta^2
+o(1).
\label{app-phase-leading-action}
\end{equation}
The self-coupling term in Eq.~(\ref{app-phase-propagation-parts}) is
\begin{equation}
4\alpha_j^2
\int_{\tau_j^-+\eta}^{\tau_j^+-\eta}
\frac{d\tau}
{4\tau^2-1-\varepsilon_j/|x|}
=
2\alpha_j^2\ln\eta+o(1).
\label{app-phase-self-action}
\end{equation}

For $k\ne j$, the contribution of level $k$ is
\begin{equation}
{\cal S}_{jk}
=
2\alpha_k^2\,{\rm P}\!\int_{\tau_j^-}^{\tau_j^+}
\frac{d\tau}
{4\tau^2-1-\varepsilon_k/|x|}
+o(1).
\label{app-phase-pair-action}
\end{equation}
Then
\begin{eqnarray}
{\rm P}\!\int_{\tau_j^-}^{\tau_j^+}
\frac{d\tau}{4[\tau^2-(\tau_k^+)^2]}
=
\frac{1}{4\tau_k^+}
\ln\left|\frac{\tau_j^+-\tau_k^+}
{\tau_j^++\tau_k^+}\right|
=
\frac{1}{2}
\ln\!\left(
\frac{|\varepsilon_j-\varepsilon_k|}{4|x|}
\right)
+o(1).
\label{app-phase-pair-integral}
\end{eqnarray}
Consequently,
\begin{equation}
{\cal S}_{jk}
=
\alpha_k^2
\ln\!\left(
\frac{|\varepsilon_j-\varepsilon_k|}{4|x|}
\right)
+o(1).
\label{app-phase-pair-result}
\end{equation}
If the poles $\tau_k^\pm$ lie inside $(\tau_j^-,\tau_j^+)$, the logarithmic
terms at the two sides of each pole cancel. Thus
Eq.~(\ref{app-phase-pair-integral}) is independent of the auxiliary
cutoffs $\eta_k$.

Combining Eqs.~(\ref{app-phase-leading-action}),
(\ref{app-phase-self-action}), and (\ref{app-phase-pair-result}) gives
\begin{eqnarray}
{\cal S}_j
=
-\frac{4}{3}|x|^{3/2}
-2\varepsilon_j\sqrt{|x|}
+8|x|^{3/2}\eta^2
+2\alpha_j^2\ln\eta
+\sum_{\substack{k=1\\k\ne j}}^n
\alpha_k^2
\ln\!\left(
\frac{|\varepsilon_j-\varepsilon_k|}{4|x|}
\right)
+o(1).
\label{app-phase-full-action}
\end{eqnarray}
The $8|x|^{3/2}\eta^2$ and $2\alpha_j^2\ln\eta$ terms in
Eq.~(\ref{app-phase-full-action}) cancel the corresponding terms in
Eq.~(\ref{app-phase-transition-sum}), so Eq.~(\ref{app-phase-sum}) is
independent of $\eta$.

Finally, Eq.~(\ref{app-lz-theta-j}) gives
\begin{eqnarray}
2\theta_j
=
\frac{4}{3}|x|^{3/2}
+2\varepsilon_j\sqrt{|x|}
+\left(
\frac{3\alpha_j^2}{2}
+\sum_{\substack{k=1\\k\ne j}}^n\alpha_k^2
\right)\ln|x|
+2\varphi_j
+o(1).
\label{app-phase-theta-expanded}
\end{eqnarray}
Its first two terms cancel the first two terms in
Eq.~(\ref{app-phase-full-action}). The logarithms of $|x|$ cancel
between Eqs.~(\ref{app-phase-transition-sum}),
(\ref{app-phase-theta-expanded}), and
(\ref{app-phase-full-action}), leaving the constant
$-3\alpha_j^2\ln2$. Therefore,
\begin{equation}
2\Phi_j^{(n)}
=
\frac{\pi}{2}
+2\,\arg\Gamma\!\left(i\frac{\alpha_j^2}{2}\right)
+2\varphi_j
-3\alpha_j^2\ln2
+\sum_{\substack{k=1\\k\ne j}}^n
\alpha_k^2
\ln\!\left(
\frac{|\varepsilon_j-\varepsilon_k|}{4}
\right).
\label{app-phase-final-doubled}
\end{equation}
Dividing by two gives Eq.~(\ref{app-phase-final}) in the main text.

\subsection{Derivation of the phases $\phi_{k0}^{(n)}$}
\label{app-phik0}

The following derives the common phase $\phi_{k0}^{(n)}$ in the amplitude
$U_{k0}^{(n)}$, where $k\geq1$. Consider first the trajectory that
changes from level $0$ to level $k$ in ${\rm Int}_k^-$, propagates on level
$k$ between the two crossing regions, and returns to level $0$ in
${\rm Int}_k^+$. Its phase relative to the trajectory that always remains on
level $0$ is $2\Phi_k^{(n)}$.

The trajectory that ends on level $k$ follows the same path up to
${\rm Int}_k^+$, but it passes through this interval on level $k$ instead of
turning from level $k$ to level $0$. After ${\rm Int}_k^+$, it propagates on
level $k$ rather than on level $0$.

Work directly with the evolution matrix in the adiabatic basis.
According to Eq.~(\ref{app-lz-UIplus-final}), its relevant matrix
elements are
\be
\left(U_{{\rm Int}_k^+}\right)_{0k}
=
\sqrt{q_k}\,e^{i\Lambda_k^+},
\qquad
\left(U_{{\rm Int}_k^+}\right)_{kk}
=
\sqrt{p_k}.
\label{app-phik0-right-elements}
\ee
Thus, replacing the transition from level $k$ to level $0$ by passing
on level $k$ changes the phase by $-\Lambda_k^+$.

At the final time, the adiabatic and diabatic states are related by
\be
|m_a(T)\ra
=
\exp\left[
-i\,{\rm P}\int_{-T}^{T}E_m^{(n)}(\tau)\,d\tau
\right]|m\ra,
\qquad m=0,k.
\label{app-phik0-final-basis}
\ee
It follows that
\be
\phi_{k0}^{(n)}
=
-\Lambda_k^+
-
{\rm P}\int_{-T}^{T}E_k^{(n)}(\tau)\,d\tau .
\label{app-phik0-definition}
\ee
Since
\be
\phi_{00}
=
-{\rm P}\int_{-T}^{T}E_0^{(n)}(\tau)\,d\tau,
\ee
Eq.~(\ref{app-phik0-definition}) can equivalently be written as
\be
\phi_{k0}^{(n)}
-
\phi_{00}
=
-\Lambda_k^+
+
{\rm P}\int_{-T}^{T}\left[E_0^{(n)}(\tau)-E_k^{(n)}(\tau)\right]d\tau .
\label{app-phik0-relative}
\ee

Evaluate the last two terms in
Eq.~(\ref{app-phik0-relative}). The adiabatic phases must be
calculated using the same excluded intervals of width $2\eta$ that
were used in the derivation of $U_{{\rm Int}_k^+}$. The leading part of
the energy difference gives
\begin{eqnarray}
{\rm P}\int_{-T}^{T}
\left[
E_0^{(n)}(\tau)-E_k^{(n)}(\tau)
\right]
d\tau
=
\frac{16}{3}|x|^{3/2}T^3
-
4|x|^{3/2}T
-
4\varepsilon_k\sqrt{|x|}\,T
+
o(1).
\label{app-phik0-leading-action}
\end{eqnarray}

The logarithmic terms must be evaluated while retaining the
displacements of the crossing points,
\be
\tau_j^\pm
=
\pm\left(
\frac12+\frac{\varepsilon_j}{4|x|}
\right)
+
o(|x|^{-1}).
\label{app-phik0-crossings}
\ee
For every $j\ne k$, the corresponding contribution to the difference
of the adiabatic phases is
\be
\left[
{\rm P}\int_{-T}^{T}
\left(E_0^{(n)}(\tau)-E_k^{(n)}(\tau)\right)d\tau
\right]_j
=
\frac{\alpha_j^2}{2}
\ln\left(
\frac{4|x|\eta}
{|\varepsilon_k-\varepsilon_j|}
\right)
+
o(1).
\label{app-phik0-pair-action}
\ee
The factor $|\varepsilon_k-\varepsilon_j|/(4|x|)$ follows from the
separation of the crossing points,
\be
|\tau_k^+-\tau_j^+|
=
\frac{|\varepsilon_k-\varepsilon_j|}{4|x|}
+
o(|x|^{-1}).
\label{app-phik0-crossing-distance}
\ee
If the integration range contains the crossing point $\tau_j^+$,
the logarithmic terms at $\tau_j^+-\eta$ and
$\tau_j^++\eta$ are combined using the principal-value prescription.
The remaining cutoff dependence is the term
$(\alpha_j^2/2)\ln\eta$ in
Eq.~(\ref{app-phik0-pair-action}).

For $j=k$, the logarithmic contributions from the two sides of the
crossing enter with opposite signs. Hence, their sum contains no
term proportional to $\alpha_k^2\ln\eta$. Combining all
contributions, one obtains
\begin{eqnarray}
{\rm P}\int_{-T}^{T}[E_0^{(n)}(\tau)-E_k^{(n)}(\tau)]d\tau
=
\frac{16}{3}|x|^{3/2}T^3
-
4|x|^{3/2}T
-
4\varepsilon_k\sqrt{|x|}\,T
+
\frac12
\sum_{\substack{j=1\\j\ne k}}^n
\alpha_j^2
\ln\left(
\frac{4|x|\eta}
{|\varepsilon_k-\varepsilon_j|}
\right)
+
o(1).
\label{app-phik0-full-action}
\end{eqnarray}

The phase $\Lambda_k^+$ was derived in
Eqs.~(\ref{app-lz-Lambda-plus-split}) and
(\ref{app-lz-Lambda-plus-zero}). Using
$T=t_0/\sqrt{|x|}$, it is
\begin{eqnarray}
\Lambda_k^+
=
\frac{8}{3}t_0^3
-
2|x|t_0
-
2\varepsilon_kt_0
+
\frac{3\pi}{4}
+
{\rm arg}\Gamma\left(i\frac{\alpha_k^2}{2}\right)
+
\varphi_k
-
\frac{3\alpha_k^2}{2}\ln2
+
\frac12
\sum_{\substack{j=1\\j\ne k}}^n
\alpha_j^2\ln|x|
+
\frac12
\sum_{\substack{j=1\\j\ne k}}^n
\alpha_j^2\ln\eta
+
o(1).
\label{app-phik0-Lambda-plus}
\end{eqnarray}

The $\eta$-dependent part of
${\rm P}\int_{-T}^{T}[E_0^{(n)}(\tau)-E_k^{(n)}(\tau)]d\tau$ is
\be
\frac12
\sum_{\substack{j=1\\j\ne k}}^n
\alpha_j^2\ln\eta,
\ee
which is identical to the $\eta$-dependent part of $\Lambda_k^+$.
Since these two quantities enter Eq.~(\ref{app-phik0-relative}) with
opposite signs, all terms proportional to $\ln\eta$ cancel.

The terms proportional to $\ln|x|$ cancel in the same way. 
Substituting Eqs.~(\ref{app-phik0-full-action}) and
(\ref{app-phik0-Lambda-plus}) into
Eq.~(\ref{app-phik0-relative})  leads to
\begin{eqnarray}
\phi_{k0}^{(n)}-\phi_{00}
=
\frac{8}{3}t_0^3
-
2|x|t_0
-
2\varepsilon_kt_0
-
\frac{3\pi}{4}
-
{\rm arg}\Gamma\left(i\frac{\alpha_k^2}{2}\right)
-
\varphi_k
+
\frac{3\alpha_k^2}{2}\ln2
+
\frac12
\sum_{\substack{j=1\\j\ne k}}^n
\alpha_j^2
\ln\left(
\frac{4}{|\varepsilon_k-\varepsilon_j|}
\right).
\label{app-phik0-relative-result}
\end{eqnarray}

Using the expression for $\Phi_k^{(n)}$ in
Eq.~(\ref{app-phase-final}), the finite part of
Eq.~(\ref{app-phik0-relative-result}) is
$-\Phi_k^{(n)}-\pi/2$. Therefore,
\be
\phi_{k0}^{(n)}-\phi_{00}
=
\frac{8}{3}t_0^3
-
2|x|t_0
-
2\varepsilon_kt_0
-
\Phi_k^{(n)}
-
\frac{\pi}{2}.
\label{app-phik0-relative-Phi}
\ee
Finally, using Eq.~(\ref{phi00-final}),
one obtains
\be
\phi_{k0}^{(n)}
=
-2\varepsilon_kt_0
-
\Phi_k^{(n)}
-
\frac{\pi}{2}.
\label{app-phik0-final}
\ee

\subsection{Origin of the  minus sign in Eq.~(\ref{ind-Uk0-general})}
\label{app-phik0-minus-sign}

The two groups of trajectories contributing to $U_{k0}^{(n)}$ use
different matrix elements of the evolution matrix in ${\rm Int}_k^+$.
A trajectory that changes from level $0$ to level $k$ in ${\rm Int}_k^-$
and then passes through ${\rm Int}_k^+$ uses
\be
\left(U_{{\rm Int}_k^-}\right)_{k0}
=
\sqrt{q_k}\,e^{i\Lambda_k^-},
\qquad
\left(U_{{\rm Int}_k^+}\right)_{kk}
=
\sqrt{p_k}.
\label{app-phik0-first-path-elements}
\ee
Using
\be
\Lambda_k^-+\Lambda_k^+
=
2\Phi_k^{(n)},
\label{app-phik0-Lambda-Phi}
\ee
the phase of this  trajectory is proportional to
\be
e^{i\phi_{k0}^{(n)}}e^{2i\Phi_k^{(n)}}.
\label{app-phik0-first-path-phase}
\ee

The second group reaches ${\rm Int}_k^+$ on level $0$ and changes there to
level $k$. According to Eq.~(\ref{app-lz-UIplus-final}), this
transition uses the matrix element
\be
\left(U_{{\rm Int}_k^+}\right)_{k0}
=
-\sqrt{q_k}\,e^{-i\Lambda_k^+}.
\label{app-phik0-negative-element}
\ee
The minus sign in this matrix element is therefore the origin of the
relative minus sign between the amplitudes of two types of trajectories: turning level when initially going down or up. 
After extracting the common phase
$e^{i\phi_{k0}^{(n)}}$ and the  Landau--Zener probability
factors, the two contributions combine as in Eq.~(\ref{ind-Uk0-general}).

\section{Derivation of $U_{00}^{(n)}$ for $x\rightarrow+\infty$}
\label{U00-plus-sec}

Consider the amplitude $U_{00}^{(n)}$ for the evolution at large
positive $x$. In the radial basis introduced in
Section~\ref{plusx-radial-sec}, the branch that coincides with the
radial state $|R\ra$ away from $t=0$ is the $n$-th adiabatic level.
Near $t=0$, the $0$-th and $n$-th adiabatic levels approach each other
and form an avoided crossing. All angular levels are separated from
these two levels by energies of order $x$. Their couplings to $|0\ra$
are of order one and, therefore, their effect on this avoided crossing
vanishes as $x\rightarrow+\infty$. Consequently, the calculation of
$U_{00}^{(n)}$ near this crossing reduces asymptotically to a two-state
Landau--Zener problem independently of $n$.

\subsection{Landau--Zener transition at the crossing of levels $0$ and $1$}
\label{LZ01}
 $2\times2$ block of the Hamiltonian (\ref{app-ra-H-plus-expanded})  for states $|0\ra$ and $|R\ra$ 
is
\be
H_{0}
=
\left[
\begin{array}{cc}
4t^2+x-2R^2
&
-4tR-2iR'
\\
-4tR+2iR'
&
-4t^2-x+2R^2
\end{array}
\right].
\label{U00-radial-H}
\ee
Near the crossing point $t=0$, introduce the states
\be
|\mp\ra
\equiv
\frac{|0\ra\mp|R\ra}{\sqrt{2}}.
\label{U00-pm-basis}
\ee
Keeping the terms that remain relevant as $x\rightarrow+\infty$, the linear-time approximation  in this basis is given by
\be
H_{\rm LZ}
=
\left[
\begin{array}{cc}
4t\sqrt{x/2}
&
2(2x)^{1/4}\rho
e^{-i[\pi+\theta_1(x)]}
\\
2(2x)^{1/4}\rho
e^{i[\pi+\theta_1(x)]}
&
-4t\sqrt{x/2}
\end{array}
\right].
\label{U00-LZ-H}
\ee
Thus, the parameters of this Landau--Zener problem are
\be
\beta=4\sqrt{\frac{x}{2}},
\qquad
|\gamma_1|=2(2x)^{1/4}\rho,
\qquad
\frac{|\gamma_1|^2}{2\beta}=\rho^2.
\label{U00-LZ-parameters}
\ee
The absolute value of the transition amplitude that connects the
positive-energy adiabatic state before the crossing to the
positive-energy adiabatic state after the crossing is therefore
\be
\left|U_{00}^{(n)}\right|
=
\sqrt{1-e^{-2\pi\rho^2}}.
\label{U00-plus-absolute}
\ee

\subsection{Phase of $U_{00}^{(n)}$}
\label{app-phase00p}
To calculate the phase, separate the evolution into the two
adiabatic intervals $(-t_0,-T)$ and $(T,t_0)$ and the nonadiabatic
interval $(-T,T)$ containing the crossing point of the $0$-th and
$n$-th adiabatic levels, where
\be
1\ll T\ll\sqrt{x}.
\label{U00-T-condition}
\ee
The positive adiabatic energy of the radial block is
\be
E(t)
=
\left[
\left(4t^2+x-2R^2\right)^2
+(4tR)^2
+4(R')^2
\right]^{1/2}.
\label{U00-E}
\ee
Averaging the rapidly oscillating terms gives
\be
E(t)
=
E_{0;(0)}(t)
+
\frac{2\rho^2\sqrt{2x}}{E_{0;(0)}(t)}
+\ldots,
\label{U00-E-expansion}
\ee
where
\be
E_{0;(0)}(t)
\equiv
\sqrt{(4t^2)^2+8xt^2}
=
4|t|\sqrt{t^2+\frac{x}{2}}.
\label{U00-E0}
\ee

The contribution of $E_{0;(0)}(t)$ to the phase accumulated in the two
adiabatic intervals is
\begin{eqnarray}
-2\int_T^{t_0}E_{0;(0)}(t)\,dt
=
-\frac{8t_0^3}{3}
-2xt_0
+\frac{2\sqrt{2}}{3}x^{3/2}
+2\sqrt{2x}\,T^2
+o(1).
\label{U00-E0-integral}
\end{eqnarray}
The correction proportional to $\rho^2$ gives
\begin{eqnarray}
-2\int_T^{\infty}
\frac{2\rho^2\sqrt{2x}}{E_{0;(0)}(t)}\,dt
=
-\rho^2
\left[
\ln2+\ln x-2\ln T
\right]+o(1).
\label{U00-rho-integral}
\end{eqnarray}

The asymptotic dynamical phases of the Landau--Zener solutions inside
the interval containing the avoideed crossing contribution
\be
\phi_{\rm nad}
=
-2\sqrt{2x}\,T^2
-\rho^2
\left[
\ln(4\sqrt{2})
+\frac12\ln x
+2\ln T
\right].
\label{U00-nad-phase}
\ee
With the phase convention of Eq.~(\ref{U00-LZ-H}), the phase of the
Landau--Zener transition amplitude is
\be
\phi_{\rm LZ}
=
-\frac{3\pi}{4}
+{\rm arg}\Gamma(i\rho^2)
-\theta_1(x).
\label{U00-LZ-phase}
\ee
Finally, the evolution along the two vertical parts of the deformed
integration path contributes the phase $4xt_0$. Hence,
\begin{eqnarray}
{\rm arg}\,U_{00}^{(n)}
=
4xt_0
-2\int_T^{t_0}E_{0;(0)}(t)\,dt
-2\int_T^\infty
\frac{2\rho^2\sqrt{2x}}{E_{0;(0)}(t)}\,dt
+\phi_{\rm nad}
+\phi_{\rm LZ}.
\label{U00-total-phase-before}
\end{eqnarray}
All terms that depend on $T$, $x^{3/2}$, and $\ln x$ cancel:
\be
-\rho^2\ln x
-\frac{\rho^2}{2}\ln x
+\frac{3\rho^2}{2}\ln x
=0,
\label{U00-logx-cancel}
\ee
where the last term follows from Eq.~(\ref{theta1p}). The remaining
logarithmic constant is
\be
-\rho^2\ln2
-\rho^2\ln(4\sqrt{2})
=
-\frac{7\rho^2}{2}\ln2.
\label{U00-log2-result}
\ee
Consequently,
\begin{eqnarray}
U_{00}^{(n)}
=
\sqrt{1-e^{-2\pi\rho^2}}
\times
\exp\left\{
i\left[
-\frac{8t_0^3}{3}
+2xt_0
-\frac{7\rho^2}{2}\ln2
-\frac{3\pi}{4}
+{\rm arg}\Gamma(i\rho^2)
-\phi_1
\right]
\right\}.
\label{U00-plus-final}
\end{eqnarray}

\section{Derivation of the phases $\phi_m$ for $m=2,\ldots,n$}
\label{app-plusx-phim}

Consider the asymptotic solution initially as
\begin{eqnarray}
u_m(x)
&=&
 A_m\cos\theta_m(x)+o(1),
\qquad m=2,\ldots,n,
\label{app-phim-um-initial}\\
\theta_m(x)
&=&
\sqrt{\varepsilon_m}\,x+\kappa_m\ln x+\phi_m+o(1).
\label{app-phim-theta-initial}
\end{eqnarray}
The coefficient $\kappa_m$ and the constant $\phi_m$ follow
by comparing the phases of $U_{m0}^{(n)}$ calculated at
$x\rightarrow\pm\infty$. At positive $x$, the contributing
trajectory passes through the central crossing, stays on
the radial state until $t=t_m$, and then turns to state $m$.

\subsection{Effective crossing and Landau--Zener phase}
\label{app-Hm-eff}

The unperturbed energies are
\begin{eqnarray}
E_R(t)=-4t\sqrt{t^2+\frac{x}{2}},
\quad
E_m(t)=-4t^2-x+2\varepsilon_m.
\label{app-phim-energies}
\end{eqnarray}
Their crossing condition,
\begin{equation}
4t_m\sqrt{t_m^2+\frac{x}{2}}
=
4t_m^2+x-2\varepsilon_m,
\label{app-phim-crossing-equation}
\end{equation}
gives
\begin{equation}
t_m
=
\frac{x-2\varepsilon_m}{4\sqrt{\varepsilon_m}}
=
\frac{x}{4\sqrt{\varepsilon_m}}
-\frac{\sqrt{\varepsilon_m}}{2}.
\label{app-phim-tm}
\end{equation}

After eliminating the remote zeroth state and subtracting
a scalar term, the effective Hamiltonian near $t=t_m$ in
the ordered basis $(|R\ra,|m\ra)$ is
\begin{equation}
H_m^+(s)=
\left[
\begin{array}{cc}
-\beta_m s&\gamma_m^*\\
\gamma_m&\beta_m s
\end{array}
\right],
\qquad s=t-t_m.
\label{app-phim-LZ-H}
\end{equation}
Since
\begin{equation}
\left.\frac{d(E_R-E_m)}{dt}\right|_{t=t_m}
=
-\frac{16\varepsilon_m^{3/2}}{x+2\varepsilon_m},
\label{app-phim-relative-slope}
\end{equation}
the radial state is the downward-sloping first state, and
\begin{equation}
\beta_m
=
\frac{8\varepsilon_m^{3/2}}{x+2\varepsilon_m}
=
\frac{8\varepsilon_m^{3/2}}{x}+O(x^{-2}).
\label{app-phim-beta}
\end{equation}
The coupling $\gamma_m=(H_m^+)_{mR}$ contains the direct
matrix element and the virtual transition through state $0$:
\begin{equation}
\gamma_m
=
\frac{2\sqrt{2}\varepsilon_m u_m}{\sqrt{x}}
+
\frac{2\sqrt{2}i\sqrt{\varepsilon_m}\,u_m'}{\sqrt{x}}
+o(x^{-1/2}).
\label{app-phim-gm}
\end{equation}
Substituting Eq.~(\ref{app-phim-um-initial}) and its derivative
gives
\begin{eqnarray}
\gamma_m
=
\frac{2\sqrt{2}\varepsilon_m A_m}{\sqrt{x}}
e^{-i\theta_m},\quad
\arg\gamma_m&=&-\theta_m, \quad
\frac{|\gamma_m|^2}{2\beta_m}
=
\frac{A_m^2\sqrt{\varepsilon_m}}{2}+o(1)
=
I_m.
\label{app-phim-Im}
\end{eqnarray}
There is no factor $\sigma$ in this coupling phase. Indeed, the
direct matrix element is $2\varepsilon_m u_m/R$ to leading order,
while $H_{m0}=2i\sigma L_m/R+o(1)=2iu_m'+o(1)$ in the chosen 
basis. Eliminating state $|0\rangle $ adds $iRu_m'/t_m$ to the coupling.
Their sum gives Eq.~(\ref{app-phim-gm}) without a sign factor.

Let the boundaries of the Landau--Zener interval be at
$s=\pm T_m$, where
\begin{equation}
\max\left(\beta_m^{-1/2},\frac{|\gamma_m|}{\beta_m}\right)
\ll T_m\ll t_m.
\label{app-phim-Tm-condition}
\end{equation}
The interval must also be small enough that curvature
corrections vanish; for example, $T_m=x^a$ with
$1/2<a<2/3$ satisfies these requirements at fixed parameters.

Interchanging the two indices in Eq.~(\ref{app-phim-LZ-H})
maps it to Eq.~(\ref{lz1}) with $g=\gamma_m$. The required
transition amplitude is then $S_{12}$ in
Eq.~(\ref{lz-scatt-mt}), with phase
\begin{equation}
\frac{\pi}{4}+\arg\Gamma(iI_m)+\arg\gamma_m.
\label{app-phim-Stokes-phase}
\end{equation}
Restoring the asymptotic phases at the interval boundaries,
the local contribution is
\begin{equation}
\frac{\pi}{4}+\arg\Gamma(iI_m)-\theta_m
-\beta_mT_m^2-I_m\ln(2\beta_mT_m^2)+o(1).
\label{app-phim-local-phase}
\end{equation}
All other crossings encountered by this trajectory contribute
real survival amplitudes in the same phase convention.

\subsection{Adiabatic phase}
\label{app-phik0-xp}
\label{app-angular-phase-correction}

First consider the unperturbed energies in
Eq.~(\ref{app-phim-energies}). Define
\begin{eqnarray}
{\cal D}_m(t_0;x)
\equiv
\int_{t_m}^{t_0}[E_m(t)-E_R(t)]\,dt
=
\left[
\frac{4}{3}
\left(\left(t^2+\frac{x}{2}\right)^{3/2}-t^3\right)
-(x-2\varepsilon_m)t
\right]_{t_m}^{t_0}.
\label{app-phim-D-definition}
\end{eqnarray}
Using
\begin{equation}
\sqrt{t_m^2+\frac{x}{2}}=t_m+\sqrt{\varepsilon_m},
\label{app-phim-root-tm}
\end{equation}
and taking the large-$t_0$ limit first gives
\begin{equation}
{\cal D}_m(t_0;x)
=
2\varepsilon_m t_0-\sqrt{\varepsilon_m}\,x
+\frac{2}{3}\varepsilon_m^{3/2}+o(1).
\label{app-phim-D-final}
\end{equation}
The corresponding contribution to the phase is $-{\cal D}_m$.

The adiabatic phase also contains energy shifts of order
$1/x$. They cannot be neglected because the relevant
time intervals are of order $x$. For $t$ of order $x$,
\begin{equation}
E_R(t)-E_k(t)
=
\frac{x^2}{8t^2}-2\varepsilon_k+O(x^{-1}).
\label{angular-corr-gap}
\end{equation}
The direct diagonal contribution from state $k$ to the
radial energy in the second order is
\begin{equation}
\delta E_{Rk;(2)}^{\rm diag}
=
\frac{2\varepsilon_k u_k^2}{R^2}
=
\frac{4\varepsilon_k u_k^2}{x}+o(x^{-1}).
\label{angular-corr-diagonal}
\end{equation}
The effective coupling, including the virtual transition
through state $0$, is
\begin{equation}
\gamma_k(t)
=
\frac{2\sqrt{2}\varepsilon_k u_k}{\sqrt{x}}
+
\frac{i\sqrt{x}\,u_k'}{\sqrt{2}\,t}
+o(x^{-1/2}).
\label{angular-corr-coupling}
\end{equation}
At $t=t_k$, $\gamma_k(t_k)$ reduces to the crossing coupling $\gamma_k$ in
Eq.~(\ref{app-phim-gm}). Away from the crossing, the total
energy shift is therefore
\begin{eqnarray}
\delta E_{Rk;(2)}
&=&
\frac{4\varepsilon_k u_k^2}{x}
+
\frac{
8\varepsilon_k^2u_k^2/x+x(u_k')^2/(2t^2)
}{
x^2/(8t^2)-2\varepsilon_k
}+o(x^{-1})
=
\frac{4[\varepsilon_k u_k^2+(u_k')^2]}
{x(1-16\varepsilon_k t^2/x^2)}
+o(x^{-1}).
\label{angular-corr-energy}
\end{eqnarray}
Both terms in the first line are necessary. Since
\begin{equation}
\varepsilon_k u_k^2+(u_k')^2
=
\varepsilon_k A_k^2+o(1)
=
2I_k\sqrt{\varepsilon_k}+o(1),
\label{angular-corr-invariant}
\end{equation}
the oscillation phase cancels without averaging, and
\begin{equation}
\delta E_{Rk;(2)}(t)
=
\frac{8I_k\sqrt{\varepsilon_k}}
{x(1-16\varepsilon_k t^2/x^2)}
+o(x^{-1}).
\label{angular-corr-energy-final}
\end{equation}
The corresponding shift of state $k$ is
$-\delta E_{Rk;(2)}$ to this accuracy.

For $k=m$, the trajectory accumulates the radial energy
shift before the transition and the opposite shift after it.
Excluding the Landau--Zener interval gives
\begin{eqnarray}
-\int_0^{t_m-T_m}\delta E_{Rm;(2)}\,dt
+\int_{t_m+T_m}^{\infty}\delta E_{Rm;(2)}\,dt
=-2I_m\ln\left(\frac{2t_m}{T_m}\right)+o(1).
\label{app-phim-outer-log}
\end{eqnarray}
Combining this with the local logarithmic term yields
\begin{eqnarray}
-I_m\ln(2\beta_mT_m^2)
-2I_m\ln\left(\frac{2t_m}{T_m}\right)
=-I_m\ln(8\beta_mt_m^2),
\label{app-phim-log-combination}
\end{eqnarray}
where
$
8\beta_mt_m^2
=
4\sqrt{\varepsilon_m}\,x[1+o(1)].
$ Thus the dependence on $T_m$ cancels.

For $k\ne m$, the additional phase is accumulated while
the trajectory stays on the radial state:
\begin{eqnarray}
\Delta_k\phi_{m0}^+
=
-{\rm P}\int_0^{t_m}\delta E_{Rk;(2)}(t)\,dt
=
-I_k\ln\left|
\frac{1+4\sqrt{\varepsilon_k}t_m/x}
     {1-4\sqrt{\varepsilon_k}t_m/x}
\right|
=
I_k\ln
\frac{|\sqrt{\varepsilon_m}-\sqrt{\varepsilon_k}|}
{\sqrt{\varepsilon_m}+\sqrt{\varepsilon_k}}.
\label{angular-corr-integral}
\end{eqnarray}
The principal value applies when the trajectory passes
the $k$th crossing, whose local solution is matched to
the real Landau--Zener survival amplitude. Extending the
outer expressions to $t=0$ changes these integrals only
by a vanishing contribution from the central region.

After the transition to state $m$, its couplings to the
other states $k>1$ are of order $1/x$. For fixed distinct
$\varepsilon_k$, their second-order energy shifts are
of order $1/x^2$ and produce no further finite phase.

\subsection{Total phase and cancellation of $\ln x$}
\label{app-log}

The quadratic boundary term in
Eq.~(\ref{app-phim-local-phase}) cancels the corresponding
term from the adiabatic intervals. Adding the remaining
contributions gives
\begin{eqnarray}
\phi_{m0}^+
=
-2\varepsilon_m t_0
+\frac{\pi}{4}+\arg\Gamma(iI_m)
-\theta_m(x)+\sqrt{\varepsilon_m}\,x
-\frac{2}{3}\varepsilon_m^{3/2}
-I_m\ln(4\sqrt{\varepsilon_m}\,x)
+\sum_{\substack{k=2\\k\ne m}}^n
I_k\ln
\frac{|\sqrt{\varepsilon_m}-\sqrt{\varepsilon_k}|}
{\sqrt{\varepsilon_m}+\sqrt{\varepsilon_k}}.
\label{app-phim-Uplus-before}
\end{eqnarray}
Substitution of Eq.~(\ref{app-phim-theta-initial}) yields
\begin{eqnarray}
\phi_{m0}^+
=
-2\varepsilon_m t_0
+\frac{\pi}{4}+\arg\Gamma(iI_m)-\phi_m
-\frac{2}{3}\varepsilon_m^{3/2}
-(\kappa_m+I_m)\ln x
-I_m\ln(4\sqrt{\varepsilon_m})
+\sum_{\substack{k=2\\k\ne m}}^n
I_k\ln
\frac{|\sqrt{\varepsilon_m}-\sqrt{\varepsilon_k}|}
{\sqrt{\varepsilon_m}+\sqrt{\varepsilon_k}}.
\label{app-phim-Uplus-expanded}
\end{eqnarray}

The amplitude calculated at $x\rightarrow-\infty$ has
no residual term proportional to $\ln|x|$. Path independence
therefore requires
\begin{equation}
\kappa_m=-I_m,
\label{app-phim-kappa}
\end{equation}
which gives
\begin{equation}
\theta_m(x)
=
\sqrt{\varepsilon_m}\,x-I_m\ln x+\phi_m+o(1).
\label{app-phim-theta-final}
\end{equation}
Matching the constant terms then gives
Eq.~(\ref{phi2-fin2}). In particular, the sum in
Eq.~(\ref{app-phim-Uplus-expanded}) enters the connection
formula for $\phi_m$ with the same sign because the
amplitude phase contains $-\phi_m$. This sum is empty
for $n=2$ and first contributes at $n=3$.

\section{Proof of Eq.~(\ref{actions-W-minus2})}
\label{app-proof-actions-W-minus2}

The proof uses the definition
\[
C_j^{(n)}
=
\prod_{k=1}^{j}p_k
+
\sum_{m=1}^{j}
q_m
\left(
\prod_{s=m+1}^{j}p_s
\right)
e^{2i\Phi_m^{(n)}} ,
\qquad
C_0^{(n)}=1 ,
\]
a  relation
\be
C_j^{(n)}
=
p_jC_{j-1}^{(n)}
+
q_j e^{2i\Phi_j^{(n)}} ,
\label{actions-C-recursion}
\ee
and  that $p_j+q_j=1$. Then,
\begin{eqnarray}
\nonumber
1-\left|C_j^{(n)}\right|^2
=
1-
\left|
p_jC_{j-1}^{(n)}
+
q_j e^{2i\Phi_j^{(n)}}
\right|^2
=
p_j
\left(
1-\left|C_{j-1}^{(n)}\right|^2
\right)
+
p_jq_j
\left|
e^{2i\Phi_j^{(n)}}-C_{j-1}^{(n)}
\right|^2 .
\label{actions-C-identity}
\end{eqnarray}

Multiplication of both sides in Eq.~(\ref{actions-C-identity}) by $\prod_{k=j+1}^{n}p_k$ gives
\begin{eqnarray}
\nonumber
\left(
\prod_{k=j+1}^{n}p_k
\right)
\left(
1-\left|C_j^{(n)}\right|^2
\right)
=
\left(
\prod_{k=j}^{n}p_k
\right)
\left(
1-\left|C_{j-1}^{(n)}\right|^2
\right)
+
p_jq_j
\left(
\prod_{k=j+1}^{n}p_k
\right)
\left|
e^{2i\Phi_j^{(n)}}-C_{j-1}^{(n)}
\right|^2 .
\label{actions-C-identity-mult}
\end{eqnarray}
The second term on the right-hand side is
$\left|U_{j0}^{(n)}\right|^2$, because the transition amplitude to
level $j$ has the form
\[
U_{j0}^{(n)}
=
e^{i\phi_{j0}^{(n)}}
\sqrt{
p_jq_j
\prod_{k=j+1}^{n}p_k
}
\left(
e^{2i\Phi_j^{(n)}}-C_{j-1}^{(n)}
\right).
\]
Hence,
\[
\left(
\prod_{k=j+1}^{n}p_k
\right)
\left(
1-\left|C_j^{(n)}\right|^2
\right)
=
\left(
\prod_{k=j}^{n}p_k
\right)
\left(
1-\left|C_{j-1}^{(n)}\right|^2
\right)
+
\left|U_{j0}^{(n)}\right|^2 .
\]
This is the recursion
\[
W_j=W_{j-1}+\left|U_{j0}^{(n)}\right|^2,
\]
provided that
\[
W_j=
\left(
\prod_{k=j+1}^{n}p_k
\right)
\left(
1-\left|C_j^{(n)}\right|^2
\right).
\]
The initial value is correct because $C_0^{(n)}=1$, and hence
\[
W_0=
\left(
\prod_{k=1}^{n}p_k
\right)
\left(
1-\left|C_0^{(n)}\right|^2
\right)
=0.
\]
Therefore, by induction,
\[
W_j
=
\sum_{m=1}^{j}\left|U_{m0}^{(n)}\right|^2
=
\left(
\prod_{k=j+1}^{n}p_k
\right)
\left(
1-\left|C_j^{(n)}\right|^2
\right),
\]
which proves Eq.~(\ref{actions-W-minus2}).

\begin{acknowledgements}
Author thanks Prof.~Alexander~Its for useful discussions. This work was carried out under the
auspices of the U.S. DoE through the Los Alamos National Laboratory, operated by Triad
National Security, LLC (Contract No. 892333218NCA000001). 
\end{acknowledgements}

\bibliography{ref2}

\end{document}